\documentclass[apsprl,showpacs,floatfix,superscriptaddress,twocolumn]{revtex4-2}
\usepackage[usenames,dvipsnames]{color} 
\usepackage{graphicx}
\usepackage{amsfonts}
\usepackage[figuresright]{rotating}  
\usepackage{amssymb}
\usepackage{bm}
\usepackage{amsmath}
\usepackage{mathtools}
\usepackage{psfrag}
\usepackage{multirow}
\usepackage{tabularx}
\usepackage{textcomp}
\usepackage{units}
\usepackage{lipsum}
\usepackage{xr}
\usepackage{physics}
\usepackage{soul}
\usepackage{titlesec}
\usepackage{times}
\usepackage{enumitem}

\usepackage{subfigure}
\renewcommand{\selectlanguage}[1]{}

\DeclareMathAlphabet\mathbfcal{OMS}{cmsy}{b}{n}

\titlespacing*{\section}
{0pt}{1ex plus .25ex}{1ex plus 1ex}
\titlespacing*{\subsection}
{0pt}{1ex plus .25ex}{1ex plus 1ex}
\titlespacing*{\subsubsection}
{0pt}{1ex plus .25ex}{1ex plus 1ex}

\titleformat{\section}
 {\fontsize{10}{15}\bfseries }
  {\thesection}
  {1em}
  {}

\titleformat{\subsection}
  {\bfseries }
  {\thesubsection}
  {2em}
  {}

\titleformat{\subsubsection}
  {\itshape}
  {\thesubsubsection}
  {2em}
  {}

\makeatletter

\titleclass{\subsubsubsection}{straight}[\subsection]

\newcounter{subsubsubsection}[subsubsection]
\renewcommand\thesubsubsubsection{\thesubsubsection.\arabic{subsubsubsection}}
\titleformat{\subsubsubsection}
   {\normalfont\normalsize\itshape \centering}{\thesubsubsubsection}{1em}{}
\titlespacing*{\subsubsubsection}
{0pt}{1ex plus .3ex}{1ex plus .3ex}
\def\mbf{\mathbf}
\def\t{\text}

\def\beit{\begin{itemize}}
\def\eit{\end{itemize}}
\def\mbf{\mathbf}

\begin{document}

\title{Polaritonic Control of Blackbody Infrared Radiative Dissociation}

\author{Enes Suyabatmaz}
\affiliation{Department of Physics, Emory University, Atlanta, GA, 30322}
\author{Gustavo J. R. Aroeira}
\affiliation{Department of Chemistry and Cherry Emerson Center for Scientific Computation, Emory University, Atlanta, Georgia 30322, United States}
\author{Raphael F. Ribeiro}
\email{raphael.ribeiro@emory.edu}
\affiliation{Department of Chemistry and Cherry Emerson Center for Scientific Computation, Emory University, Atlanta, Georgia 30322, United States}
\date{\today}

% Abstract
\begin{abstract}
Vibrational strong light-matter coupling offers a promising approach for controlling chemical reactivity with infrared microcavities. While recent research has examined potential mechanisms for this phenomenon, many important questions remain, including what type of reactions can be modified and to what extent this modification can be achieved. In this study, we explore the dynamics of Blackbody Infrared Radiative Dissociation (BIRD) in microcavities under weak and strong light-matter interaction regimes. Using a Master equation approach, we simulate the effects of infrared field confinement and polariton formation on BIRD rates for diatomic molecules weakly coupled to the radiation field. We present a framework explaining how infrared microcavities influence BIRD kinetics, highlighting the importance of overtone transitions in the process. Our findings outline conditions under which significant enhancement or mild suppression of BIRD rates can be achieved, offering insights into practical limitations and new strategies for controlling chemistry within infrared resonators. 
\end{abstract}

\maketitle

% Main Content with Regular Citations

\section{Introduction}
Chemical reaction control is critical in applications ranging from industrial catalysis to medicine. Conventional methods for manipulating reaction rates involve varying temperature, pressure, or catalyst composition. An unconventional approach for controlling thermal reaction kinetics has been recently reported \cite{thomas2016, thomas2019, hirai2020, ahn2023} employing infrared (IR) microcavities \cite{kavokin2017microcavities} consisting of two moderate-quality mirrors separated by a distance $L_C$ of the order of IR wavelengths resonant with typical molecular vibrations \cite{nagarajan2021}. When a collection of molecules with sufficiently large IR oscillator strength is embedded in a resonant microcavity, vibrational strong coupling occurs as signaled by the emergence of hybrid quasiparticles (vibrational polaritons) consisting of a superposition of molecular and electromagnetic modes of the confined device \cite{vinogradov1992,shalabney_coherent_2015, simpkins_spanning_2015}. Recent theoretical research has probed potential mechanisms for the observed polariton effects on chemical reaction rates\cite{galego2019cavity, campos2019resonant, zhdanov2020vacuum, wang2022chemical, lindoy2022resonant, schafer2022shining, sun2022suppression, ahn2023, Vega2025, lai2024non, moiseyev2024conditionsenhancementchemicalreactions}, and experimental progress towards the characterization of gas-phase reactivity under strong light-matter coupling has also been reported \cite{weichman2023gas, weichman2024gas}. Still, several fundamental questions remain\cite{simpkins2021mode, campos2023swinging}, including what types of reactions can be controlled with an optical microcavity and to what extent this control can be exerted. In this study, we investigate gas-phase unimolecular dissociation activated by the absorption of microcavity blackbody infrared radiation \cite{dunbar2004bird} under weak and strong coupling with a suitable material.

Blackbody infrared radiative dissociation (BIRD) was proposed by Perrin in 1913 as a mechanism for gas phase unimolecular dissociation \cite{perrin1970atomes}. However, experimental observations in the 1920s ruled BIRD out in favor of the collisional activation mechanism \cite{lindemann1922discussion, daniels1928radiation, king1984chemical}. Several decades later, the BIRD hypothesis was proven appropriate at sufficiently low pressures\cite{thoelmann1994spontaneous,dunbar1995zero,schnier1996blackbody}. Since then, BIRD has become a valuable tool for investigating the thermal dissociation kinetics of molecular ions and clusters and has been observed in several systems, including ion clusters, transition-metal complexes, and biopolymers \cite{dunbar2004bird}. 

The thermal radiation spectrum heavily influences BIRD rates. Given that free space and microcavity electromagnetic density of states and thermal radiation density can be significantly different \cite{kavokin2017microcavities}, microcavity BIRD is expected to proceed with different rates compared to free space BIRD. However, the extent of this change, whether suppression or enhancement can be achieved, and if strong light-matter coupling introduces new possibilities relative to the weak coupling regime remain open questions. In this study, we address these questions by examining the dissociation of diatomic molecules driven by multiphoton or multipolariton absorption in microcavities. The qualitative conclusions drawn from this analysis are expected to be generalizable to polyatomic systems. 

The primary aim of this work is to characterize thermal radiative dissociation in microcavities. \textcolor{black}{Whereas prior studies employed hierarchical equations of motion and analytical rate theories based on Fermi's golden rule to explore energy-diffusion-limited reactivity modeled by double-well potentials \cite{lindoy2022resonant,sun2022suppression,Lindoy2023,Ying2024a,Ying2024b,Vega2025}, here we employ a Pauli master equation to investigate irreversible bond breaking with a Morse potential in a scenario where the microcavity or polaritons act solely as passive modifiers of the electromagnetic environment. Our methodology further differs by its 
microscopically detailed treatment of mechanical and electrical anharmonicity inherent to reactivity processes, and by its inclusion of the coupling of both fundamental and overtone vibrational transitions to the modified photon density of states, thereby providing a detailed description of the microcavity effect on the multiple reactive pathways available in BIRD.}

\par We employ a Pauli Master equation to investigate BIRD in (i) free-space, (ii) under weak coupling with a nearly empty microcavity, and (iii) under weak coupling with a polaritonic material. Our results give upper bounds for BIRD enhancement and suppression in weakly and strongly coupled microcavities \textcolor{black}{with perfect mirrors (see SI Sec. 8 for a discussion of leaky microcavities)} and identify overtone transitions as crucial ingredients for the observed modulation of BIRD. In Sec. II, we describe our methodology, Sec. III provides the main results and related discussion, and Sec. IV summarizes our main results and conclusions. 

\section{Methods}
\begin{figure*}
    \centering
    \includegraphics[width=1\linewidth]{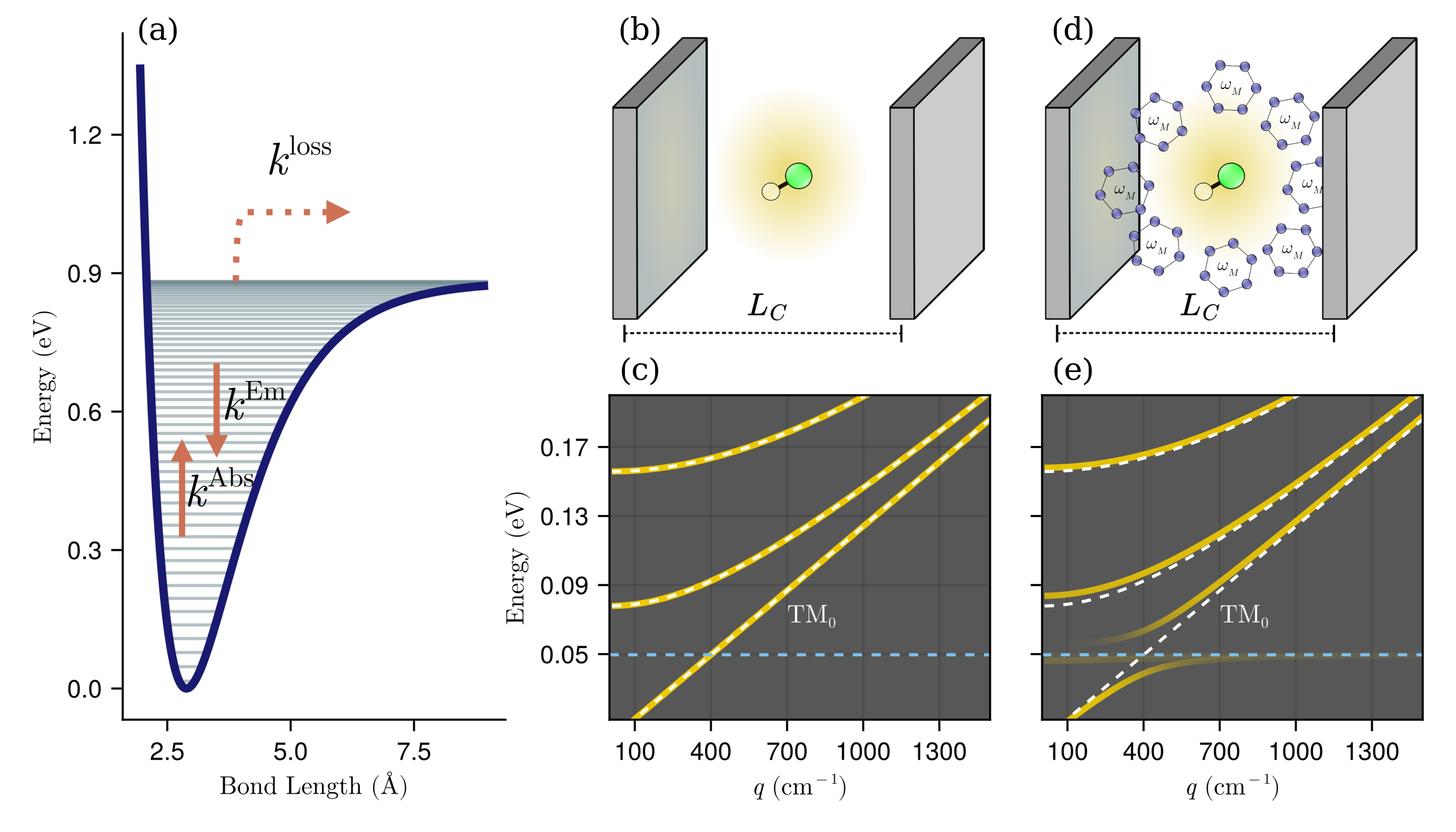}
    \caption{(a) Illustration of the Morse potential describing the vibrational states and considered transitions of the NaLi molecule. (b) Illustration of a diatomic molecule inside a high quality planar microcavity with longitudinal length $L_C$. (c) Dispersion curves of transverse-electric (TE) and transverse-magnetic (TM) microcavity modes. (d) Scheme illustrating a reactive diatomic molecule inside a planar resonator strongly coupled to a host material system with significant oscillator strength and transition frequency $\omega_M$ \textcolor{black}{(dashed line at $\omega = 0.05~ \text{eV}$)}. In this setup, the diatomic dissociation is mediated by absorption of thermal polaritons. (e) Illustration of polariton dispersion curves.}
    \label{fig:scheme}
\end{figure*}

\subsection{Pauli Master Equation for diatomic BIRD}
\textcolor{black}{Our analysis of BIRD in microcavities employs the same kinetic framework that has proven successful for BIRD studies in free space \cite{dunbar1995zero, dunbar1998activation, stevens2002blackbody, dunbar2004bird}. Specifically, we propagate the populations of bound vibrational levels with a Pauli master equation\cite{pauli1928festschrift} including transition rates obtained from Fermi’s golden rule \cite{dirac1927}. This choice is justified because the infrared absorption and emission events of interest ($> $ 100 ns) are orders of magnitude slower than rovibrational decoherence, which randomizes vibrational coherences within on a much faster timescale. The resulting separation of time scales renders the diatomic reduced density matrix effectively diagonal (in the Morse potential Hamiltonian eigenstate basis) well before any population transfer occurs, allowing the Liouville–von Neumann equation to be coarse-grained to the Pauli form \cite{pauli1928festschrift, vanhove1954quantum, vankampen1954quantum}}.

\textcolor{black}{We neglect collisional activation and assume that dissociation is driven exclusively by photon absorption, consistent with the conditions of BIRD experiments\cite{dunbar1995zero, dunbar2004bird}. In such experiments, the use of dilute gas samples at low pressures ensures that intermolecular collisions are negligible. External fields are applied to trap and stabilize molecular trajectories within the interior of the reaction vessel. This prevents molecules from exchanging energy with walls and allows radiative processes to dominate \cite{Schuster2011, Krems2018, Williams2018}.} Rotational degrees of freedom are ignored since they are weakly coupled to vibrational transitions, and the rotational dynamics are essentially classical at the temperatures of interest to BIRD. We also neglect Doppler and lifetime broadening throughout.

For the sake of simplicity, we assumed that the thermalization of the background radiation is faster than all the considered radiative transitions. We analyzed microcavity effects on BIRD occurring via two distinct mechanisms. In the first, denoted, dissociation via the doorway state, the dissociation proceeds only via the highest-energy bound state (the doorway state). In this case, the rate of change of the population in the $i$th vibrational state ($N_i$) is given by
\begin{equation}
    \frac{dN_i(t)}{dt} = \sum_{j\neq i}^{i_{\t{max}}} \left[ k_{ij} N_{j}(t) - k_{ji} N_{i}(t) \right] - k_{\t{loss}}N_{i}(t)\delta_{i,i_{\t{max}}}\;, \label{dndt}
\end{equation}
where $k_{ij}$ denotes rate of absorption when $i > j$ and emission when $i < j$ and  $i_\t{max}$ is the highest energy bound state (Fig. \ref{fig:scheme}a). {As mentioned}, this state is only loosely bound and is assumed to undergo irreversible dissociation at a rate $k_\t{loss} \gg k_{ij} \, \forall ~i,j$. Equation \ref{dndt} can be written in matrix notation as
\begin{equation}
    \frac{d[\textbf{N}]}{dt} = - \textbf{J}\cdot [\textbf{N}],
 \end{equation}
where $\textbf{N}$ is a column vector containing the vibrational populations and $\mathbf{J}$ is the transition matrix. The lowest eigenvalue $\lambda_{0}$ of $\textbf{J}$ is the rate constant of the unimolecular reaction \cite{valance1966theoretical, gilbert1990theory} of the described mechanism. Note that this mechanism does not consider dissociation from intermediate vibrational levels ($i < i_{\t{max}}$). Thus, our results are expected to be valid for BIRD promoted by multiphoton dissociation through a particular doorway state, which in our case is the highest-energy bound state. 

{The second dissociation mechanism considered allows for direct radiative dissociation from any bound state of the diatomic molecule. Compared to Eq.~\ref{dndt}, this introduces a dissociation channel for each bound state, characterized by a loss rate $k^i_{\text{loss}}$. The corresponding population rate equations are given by}
\begin{equation}
    \frac{dN_i(t)}{dt} = \sum_{j\neq i}^{\nu_{\t{max}}} \left[ k_{ij} N_{j}(t) - k_{ji} N_{i}(t) \right] - k^i_{\t{loss}}N_{i}(t). \label{dndt_continuum}
\end{equation}
{The transition matrix $\mathbf{J}$ now includes contributions from bound-continuum decay rates in all diagonal elements. The presence of $k_i^{\t{loss}}$ for all $i$ ensures that the vibrational population can be irreversibly removed from each level due to dissociation. The matrix $\mathbf{J}$ remains positive-definite, and its smallest eigenvalue $\lambda_0$ continues to govern the long-time dynamics of the system.} %Note the dispersion of bound-continuum transitions over a broad frequency spectrum constrains the utility of resonant electromagnetic field enhancement at a few selected frequencies, suggesting this mechanism will be less sensitive to cavity effects relative to the discrete, resonant nature of the doorway-state process.}%Bound-continuum transitions occur within a broad range of frequencies. This limits the effectiveness of accelerating a reaction via selective enhancement of the electromagnetic field at particular frequencies and suggests a reduced sensitivity to microcavity effects in comparison to the doorway state mechanism.}

\subsection{Radiative transition rates}
Assuming the molecular system that undergoes BIRD is weakly coupled to the radiation field, the transition rates $k_{ij}$ can be obtained from perturbation theory (Fermi's Golden Rule \cite{fermi1932quantum, sakurai2017modern}) and written in terms of Einstein coefficients in free space \cite{einsteinradiation, cohen2019quantum} and their generalizations in a weakly coupled and polaritonic microcavity. Rotational dynamics is much faster than the thermal infrared radiative transition, so we employ isotropically averaged radiative rates in the long-wavelength limit (electrical dipole approximation) throughout this work.

In free space, the spontaneous emission rate is given by the Einstein $A$ coefficient\cite{einsteinradiation, cohen2019quantum}. In the case where spontaneous emission induces the $i\rightarrow j$ transition with $i > j$ this coefficient is given by
\begin{equation}
    k_{ij}^{\text{SpEm}}  = A_{ij} = \frac{\omega_{ij} |\mu_{ij}|^2\pi}{3\epsilon_{0}\hbar} D_0(\omega_{ij}), \label{eq:Aij}
\end{equation}
where $\epsilon_0$ and $\hbar$ are the vacuum electrical permittivity and reduced Plank's constant, respectively, $\omega_{ij} = (E_i-E_j)/\hbar$ is the transition frequency, $\mu_{ij}$ is the corresponding transition dipole moment and $D_0(\omega) = \omega^2/\pi^2c^3$ is the free-space radiative density of states (DOS). The Einstein $B$ coefficient associated with stimulated emission and absorption processes involving states $i$ and $j$ is given by $B_{ij} = \pi |\mu_{ij}|^2/(3\epsilon_{0} \hbar^2)$. The corresponding thermal stimulated emission and absorption rate constants are given by
\begin{equation}
  k_{ij}^{\text{Abs}}(T)  =  k_{ij}^{\text{StEm}}(T) = B_{ij}\rho_{0}(\omega_{ij}, T) \;, \label{eq:kabs}
\end{equation}
where $\rho_0(\omega,T)$ is the free space blackbody radiation energy density
\begin{align}
    \rho_0(\omega,T) = \hbar\omega ~n_\text{BE}(\omega,T) D_0(\omega), \label{eq:rho0}
\end{align}
and $n_{\text{BE}}(\omega,T)$ is the Bose-Einstein distribution thermal occupation number $n_{\text{BE}}(\omega,T) = 1/[\text{exp}(\hbar\omega/k_B T)-1]$\cite{bose1924plancks} and $k_B$ is the Boltzmann constant. 

The final pieces necessary to compute the rate constants above are the transition dipole matrix $\mu_{ij}$ and the frequencies $\omega_{ij}$. We estimate the latter assuming the electronic ground-state potential energy curve of the examined diatomic is given by the Morse potential [Fig. \ref{fig:scheme}(a)]\cite{morse1929diatomic}. The bound wave functions associated with the Morse potential are employed in the computation of the vibrational transition dipole moments $\mu_{ij}$ via 
\begin{align}
\mu_{ij} &= \int \psi^*_i(r) \mu(r) \psi_j(r) \mathrm{d}r \label{braketdip},
\end{align}
where $\psi_i(r)$ represents the $i$th vibrational level wave function, and the dipole function $\mu(r)$ models the nonlinear change in the (electronic ground-state) molecular electrical dipole with the bond length $r$ \cite{bishop1990vibrational}. 

{The rates $k_{\text{loss}}^{i}$ for radiative absorption into the continuum by each bound state $i$ are obtained by a simple generalization of Eqs. \ref{eq:kabs} as discussed in the Supplementary Material.}

\textcolor{black}{We investigated the thermal IR radiative dissociation of the NaLi molecule. The electrical dipole function and Morse potential were obtained via interpolation from previous CCSDT/cc-pCVQZ calculations available in the literature \cite{Fedorov2014} (see Supplementary Information for further details). Morse potential parameters are listed in Table \ref{tab:morseparameters}.}
\renewcommand{\arraystretch}{2}
\begin{table}[h]
    \caption{\textcolor{black}{Parameters characterizing the Morse potential associated to NaLi.}}
    \centering
    \begin{tabular}{c|c|c}
        \textbf{Parameter} & \textbf{Notation (unit)} & \textbf{Value}  \\
        \hline
        Electronic ground-state dissociation energy & $D_e$ (eV)           & 0.882       \\
        \hline
        Equilibrium bond length & $r_e$ (\AA)          & 2.895     \\
        \hline
        Harmonic frequency & $\omega_e~\left(\text{cm}^{-1}\right)$  & 257.4     \\
        \hline
        Anharmonicity  & $\omega_e \chi_e~\left(\text{cm}^{-1}\right)$ & 2.33   \\
        \hline
        Reduced mass & $m_r$ (a.u)          & 5.331      \\
    \end{tabular}
    \label{tab:morseparameters}
\end{table}
The transition rates $k_{ij}$ are proportional to the electromagnetic density of states $D(\omega)$, suggesting BIRD rates can be controlled by placing the reactive molecule inside a microcavity. To examine this scenario, we adapt the free space formalism described above to arbitrary photonic devices by replacing $D_0(\omega)$ in Eqs. \ref{eq:Aij} and \ref{eq:rho0} with the appropriate DOS for a weakly coupled microcavity $D_\text{C}(\omega)$. 

In a planar empty Fabry-Perot microcavity composed of two parallel mirrors (with unit reflectivity for simplicity) separated by a distance $L_C$, [Fig. \ref{fig:scheme}(b)], the frequencies of the electromagnetic modes are given by
\begin{equation}
   \omega_{\t{C}}(m, q)= c\sqrt{q^2+\frac{m^2\pi^2}{L_C^2}}\;,\label{eq:omega_c}
\end{equation}
where $\mbf{q}=(q_x,q_y) \in \mathbb{R}^2$ is the in-plane projection of the mode wave vector, $q^2 = q_x^2 + q_y^2$, and $m$ is an integer quantum number that identifies the longitudinal wave vector $k_z = m\pi/L_C$ and thus the photonic energy bands in Fig. \ref{fig:scheme}(c). For transverse-magnetic (TM) modes $m = 0,1,2,...$ whereas $m=1,2,3...$ for transverse-electric (TE) modes \cite{zoubi2005microscopic, jackson2021classical}. Including the TM$_0$ mode here is crucial, as otherwise, transitions with energy below the cavity cutoff would be impossible.
 
Using Eq. \ref{eq:omega_c}, it can be shown (see Supporting Information) that the spatially-averaged electromagnetic DOS inside a microcavity in the weak coupling regime can be expressed by
\cite{barnes_classical_2020}
\begin{align}
    D_C(\omega) = \frac{\omega}{\pi c^2 L_C} \left\lfloor \frac{\omega L_C}{\pi c} \right\rfloor + \frac{\omega}{2\pi c^2 L_C}\;,\label{eq:Dc}
\end{align}
where $\lfloor x \rfloor$ denotes the floor function. \textcolor{black}{This work focuses on establishing an upper limit for microcavity effects on BIRD by assuming perfectly reflecting mirrors. Real microcavities are leaky, so to assess the robustness of our predictions, we also computed the photon density of states and corresponding quantum state transition rates in imperfect microcavities with finite reflectivity and absorbing metallic mirrors using the electromagnetic field dyadic Green function \cite{barnes_classical_2020} (see Supplementary Information for details)}.

Lastly, we considered the BIRD scenario in which the diatomic interacts weakly with the electromagnetic component of polariton modes emergent from vibrational strong coupling between the microcavity modes and a material system (solid-state or molecular ensemble) with sufficiently large IR oscillator strength.  \textcolor{black}{We emphasize that the polariton modes arise from collective strong coupling between the cavity and a host material with substantial infrared oscillator strength. The reactive molecule experiences modified radiative dynamics due to changes in the photon density of states induced by the formation of polaritons, as illustrated in Figs. \ref{fig:scheme}(d) and (e). Consequently, in this study, the interaction strength between our reactive diatomic and the radiation field is obtained from first principles without any scaling parameters.}

\textcolor{black}{While this polariton-assisted scenario poses challenges for experimental realization as BIRD generally requires low-density, collision-free conditions, it is introduced here as a theoretical construct that isolates the effect of polariton-modified photonic environments on molecular infrared radiative dissociation. Nonetheless, one could envision suppressing collisions by confining reactive molecules using transparent partitions or external fields, while maintaining exposure to the confined electromagnetic modes\cite{Schuster2011, Krems2018, Williams2018}. A detailed treatment of these experimental considerations, as well as the role of direct interactions between host and reactive molecules, lies beyond the scope of the present work and warrants future investigation.}

The formalism discussed above for BIRD in a weakly coupled microcavity can be straightforwardly adapted to model polariton-mediated BIRD (see SI) by replacing the photon density of states $D_0(\omega)$ with the \textit{photon-weighted polariton density of states} $D_P(\omega)$.

The Hamiltonian of the strongly coupled molecular subsystem and the microcavity is given in the Power-Zienau-Woolley (dipole) gauge \cite{power1959coulomb, woolley1971quantum}
\begin{equation}
    H = H_{\t{SM}} + H_{\t{L}}+H_{\t{INT}}\;,
\end{equation}
where the interaction $H_{\t{INT}}$ is given by 
\begin{equation}
    H_{\t{INT}} = \frac{1}{\epsilon_{0}}\int \left[ -\textbf{D}(\textbf{r})\cdot\textbf{P}(\textbf{r})+ \frac{1}{2}\textbf{P}^2(\textbf{r})\right]d^3\textbf{r},
\end{equation}
where $\textbf{P}(\textbf{r})$ is the matter polarization density and $\textbf{D}(\textbf{r})$ is the electrical displacement field \cite{power1982quantum, craig1998molecular}. The pure matter and bare electromagnetic Hamiltonians $H_{SM}$ and $H_{L}$ generate the dynamics of a collection of free harmonic oscillators corresponding to isotropic uniformly distributed matter vibrations (e.g., from a dispersionless paraelectric material \cite{ashida2020quantum, curtis2023local}) and planar microcavity modes with frequency-momentum dispersion given by Eq. \ref{eq:omega_c}. Assuming a strongly coupled 3D isotropic material system with uniform spatial distribution, it follows in the mean-field limit using the standard Bogoliubov transformations  \cite{bogoliubov1958superconductivity, hopfield_theory_1958, agranovich1984crystal} that the photon-weighted polariton density of states $D_P(\omega)$ can be written as (see SI)
\begin{equation}
    D_{P}(\omega) = \sum_{m=0}^{\infty}\left(1-\frac{\delta_{0,m}}{2}\right)\frac{q(m,\omega)P_C(\omega)}{\pi  v_g(\omega, q)L_{C}} \theta[\omega_\t{C}(\omega)-m\pi c/L_C], 
    \label{eq:dpomega}
\end{equation}
where $\theta(x) = 1$ for $x \geq 0$ and vanishes elsewhere, $\omega_\t{C}(\omega)$ is the frequency of the photon mode that under strong coupling leads to the formation of a polariton with frequency $\omega$, $P_C(\omega)$ is the photon content of the polariton modes with frequency $\omega$ 
\begin{equation}
    P_{C}(\omega)= \frac{(\omega^2-\omega_M^2 -\Omega_R^2)^2 }{(\omega^2-\omega_M^2 -\Omega_R^2)^2+\omega^2_{C}(\omega) \Omega_{R}^2},
\end{equation}
$\omega_M$ is the matter oscillator frequency,  $\Omega_R$ is the collective light-matter interaction strength, and 
the polariton group velocity ($v_g$) is given by
\begin{align}
v_g(\omega,q) = \frac{q c^2}{2\omega} \left( 1 \pm \frac{\omega_C^2 - \omega_M^2 + \Omega_R^2}{\sqrt{ \left( \omega_C^2 + \omega_M^2 + \Omega_R^2 \right)^2 - 4\omega_C^2  \omega_M^2}} \right) \;, \label{eq:vg}
\end{align}
where the positive sign applies when $\omega$ belongs to an upper polariton branch, i.e., $\omega > \omega_C$, and the negative sign is used if $\omega < \omega_C$. In this expression, we have omitted the frequency dependence of $\omega_C$ for simplicity. 
\par Our model shows singular behavior for $D_P(\omega)$ as $\omega \rightarrow \omega_M$. In this limit, both the photon content $P_C(\omega)$ and the group velocity $v_g(\omega,q)$ approach zero, but the ratio $P_C(\omega)/v_g(\omega,q) \rightarrow \infty$ resulting in $D_P(\omega)$ diverging to positive infinity. This singularity is due to the formation of the well-known stopgap region\cite{litinskaya2009gap, huang2003theory, gomez2004propagation, todorov2012intersubband} where $D_P(\omega) = 0$. As previous work shows \cite{satanin2005localization}, introducing an ultraviolet cutoff and material disorder eliminates the singularity of $D_P(\omega)$ at $\omega_M$. Nevertheless, there remains significant enhancement of $D_P(\omega)$ in a neighborhood of $\omega_M$, which depends in a non-universal way on disorder and coherence loss mechanisms. \textcolor{black}{Similarly, the stopgap is known to contain weakly coupled states in disordered systems causing $D_P(\omega)$ to become nonzero in this region. However, the photon density of states remains lower in the stopgap in comparison to the weak coupling regime.}

\textcolor{black}{The robustness of the enhanced and reduced photon-weighted polariton density of states at frequencies $\omega$ approaching $\omega_M$ from the left and right, respectively, support our choice to to use the formally exact expression presented in Eq. \ref{eq:dpomega} and interpret our results as upper bounds on the effect of polaritons on BIRD. The relevance of these bounds to experiments will be shown to depend on the collective light-matter interaction strength and the detuning between transitions of interest and $\omega_M$ as discussed in detail in \textbf{Results and Discussion}.}

\section{Results and Discussion}
\subsection{Microcavity-assisted blackbody infrared radiative dissociation via doorway state} 

Our analysis of radiative dissociation in weakly coupled microcavities assumes a single diatomic molecule (or a dilute sample at low enough pressure that collisions can be neglected) in a planar microcavity. The tunable electromagnetic environment afforded by the microcavity allows BIRD rates to be either enhanced or suppressed relative to the free space, depending on which transitions are affected by the modified photon DOS.

\begin{figure}
    \centering
    \includegraphics[width=\linewidth]{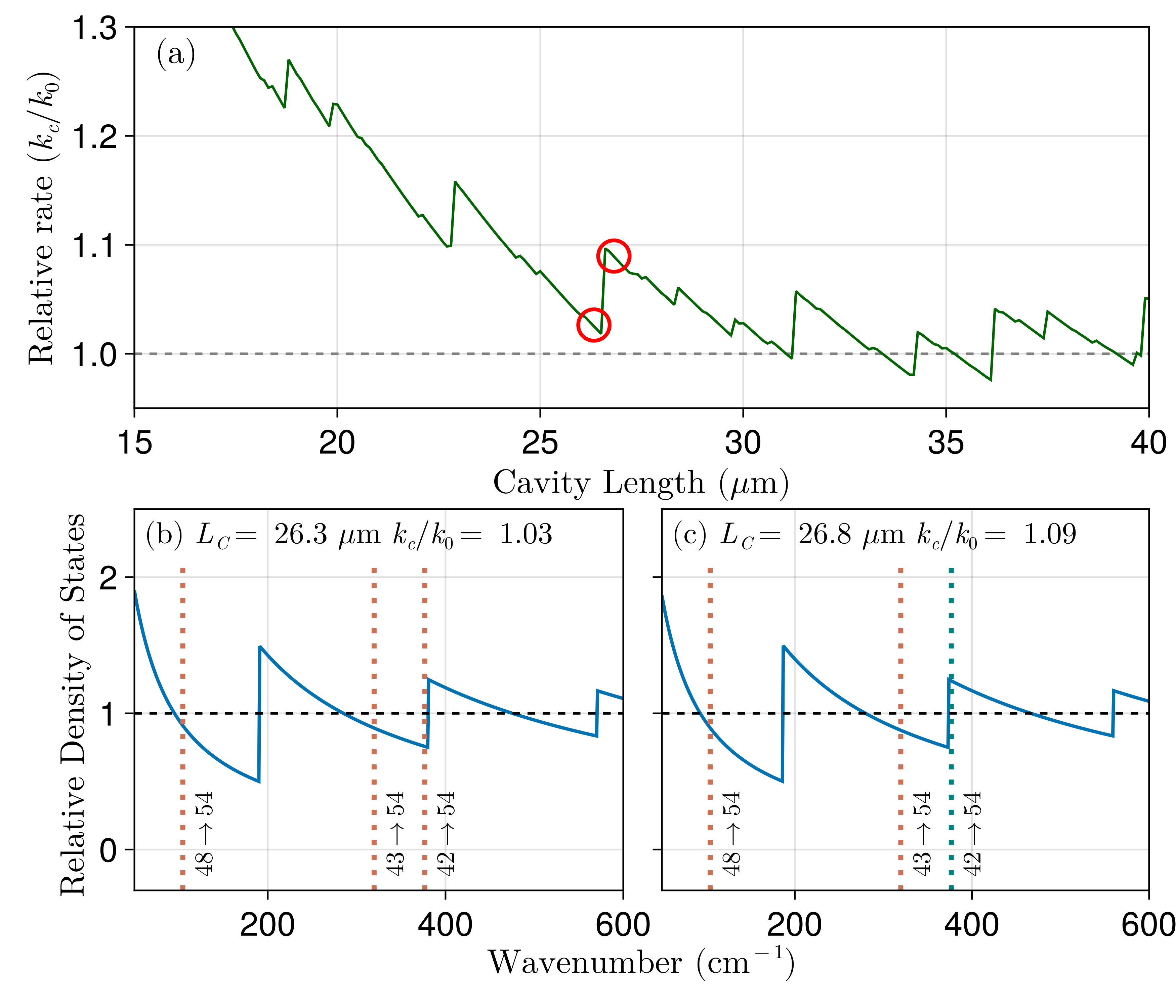}
    \caption{(a) \textcolor{black}{Ratio of BIRD rate inside a nearly empty microcavity (weak coupling regime) to the free-space rate as a function of microcavity length for NaLi. Red circles highlight specific lengths further discussed in (b) and (c). Microcavity photon DOS normalized to the free space DOS for $L_C = 26.3$ and $26.8$ $\mu$m, in (b) and (c), respectively.  The dotted vertical lines indicate the location of the molecular transition frequencies corresponding to the most relevant overtone transitions, with color coding to distinguish between suppression (red) and enhancement (green). }}
    \label{fig:weak_coupling}
\end{figure}

In Fig. \ref{fig:weak_coupling}(a), BIRD rates in the weak coupling regime are shown relative to the free space rate at \textcolor{black}{$T = 400~\t{K}$. For small microcavity lengths $L_C < 25$ $\mu$m, a mild but consistent enhancement of BIRD rates is observed.}  At these lengths, an overall increase in the microcavity electromagnetic DOS over a wide frequency range covering several important transitions is associated with the observed enhancement. \textcolor{black}{The ratios of microcavity BIRD rates to free space BIRD rates approach 1 as $L_C$ increases beyond a few tens of microns.} This is expected since the microcavity DOS converges to the free space limit when $L_C \rightarrow \infty$.
\par On top of the main feature described above, we also observe an oscillatory pattern in the BIRD rates relative to free space shown in Fig. \ref{fig:weak_coupling}(a). This pattern arises from the interplay of discrete vibrational transitions enhanced or suppressed depending on $L_C$. As we demonstrate below, this complexity is caused by overtones, which play an essential role in our analysis. 

We conducted a sensitivity analysis of the dissociation rate to interpret the origin of the oscillatory behavior in Fig. \ref{fig:weak_coupling}(a) and identify the most consequential radiative transitions involved in the BIRD of \textcolor{black}{NaLi. This analysis measures how impactful an enhancement in the photon DOS at a specific transition frequency is on the dissociation process (see Supporting Information for more details). For NaLi dissociation, changes in the photon density of states corresponding to the frequency of overtone transitions $i \rightarrow 54$, with $i \in \{42,43,48\}$ are the most consequential for the BIRD rate.} The identified overtones take the vibrational mode to the highest energy-bound state, from which dissociation rapidly occurs. We emphasize that overtones are crucial because, at high excitation levels, fundamental transitions ($i\rightarrow i+1$)  involve very small energies and, therefore, have associated small thermal photon populations and reduced oscillator strengths. See Supporting Information for a brief discussion of results obtained without overtone transitions.

Equipped with knowledge of the most relevant transitions for the investigated BIRD of \textcolor{black}{NaLi via the most energetic bound state, we can explain the microscopic origin of the intricate pattern in Fig. \ref{fig:weak_coupling}(a) considering how the microcavity DOS changes around the previously mentioned overtones. Figs. \ref{fig:weak_coupling}(b) and \ref{fig:weak_coupling}(c) present the ratio of microcavity DOS to the free space for selected values of $L_C$ where suppression or enhancement of BIRD was verified in Fig. \ref{fig:weak_coupling}(a) (red circles). Overtone frequencies are highlighted with dotted vertical lines, color coded by whether they are enhanced (green) or suppressed (red) in the corresponding microcavity. In Fig. \ref{fig:weak_coupling}(b) ($L_C = 26.3$ $\mu$m), the most relevant NaLi overtones are suppressed, and this leads to the local minimum observed in $k_c/k_0$ in Fig. \ref{fig:weak_coupling}. In contrast, when the microcavity length slightly increases to $L_C = 26.8$ $\mu$m [Fig. \ref{fig:weak_coupling}(c)] the photon DOS is enhanced at the $42\rightarrow 54$ transition frequency, and we observe a corresponding leap in the BIRD rate in Fig. \ref{fig:weak_coupling}(a). Introducing a finite linewidth to the considered radiative transitions would smooth out the oscillations observed in Fig. \ref{fig:weak_coupling}(a) and dampen their amplitude. Consequently, the weakly coupled microcavity effect on BIRD would be significant only at lengths less than a few tens of microns.}

\par Note that suppression effects are generally minor in Fig. \ref{fig:weak_coupling}(a) because the dissociative process has many pathways. Therefore, blocking or slowing down a particular transition may reduce the dissociation rate, but the effect is almost always negligible as the molecular system explores alternative dissociation pathways. Our analysis of BIRD in weak coupling reveals two main results that we expect to hold generically for chemistry in microcavities: (i) overtone transitions are crucial and should not be ignored, and (ii) blocking a single reactive pathway is likely inconsequential.

\textcolor{black}{We conclude this section by noting that BIRD simulations with lossy mirrors (SI Sec. 8) show (i) broadening and suppression of the sharp features in Fig. \ref{fig:weak_coupling} and (ii) enhanced dissociation rates relative to Fig. \ref{fig:weak_coupling}(a) at short $L_C$ due to the presence of evanescent modes with significant amplitude near the imperfect metal interfaces\cite{vinogradov1992, ashida2020quantum}. Overall, lossy mirrors suppress the oscillatory behavior of the relative BIRD rate with microcavity length while preserving the order-of-magnitude effects predicted for perfect mirrors. This confirms that, modulo the fast oscillations in Fig. \ref{fig:weak_coupling}(a), the results presented in this Section are representative of experimentally accessible photon resonators.}

\subsection{Polariton-assisted blackbody infrared radiative dissociation via doorway state} 

To analyze polariton-assisted BIRD, we considered a microcavity containing, in addition to the reactive diatomic molecule, a host material strongly coupled with the resonator modes. Fig. \ref{fig:scheme}(d) illustrates this scenario, emphasizing the formation of polariton states, which effectively change the photon frequency-momentum dispersion relation [Fig. \ref{fig:scheme} (e)] and the electromagnetic DOS. Importantly, the reactive NaLi molecule remains weakly coupled to the confined electromagnetic field. We explore the implications for diatomic BIRD via the most energetic bound state in the following.

\begin{figure}[t]
    \centering
    \includegraphics[width=\linewidth]{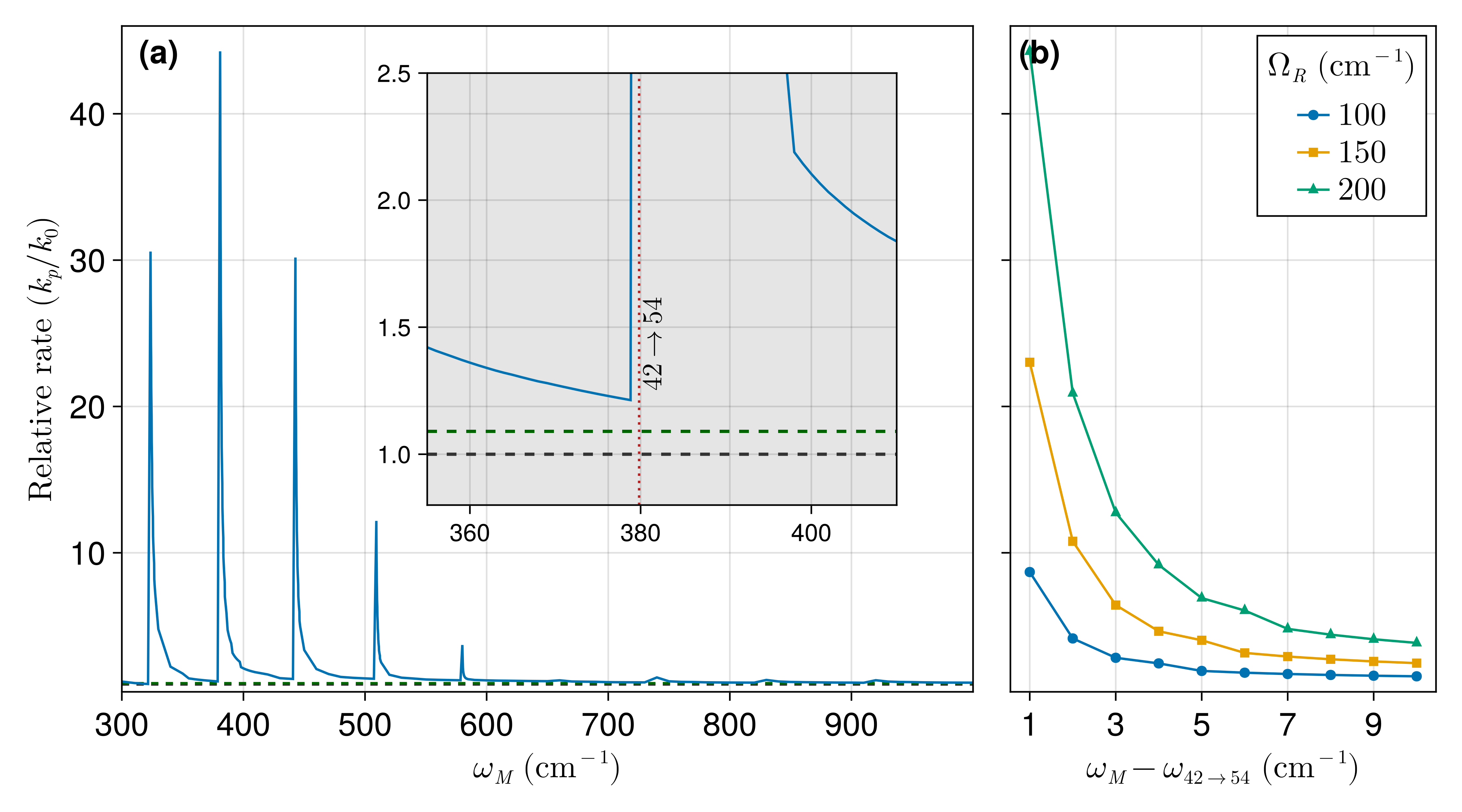}
    \caption{Ratio of polariton-assisted BIRD rates to free space BIRD rates. In the polaritonic case, the diatomic molecule is embedded in a strongly coupled microcavity with variable host material with frequency $\omega_M$.  \textcolor{black}{In (a), the collective light-matter interaction strength, $\Omega_R$, is fixed at 200 cm$^{-1}$.  The horizontal dashed gray and green lines indicate where the dissociation rates equal $k_0$ and $k_c$, respectively. The minimum difference between $\omega_M$ and $\omega_{i\rightarrow j}$ used here is 1 cm$^{-1}$ to match panel (b). The inset in (a) shows a zoomed-in view around the overtone $42 \rightarrow 54$. In (b), relative BIRD rates are shown for different Rabi frequencies ($\Omega_R$), depicting the variation of the relative rate enhancement with the detuning between the host molecule and the diatomic transition energies.}}
    \label{fig:strong_coupling}
\end{figure}

\begin{figure}[t]
    \centering
    \includegraphics[width=\linewidth]{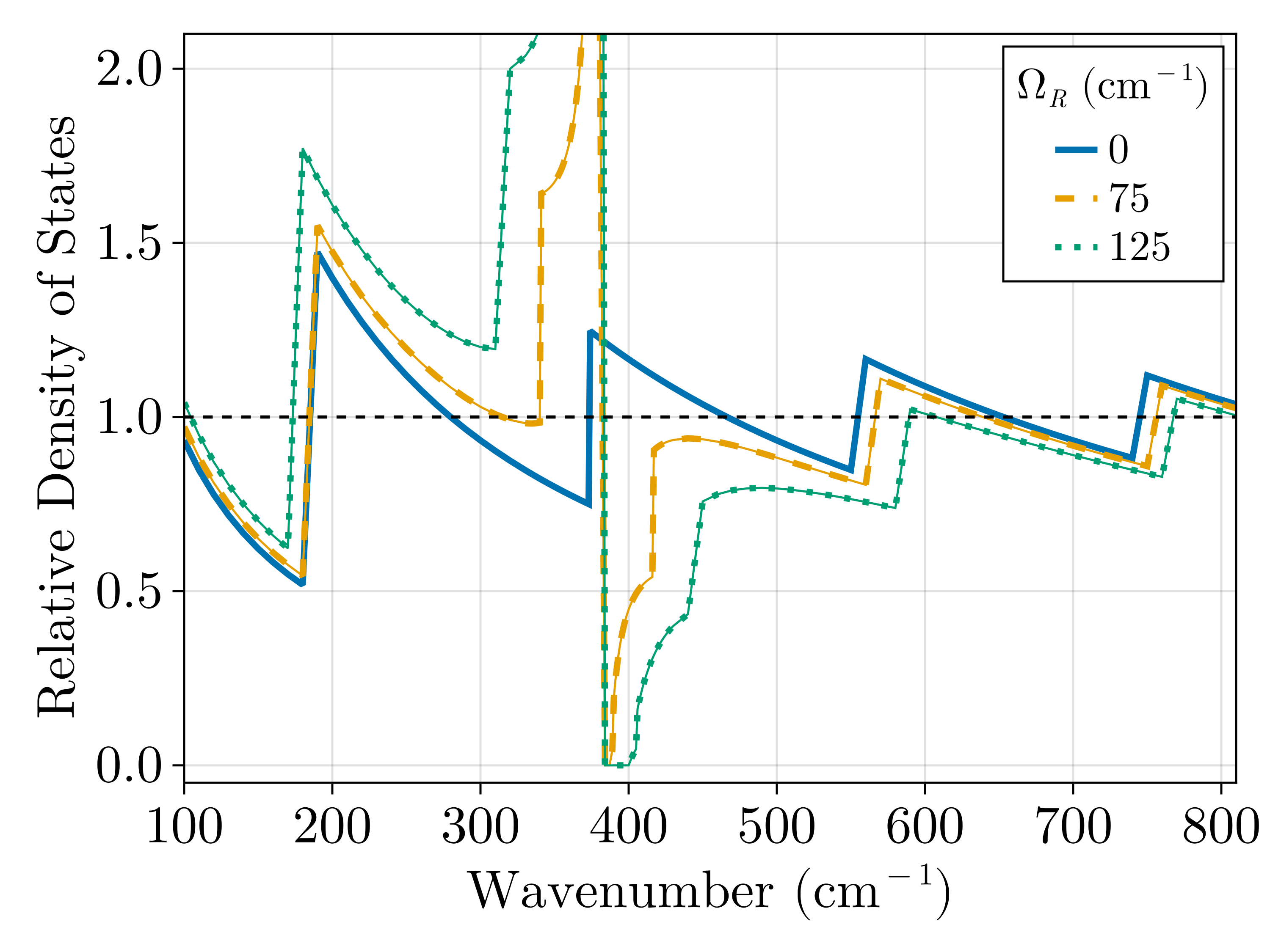}
    \caption{Photon-weighted polariton density of states $D_P(\omega)$, normalized to the free space DOS, for a microcavity with length $L_{C} = 26.8 ~\mu$m strongly coupled to a material with a bright transition at frequency $\omega_M = 380$ cm$^{-1}$. The curves show the effect of different Rabi splitting values ($\Omega_R$) on the photon-weighted polariton DOS, which is relevant for radiative processes mediated by polaritonic systems. As $\Omega_R$ increases, the frequency range below $\omega_M$, where the photon-weighted polariton DOS significantly exceeds that of free space, widens. This trend explains the impact of the collective light-matter interaction strength on the observed polariton-assisted bond infrared dissociation (BIRD) enhancement shown in Fig. \ref{fig:strong_coupling}.}
    \label{fig:DOSp}
\end{figure}

\par Fig. \ref{fig:strong_coupling} presents the ratio of polariton-assisted BIRD to the corresponding free space rates ($k_p/k_0$) as a function of the host matter frequency, $\omega_M$ and fixed microcavity length \textcolor{black}{$L_C$ at 26.8 $\mu\t{m}$ where we observed enhanced rates.} In Fig. \ref{fig:strong_coupling}(a), we observe prominent peaks where the relative rates are notably enhanced. These peaks occur when the matter frequency ($\omega_M$) is greater than but sufficiently close to the specific overtone transition frequencies ($\omega_{i\rightarrow j}$). \textcolor{black}{The asymmetry in the relative dissociation rate profile around the observed maxima, shown in more detail in the inset of Fig. \ref{fig:strong_coupling}(a), emerges as a result of the formation of the polariton stopgap region. When $\omega_M$ is close but smaller than $\omega_{i\rightarrow j}$, this transition falls within the stopgap region and is completely suppressed as the density of states becomes zero at that frequency. Nonetheless, as discussed for the weak coupling regime, a strong suppression of reaction rates is not observed due to the many alternative pathways the reaction can access.}

\textcolor{black}{Focusing on the $42\rightarrow54$ overtone ($\omega_{42\rightarrow 54}= 380 ~\text{cm}^{-1}$) which is the most consequential transition for the reaction, the effect of a variation in $ \omega_M - \omega_{42\rightarrow 54}$ at various collective coupling strengths ($\Omega_R$) on BIRD rates is examined in Fig. \ref{fig:strong_coupling}(b). Increasing $\Omega_R$ leads to a significant enhancement of the polariton-assisted BIRD rates, reaching a factor of 40 at the large Rabi frequency $\Omega_R = $200 cm$^{-1}$ and small $\omega_M - \omega_{42\rightarrow 54}$. As this frequency difference increases, the enhancement drops: at 3 cm$^{-1}$, the enhancement is about 12-fold; at 5 cm$^{-1}$ detuning, around 7-fold; and beyond 7 cm$^{-1}$, the enhancement persists around 4-fold. This reduction in dissociation rate enhancement with increasing $\omega_M - \omega_{42\rightarrow 54}$ follows from the behavior of the photon-weighted polariton DOS near the stopgap as we discuss next.}

In Fig. \ref{fig:DOSp}, the photon-weighted polariton density of states $D_P(\omega)$ is shown for selected values of \textcolor{black}{$\Omega_R$ at $L_C = 26.8$ $\mu$m and $\omega_M = \omega_{42\rightarrow 54} = 380$ cm$^{-1}$.} The case where $\Omega_R = 0$, representing the weak coupling limit, is included here for direct comparison with $D_C(\omega)$. It can be seen from Fig. \ref{fig:DOSp} that the photon-weighted polariton DOS is only significantly different from the bare microcavity DOS \textcolor{black}{in a frequency range around $\omega_M$ defined as $\{\omega_M - \alpha(\Omega_R) <\omega < \omega_M +\alpha(\Omega_R)\}$,  where $\alpha(\Omega_R)$ is a positive frequency that depends on $\Omega_R$ and satisfies $D_P[\omega_M \pm \alpha(\Omega_R)] \approx D_C[\omega_M \pm \alpha(\Omega_R)]$. Outside this range, $D_P(\omega)$ retains features of the weak coupling case, such as oscillations in $\omega$, which can be understood from Eq. \ref{eq:Dc}. Fig. \ref{fig:DOSp} further illustrates how increasing $\Omega_R$ broadens the stopgap region and expands $\alpha(\Omega_R)$, leading to a wider frequency range for which $D_P(\omega) \gg D_C(\omega)$.} As a result, depending on the Rabi splitting, strong coupling can induce changes in the rates of multiple vibrational transitions.

\textcolor{black}{The singular behavior of $D_P(\omega)$ at $\omega_M$ is originated by the implicit assumption of our model that off-resonant photons with arbitrarily large energy mix with the strongly coupled matter system excitations and form polaritons. These interactions, albeit weak, lead to the formation of lower polariton (LP) modes with very small photon contents and frequencies arbitrarily close to $\omega_M$. This behavior is reflected in the expression for the photon-weighted polariton DOS in Eq. \ref{eq:dpomega}, where the photon content appears in the numerator and the polariton group velocity in the denominator. As the frequency $\omega $ approaches $ \omega_M $, the group velocity vanishes faster than the photon content, resulting in an arbitrarily large photon DOS. This divergence in the DOS manifests itself as an accumulation of LP states near $ \omega_M $, driven by the off-resonant interaction between matter and the microcavity radiation field. In other words, as the in-plane wave vector $k$ increases, the LP transition frequencies approach $ \omega_M $, yet they never become exactly equal in our model due to the presence of the stopgap. Consequently, within any arbitrarily small frequency interval $ \omega_M - \delta < \omega < \omega_M, ~~ \delta > 0 $, a large number of LP states exist, leading to a singularity in the photon-weighted polariton DOS at $\omega_M$.}

\textcolor{black}{The off-resonant light-matter interactions responsible for the singularity of $D_P(\omega)$ at $\omega_M$ can be eliminated by imposing an effective frequency cutoff for polaritons formed from high-energy microcavity modes by setting a minimum nonvanishing $\delta = \omega_M - \omega_{42\rightarrow 54}$. For example, when $\delta =  10$ cm$^{-1}$ photon modes across 6 branches with $\omega_C(m,q) = 945$ cm$^{-1}$ contribute to $D_P(\omega)$, while when $\delta = 1$ cm$^{-1}$, $D_P(\omega)$ contains contributions from 15 microcavity branches with $\omega_C(m,q) = 2780$ cm$^{-1}$. Therefore, by choosing a nonvanishing threshold for $\delta$ we also set a maximum energy (cutoff) for radiation states that contribute to the photon-weighted polariton DOS. Fig. \ref{fig:strong_coupling} shows the removal of an increasing number of polaritons formed from highly off-resonant photons reduces the relative BIRD rate enhancement to an extent that depends on the Rabi frequency and the minimum $\delta$.}

In experiments, imperfections in the strongly coupled material system and in the electromagnetic device \textcolor{black}{(SI, Sec. 8)} smooth the photon-weighted DOS near the stopgap. Nonetheless, previously reported simulations show the enhancement of $D_P(\omega)$ persists around $\omega_M$ in the presence of an ultraviolet (UV) cutoff and material disorder \cite{satanin2005localization}. The introduction of a UV cutoff, cavity loss, and dephasing pathways for the matter polarization field \b{are therefore expected to} lead to reduced BIRD enhancements relative to those in Fig. \ref{fig:strong_coupling} depending in a non-universal way on $\Omega_R$ and parameters characterizing the quality of the strongly coupled subsystems (e.g., microcavity quality factor, matter homogeneous linewidth, etc). \textcolor{black}{Still, given the modest impact of leaky mirrors on the order of magnitude of the microcavity effect on the BIRD rates discussed in Sec. III.A and SI Sec. 8, we expect the order of magnitude of polariton-assisted BIRD enhancement to be similarly robust to the introduction of microcavity losses characteristic of experimentally accessible devices.}

Note the singularity in the photon-weighted polariton DOS at $\omega_M$ enables a particular radiative transition to become much faster than the corresponding free space rate. However, it does not lead to arbitrarily fast reaction rates, since the complex multistep reactive process ensures that other transitions quickly become rate-limiting (see the Supporting Information for a saturation analysis). Therefore, the reported BIRD rates obtained when $\omega_M$ approaches selected $\omega_{i\rightarrow j}$ (Fig. \ref{fig:strong_coupling}) provide a \textit{finite} upper bound for polariton-enhanced BIRD rates in the lossless limit discussed here.

Before ending we emphasize our observation of the same qualitative behavior for polariton-mediated BIRD in other diatomic species (see SI). Since these chemical species have significantly different energetics, we anticipate the main trends reported in this work also apply to other diatomic and polyatomic molecules, particularly in cases where dissociation occurs through a specific doorway state.

\subsection{Dissociation via radiative bound-continuum transitions}

{In the previous sections, we examined BIRD mediated by the most energetic bound-state. Gas-phase dissociation may also be promoted by direct decay from any bound-state $i \leq i_{\text{max}}$ into the dissociative continuum. This scenario is examined in this section, where we generalize our previous treatment to include direct (irreversible) dissociation pathways for each bound-state $i$ with rate $k^{i}_\text{loss}$ as described in \textbf{Methods} and Supporting Information.}

\begin{figure}[t]
    \centering
    \includegraphics[width=\linewidth]{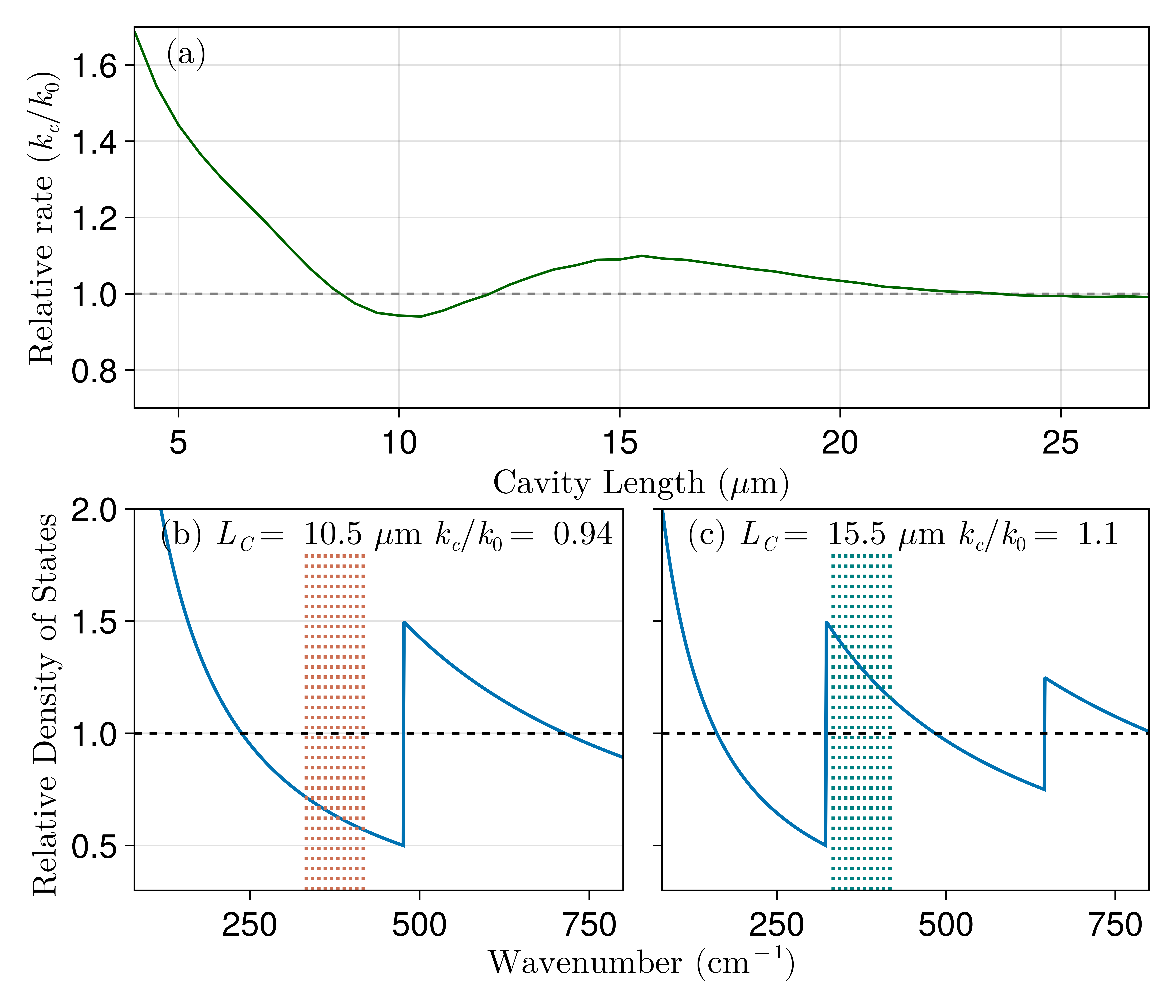}
    \caption{(a) {Ratio of BIRD rate inside a nearly empty microcavity (weak coupling regime) to the free-space rate as a function of microcavity length for NaLi. Microcavity photon DOS normalized to the free space DOS for $L_C = 10.5$ and $15.5$ $\mu$m, in (b) and (c), respectively.  The dotted vertical lines indicate the location of the molecular transition frequencies corresponding to the most relevant overtones, with color coding to distinguish between suppression (red) and enhancement (green). }}
    \label{fig:wc_continuum}
\end{figure}

{Fig.~\ref{fig:wc_continuum}(a) shows the relative dissociation rate, $k_C / k_0$, for a diatomic molecule inside a nearly empty microcavity as a function of its length. In contrast to the doorway mechanism discussed in Sec.~IIIA, which exhibits pronounced oscillations, the behavior here is noticeably simpler, yet may be understood with the same tools previously employed to rationalize the observed trends in Fig. \ref{fig:weak_coupling}. For example, from the normalized photon DOS at microcavity lengths $L = 10.5~\mu\text{m}$ and $L = 15.5$ $\mu$m  presented in Figs. \ref{fig:wc_continuum}(b,c) corresponding to local minimum and maximum in Fig. \ref{fig:wc_continuum}(a), respectively, we observe the ten most influential transitions, identified through sensitivity analysis, lie between 330–420 cm$^{-1}$. At $L = 10.5~\mu\text{m}$, these transitions fall within a frequency range where the DOS is suppressed relative to free space, whereas at $L = 15.5~\mu\text{m}$, they match with a region of enhanced DOS. This variation accounts for the observed changes and local extrema in the microcavity dissociation rates in Fig. \ref{fig:wc_continuum}(a).}
{Note when the diatomic dissociation occurs exclusively via the doorway state (as in Sec. III.A.), a few overtone transitions transferring population to the highest energy bound state control the reaction rate. However, when dissociation can proceed directly from each bound state, sensitivity analysis shows that numerous $\Delta v = +2$ transitions (e.g., $12 \rightarrow 14$, $13 \rightarrow 15$, and $14 \rightarrow 16$) can have a significant impact in the reaction rate. This feature challenges the assignment of dominant reaction pathways, yet relaxes the conditions for  enhancement or minor suppression of BIRD in a weakly coupled microcavity.} 

\begin{figure}[t]
    \centering
    \includegraphics[width=\linewidth]{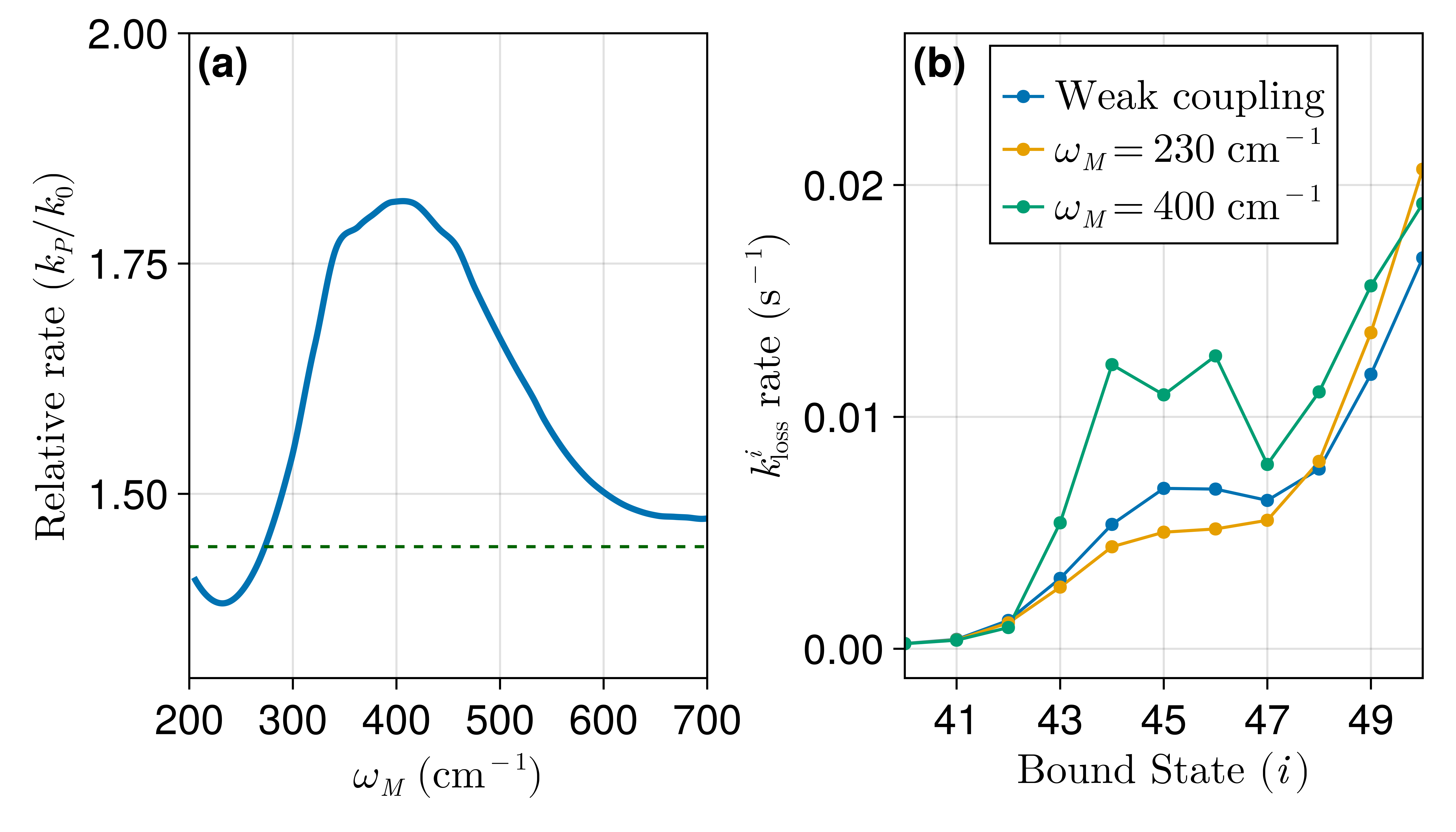}
    \caption{{(a) Ratio of polariton-assisted BIRD rates to free-space BIRD rates for $\Omega_R = 125~\text{cm}^{-1}$. The horizontal dashed green line corresponds to the weakly coupled microcavity relative rate $k_C / k_0$. The solid line was computed using a LOESS moving average\cite{Cleveland1979,Cleveland1988,Cleveland1991}, which highlights the overall trend while smoothing out sharp resonance features. (b) Decay rates from bound vibrational states ($i = 40$--50) to the continuum under weak coupling ($L_C = 5~\mu\text{m}$), and under strong coupling for different $\omega_M$ values in the polaritonic environment.}}
    \label{fig:sc_continuum}
\end{figure}

{Fig.~\ref{fig:sc_continuum}(a) presents the relative polariton-mediated BIRD rate, \(k_P / k_0\), as a function of the host material frequency \(\omega_M\) under strong coupling conditions with a collective light-matter interaction strength of $\Omega_R$ = 125 cm$^{-1}$ . Here, $L_C$ is fixed at 5~\(\mu\)m, and decay contributions from all vibrational states are included (unlike the results in Sec. III.B which considered only decay from the highest bound state). In contrast to the isolated sharp peaks observed when the reaction proceeds via the loosest bound state (Fig. \ref{fig:strong_coupling}), the introduction of multiple direct dissociative channels via bound-continuum transitions leads to a broader distribution of rate enhancements over a wide range of \(\omega_M\). While resonant features persist in the reaction rate profile corresponding to particular overtone transitions (Fig. \ref{fig:strong_coupling}a), the observed enhancements largely reflect cumulative contributions of decay channels proceeding directly from many intermediate vibrational bound states.}

{To estimate the effect of broad vibrational linewidths on the relative polariton-assisted BIRD rates, we applied locally estimated scatterplot smoothing (LOESS)\cite{Cleveland1979,Cleveland1988,Cleveland1991}, a generalized moving average method. The resulting LOESS curve, shown as a solid line in Fig.~\ref{fig:sc_continuum}, reveals the underlying polaritonic enhancement of the BIRD rates while smoothing over sharp features that arise from approximating the continuum states of the Morse potential using a dense but discrete set of box states. This averaging is also supported by the presence of physical broadening mechanisms, including thermal rotational population and weak rovibrational interactions, which naturally contribute to infrared linewidths.}

{The smoothed polariton-assisted BIRD rates reveal that at $\Omega_R$ = 125~cm\(^{-1}\), enhancement is possible when the host material frequency falls within the 300--500~cm\(^{-1}\) range. A mild suppression compared to the weak coupling case (\(k_P / k_C < 1\)) is also possible when the host material frequency falls within the 200--300~cm\(^{-1}\) range. We attribute these features to the modification of bound-continuum transition rates ($k_{\t{loss}}^{i}$) as demonstrated in Fig.~\ref{fig:sc_continuum}(b) which shows these for bound vibrational levels near the dissociation threshold. When \(\omega_M = 230~\text{cm}^{-1}\), the decay rates for intermediate bound states are slightly suppressed compared to weak coupling. In contrast, when \(\omega_M = 400~\text{cm}^{-1}\), decay rates for states \(i = 43\)--47 are enhanced due to the modified photon weighted polariton DOS. Therefore, in Fig.~\ref{fig:sc_continuum}(a), the minimum near \(\omega_M = 230~\text{cm}^{-1}\) arises from suppressed decay channels due to a locally depleted polariton DOS, while the maximum near \(\omega_M = 400~\text{cm}^{-1}\) results from enhanced decay rates promoted by enhanced polariton DOS. When the reaction is allowed to proceed through this mechanism, which is sensible for gas phase systems, maximum enhancements are reduced compared to the BIRD mediated by the highest energy bound state (or any particular specific doorway state). However, demanding resonance conditions are also lifted affording a more distributed enhancement profile that may offer a practical advantage for experimental realizations.}

\section{Conclusions}
This work presented a detailed analysis of BIRD in infrared microcavities. Our analysis revealed mechanisms for controlling dissociation phenomena in microcavities, particularly by manipulating the electromagnetic density of states (DOS) in weak and strong light-matter coupling regimes. Overtones play a critical role in the dissociation process under both weak and strong coupling cases, as the corresponding transitions act as rate-limiting steps for the dissociative process occurring via the highest energy-bound state. The radiative dissociation rate changes induced by a microcavity are essentially tied to how the photonic density of states is modified around specific overtone transitions driving the system into the dissociating (doorway) state.

We reported microcavity-assisted dissociation rate enhancements of $O(1)$ in the weak coupling regime. In microcavities where $L_C \leq 25 ~\mu\t{m}$, the photon DOS is greater than in free space for most relevant diatomic transitions, leading to a mild increase in dissociation rates. As $L_C$ increases beyond a few tens of microns, a complex pattern emerges from the interplay between enhanced and suppressed vibrational transitions, which can be rationalized by analyzing a few important overtones. Further increases in $L_C$ causes the BIRD rates to approach the free-space limit.

In the strong coupling regime, where polaritons mediate BIRD, we reported $O(10)$ enhancements when the host matter frequency is resonant with rate-limiting overtones of the dissociative process. This enhancement is due to the emergence of a Rabi frequency dependent enhanced photon-weighted polariton DOS at a targeted frequency, which enables the increase of transition rates between levels with energy differences that approach the low-energy boundary of the polariton stopgap from below. The largest enhancements are obtained when photon modes highly off resonant with the host molecular system are assumed to form polaritons, yet as we showed, sizable polariton-assisted BIRD rate enhancements remain achievable when these highly off resonant interactions are cutoff to an extent that depends on $\Omega_R$ and parameters controlling losses and disorder.
Ultimately, the BIRD enhancement factors here reported must be viewed as upper bounds for selected polariton-induced diatomic thermal radiative dissociation via the most energetic bound state.

The inclusion of bound-to-continuum transitions in our analysis alters the dissociation dynamics compared to the case where the reaction proceeds via a doorway state (the highest energy bound state). Dissociation via radiative bound-continuum transitions reveals a broader and smoother enhancement profile under microcavity length variation in the weak coupling regime. While the magnitude of the enhancement remains similar, the key difference lies in the involvement of a much greater number of vibrational transitions relative to the doorway mechanism. In the polariton-assisted case,  relative rate enhancement is generally smaller, but it no longer requires precisely matching a particular diatomic transition frequency to the host (strongly coupled) material. Instead, the multiplicity of direct dissociation pathways makes polariton-enhanced BIRD inherently more robust, enabling broadband enhancement across a wide spectrum of host material frequencies $\omega_M$.

Our work focused on how optical microcavities can modify diatomic blackbody infrared radiative dissociation. \textcolor{black}{While our analysis ignores collisional energy transfer, and therefore may be viewed as providing upper bounds to the microcavity effect on diatomic BIRD rates, the main qualitative conclusions expressed above are expected to apply broadly to systems where thermal radiation acts as a relevant energy source.} We anticipate that the framework developed here will aid in understanding and controlling thermal radiative-induced dissociation processes in more complex, polyatomic systems.\\

\textbf{Supporting Information.} Results for the HF and LiH molecules, temperature analysis, and additional computational details, including derivations, model parameters, and sensitivity analysis. \\

\textbf{Acknowledgments.} This work was supported by the donors of ACS Petroleum Research Fund under Doctoral New Investigator Grant 66399-DNI6. R.F.R. served as Principal Investigator on ACS PRF-66399-DNI6 that provided support for E. S. R.F.R also acknowledges support from NSF CAREER award Grant No. CHE-2340746, and startup funds provided by Emory University.\\

\textbf{Author Contributions.} G. J. R. A. and E. S. contributed equally to this work.

\bibliography{lib.bib}

\end{document}

\title{Supplementary Material:\\Polaritonic Control of Blackbody Infrared Radiative Dissociation}

\author{Enes Suyabatmaz}
\affiliation{Department of Physics, Emory University, Atlanta, GA, 30322}
\author{Gustavo J. R. Aroeira}
\affiliation{Department of Chemistry and Cherry Emerson Center for Scientific Computation, Emory University, Atlanta, Georgia 30322, United States}
\author{Raphael F. Ribeiro}
\email{raphael.ribeiro@emory.edu}
\affiliation{Department of Chemistry and Cherry Emerson Center for Scientific Computation, Emory University, Atlanta, Georgia 30322, United States}
\email{raphael.ribeiro@emory.edu}
\date{\today}

\maketitle
% Supporting Information Section in single column
\onecolumngrid
% Custom SI Formatting
\setcounter{figure}{0}
\setcounter{equation}{0}
\setcounter{table}{0}
\renewcommand{\thefigure}{S\arabic{figure}}
\renewcommand{\thetable}{S\arabic{table}}
\renewcommand{\theequation}{S\arabic{equation}}
\renewcommand\thesubsection{S\arabic{subsection}}
\renewcommand\thesubsubsection{\thesubsection.\arabic{subsubsection}}

\begin{spacing}{1.3}

\subsection{Model parameters for Diatomic Molecules}
We use a Morse potential to describe the potential energy of the examined diatomic molecules, 
\begin{align}
    V(r) = D_e \left(1 - e^{-\alpha (r - r_e)}\right)^2 \;,
\end{align}
where
\begin{align}
    \alpha &= \sqrt{\frac{k_e}{2D_e}}\;, \\[2mm]
    k_e &= m_r (2\pi c \nu_e)^2 \;,
\end{align}
with $c$ being the speed of light in the vacuum, $m_r$ the reduced mass and $\nu_e$ the harmonic frequency of the system. The energy levels are calculated as
\begin{align}
    E(n)/hc = \nu_e \left(n+\frac{1}{2}\right) - \nu_e \chi_e \left( n+\frac{1}{2}\right)^2 \,,
\end{align}
where $n$ is the vibrational quantum number, $h$ is Planck's constant and $\chi_e$ is the anharmonicity constant. The number of bound states is given by
\begin{align}
    n_\text{max} = \left\lfloor \frac{\sqrt{2m_r D_e}}{\hbar \alpha} - \frac{1}{2} \right\rfloor \;.
    \label{eq:nmax}
\end{align}
For NaLi molecule, the dipole function and Morse parameters were extracted from previous CCSDT/cc-pCVQZ calculations available in the literature \cite{Fedorov2014}. For the HF molecule, we obtained all Morse parameters from previous work on laser-induced dissociation \cite{Stine1979,Guldberg1991,Gross1993,kaluza_optimally_1994}. For LiH molecule, harmonic and anharmonic frequencies were obtained from experimental data\cite{Irikura2007}, and the dissociation energy was extrapolated from these parameters using the following relationship valid for a Morse potential:
\begin{align}
    D_e = \frac{hc\nu_e^2}{4(\nu_e\chi_e)} \;. \label{eq:De}
\end{align}
The dissociation energy ($D_e$) obtained using this method (2.641 eV) deviates slightly from reported experimental\cite{LiHDeE1,LiHDeE2,Irikura2007} and other \textit{ab initio} computations\cite{LiHDeT1,LiHDeT2,LiHDeT3} (approximately 2.515 eV). This discrepancy arises because the electronic ground-state LiH potential energy curve deviates from the Morse potential assumed for simplicity in Eq. \ref{eq:De}. However, we emphasize that the parameters used for NaLi, HF, and LiH are representative of these species, and a more precise description would not change the conclusions of this work in any significant way. All used Morse potential parameters are collected in Table \ref{tab:morseparameters}. 

\begin{figure}
    \centering
    \includegraphics[width=0.8\linewidth]{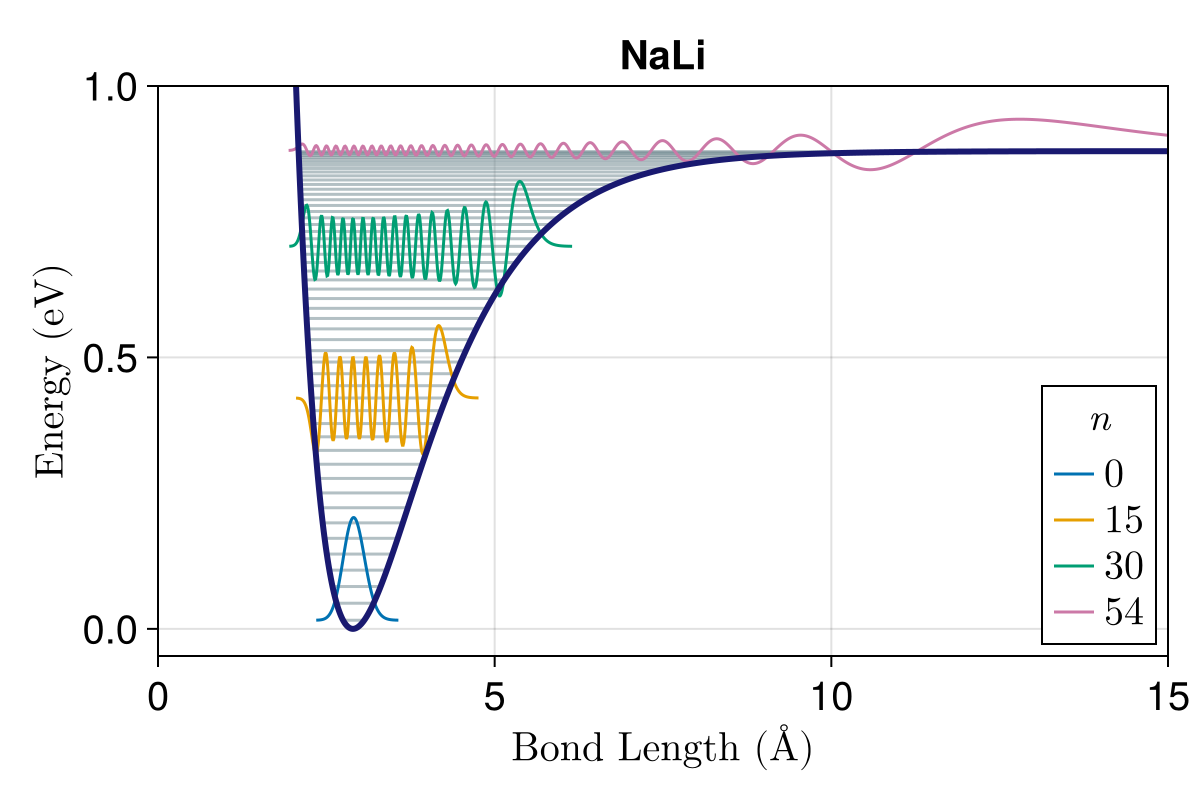}
    \caption{Morse potential and energy levels (shown as horizontal lines) used in this work for NaLi. Selected vibrational wave functions are also shown. They are shifted according to their corresponding energy for better visualization.}
    \label{fig:morseNaLi}
\end{figure}

\begin{table}[h]
    \caption{Morse parameters used for diatomic molecules.}
    \centering
    \setlength\extrarowheight{2mm}
    \begin{tabular}{|c|c|c|c|}
        \hline 
        \textbf{Constant} & \textbf{HF} & \textbf{LiH} & \textbf{NaLi} \\
        \hline
        \hline
        $D_e$ (eV)           & 6.123    & 2.641 & 0.882 \\
        \hline
        $r_e$ (\AA)          & 0.926    & 1.595 &  2.895\\
        \hline
        $\nu_e$ (cm$^{-1}$)  & 4138     & 1405  & 257.4 \\
        \hline
        $\nu_e \chi_e$ (cm$^{-1}$) & 86.70  & 23.162 & 2.33 \\
        \hline
        $m_r$ (a.u)          & 0.950     & 0.874 & 5.331  \\
        \hline
    \end{tabular}
    \label{tab:morseparameters}
\end{table}

Morse potentials corresponding to the given parameters are shown in Fig. \ref{fig:morseNaLi} along with selected wave functions. The wave functions are computed as described by Dahl and Springborg\cite{Dahl1988}. In Fig. \ref{fig:morseNaLi}, their extension is truncated when their amplitudes squared are below $1\times 10^{-8}$, giving us an idea of their spatial extent. It can be seen, for example, that in both cases, the final bound state ($n = 23, 29$ and $54$ for HF, LiH and NaLi, respectively) spans a bigger domain than lower energy states. This has consequences when choosing an integration grid as we discuss next. 

Electrical dipole transition matrix elements $\mu_{ij}$ can be computed from the vibrational wave functions as
\begin{align}
\mu_{ij} &= \int_{-\infty}^{+\infty}\psi_i^*(r)\mu(r)\psi_j(r) dr\,, \label{transitiondip}
\end{align}
where $\psi_k(r)$ represents the $k$-th vibrational state and $\mu(r)$ describes the electronic dipole moment change dependence on $r$. 

In the case of NaLi, the dipole function was constructed using a B-spline interpolation available in the \textsc{DataInterpolations.jl} package. The data used for the interpolation were obtained from previous CCSDT/cc-pCVQZ calculations \cite{Fedorov2014}. For completeness, these values are shown in Table \ref{tab:nali_dip}. 

\begin{figure}
    \centering
    \includegraphics[width=0.8\linewidth]{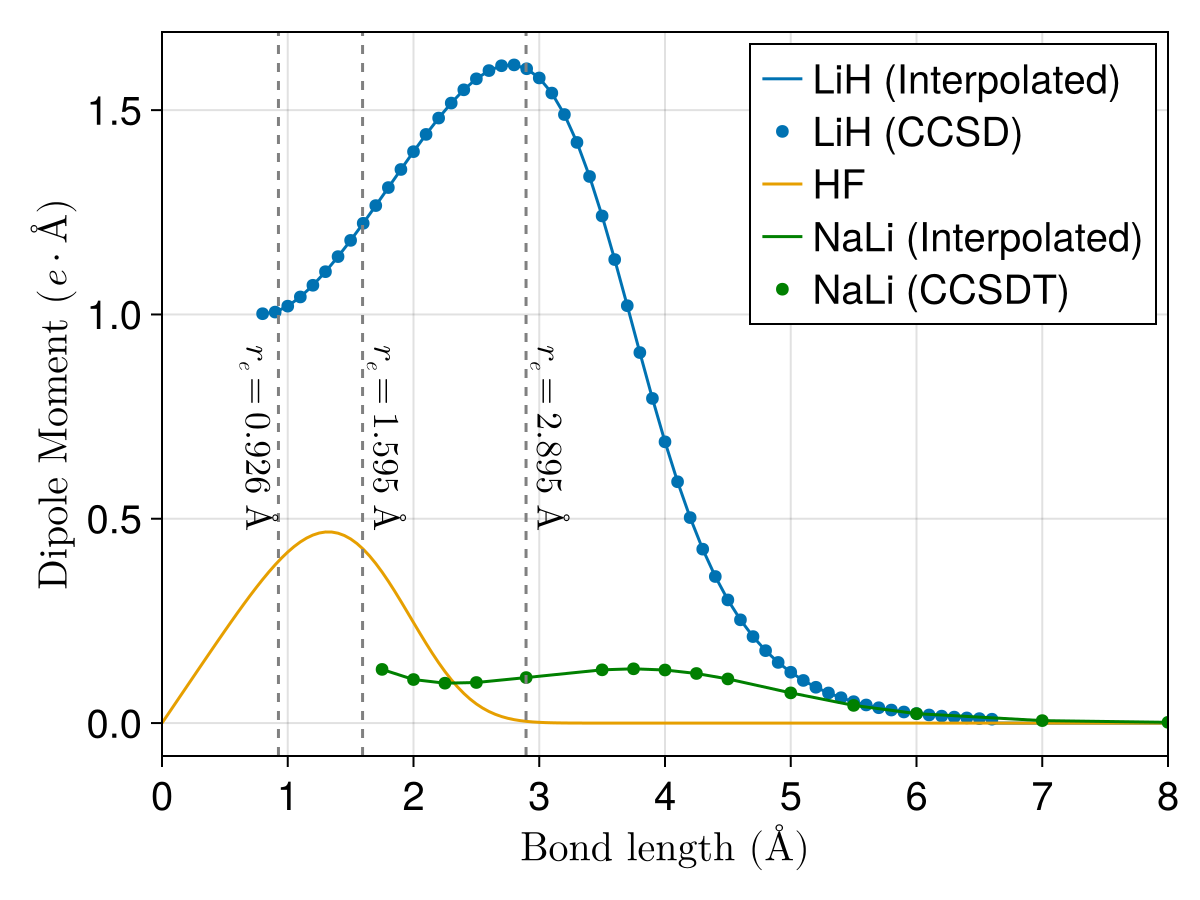}
    \caption{Dipole functions for HF, LiH, and NaLi. For HF, we use a model function ($\mu(r) = \mu_0 r e^{-\zeta r^4}$) described in ref. \citenum{kaluza_optimally_1994}. For LiH and NaLi, dipole moments were calculated using \textit{ab initio} quantum chemistry, and interpolation was used to obtain smooth functions. Dashed vertical lines indicate the equilibrium radius for HF (0.926 \AA), LiH (1.595 \AA), and NaLi (2.895 \AA).}
    \label{fig:dips}
\end{figure}

Fig. \ref{fig:dips} shows the dipole function for NaLi. Calculating dipoles for small bond lengths becomes increasingly difficult due to convergence issues. This region of configuration space is irrelevant for the computation of transition dipole matrix elements since the vibrational wave functions decay exponentially at small $r$ (see Fig. \ref{fig:morseNaLi}). Hence, we enforce $\mu \rightarrow 0$ in the interpolation as $r \rightarrow 0$. This has no impact on the computed transition dipole matrix elements. Eq. \ref{transitiondip} is evaluated numerically with integration domain spanning $r_\text{min} = 0.0$ \AA{} to $r_\text{max} = 50$ \AA{} with a grid spacing of $0.01$ \AA.

\begin{table}[h]
\centering
\caption{Computed dipole values for NaLi at the CCSDT level of theory\cite{Fedorov2014}}
\label{tab:nali_dip}
\setlength\extrarowheight{2mm}
\begin{tabular}{|c|c|}
\hline
Bond length (\AA) & Dipole moment ($e\cdot\AA$) \\
\hline
1.8 & 0.13137057268257263 \\
\hline
2.0 & 0.10659545675669918 \\
\hline
2.2 & 0.09764310394314826 \\
\hline
2.5 & 0.0993086579549717 \\
\hline
2.9 & 0.11138392454069153 \\
\hline
3.5 & 0.130329601425183 \\
\hline
3.8 & 0.13282793244291813 \\
\hline
4.0 & 0.12991321292222713 \\
\hline
4.2 & 0.12137724861163207 \\
\hline
4.5 & 0.10805281651704468 \\
\hline
5.0 & 0.07390895927466447 \\
\hline
5.5 & 0.04330440430740904 \\
\hline
6.0 & 0.023109561914050017 \\
\hline
7.0 & 0.006037633292859915 \\
\hline
8.0 & 0.0018737482633013527 \\
\hline
10.0 & 0.0006245827544337843 \\
\hline
12.0 & 0.0006245827544337843 \\
\hline
15.0 & 0.0006245827544337843 \\
\hline
\end{tabular}
\end{table}

%\begin{spacing}{1.3}

% Supplementary content

% Morse and dipole parameters
%\include{Morse}

% Smallest eigenvalue as the reaction rate
\subsection{Rate Constant from Master Equation}
The dynamics of the reactive molecule is modeled with a Markov Master equation routinely employed in studies of blackbody infrared radiative molecules in free space \cite{dunbar2004bird}. In this formalism, the time evolution of the vibrational population vector $[\mathbf{N}] = [N_1, N_2,...,N_{n}]$ is described by the equation
\begin{equation}
    \frac{d[\mathbf{N}]}{dt} = - \mathbf{J} \cdot [\mathbf{N}], \label{matrix_form}
\end{equation}
where $N_\nu$ is the population of state $\nu$, $n = n_{\t{max}}$ (Eq. \ref{eq:nmax}), and $\mathbf{J}$ is the transition matrix
\begin{equation}
\mathbf{J} = 
\begin{pmatrix}
    (k_{12} + \cdots + k_{1n}) & -k_{21} & \cdots & -k_{n1} \\
    -k_{12} & (k_{21} + \cdots + k_{2n}) & \cdots & -k_{n2} \\
    \vdots & \vdots & \ddots & \vdots \\
    -k_{1n} & -k_{2n} & \cdots & k_{\text{loss}}
\end{pmatrix}.
\end{equation}
Each matrix element $J_{ij}$ with $i \neq j$ corresponds to the transition rate $k_{ij}$ from state $i$ to state $j$. Diagonal elements $J_{ii}$ represent the total outgoing rate from state $i$. This includes all transitions from state $i$ to every other state $j$. The diagonal elements are always positive. The last row of $\mathbf{J}$ includes $k_{\text{loss}}$, which represents the irreversible loss of population due to the dissociation in the highest energy bound level.

The eigenvalues of the transition matrix $J$ are guaranteed to be positive\cite{valance1966theoretical}, and the smallest eigenvalue $\lambda_1$ of $\mathbf{J}$ is the rate constant of the unimolecular reaction. This is because, in the long-time limit, the system relaxes to the slowest decaying mode, i.e., the smallest eigenvalue of the transition matrix.

To show that the smallest eigenvalue gives the reaction rate constant, we note that the solution to Eq. \ref{matrix_form} can be expressed as \cite{gilbert1990theory}:
\begin{equation}
    [\mathbf{N}(t)] = \sum_{i=1}^{n} c_i~e^{-\lambda_i t} [\mathbf{v}_i],
\end{equation}
where $\lambda_i$ are the eigenvalues of $\mathbf{J}$, and $[\mathbf{v}_i]$ are the corresponding eigenvectors. Let the eigenvalues be ordered as
$\lambda_1 < \lambda_2 < \dots < \lambda_{n}.$ As $t \to \infty$, the term with the smallest eigenvalue $\lambda_1$ dominates the time evolution and all of the remaining $e^{-\lambda_2 t},\dots, e^{-\lambda_n t}$ approach zero rapidly because their corresponding \(\lambda_i\) values are larger than \(\lambda_1\). The term \(e^{-\lambda_1 t}\) gives the slowest decay, since \(\lambda_1\) is the smallest. This means that the slowest decaying term effectively dominates the sum after a sufficiently long induction time $t \gg 1/\lambda_{2}$ leading to

\begin{equation}
    [\mathbf{N}(t)] \sim c_1 e^{-\lambda_1 t} [\mathbf{v}_1], ~ t \gg 1/\lambda_2
\end{equation}
where $c_1$ is a constant determined by the initial conditions. This shows that the population at any energy level decays at the same rate $\lambda_1$ and thus $\lambda_1$ is the unimolecular reaction rate\cite{hanggi1990reaction}.

\subsection{Decay Rates from a Bound Vibrational Level to Continuum States}

To calculate the decay rate of a diatomic molecule from a bound vibrational level to continuum states above the dissociation threshold, we use Fermi’s Golden Rule to describe the transition from a discrete bound state to a continuum of unbound, or scattering, states:

\begin{equation}
k_{\t{loss}}^{i} = \frac{2\pi}{\hbar}\sum_{f>\nu_{\t{max}}} \left|\langle \psi_f | \hat{H}' | \psi_i \rangle\right|^2 \rho(E_f)
\label{kloss}
\end{equation}
where $|\psi_i\rangle$ is the initial bound vibrational eigenstate, $|\psi_f\rangle$ is a final continuum state with energy $E_f$, $\hat{H}'$ represents the interaction Hamiltonian (typically the dipole operator $\hat{\mu}$ in IR-driven transitions), and $\rho(E_f)$ is the density of final states per unit energy. The summation over $f$ effectively becomes an integral when the continuum is dense, and $\rho(E_f)$ accounts for the normalization of the scattering states.

We have an analytical form of the bound state wavefunction $\psi_i$, which for vibrational states can be obtained by solving the Schrödinger equation for a Morse potential. For the continuum state wavefunction $\psi_f$, scattering states that describe the free motion of the atoms after dissociation are often used. We calculate scattering states by numerical methods, such as the Numerov method\cite{pillai2012matrix}, and use them to compute transition dipole matrix elements and decay rates to the continuum. In practice, we approximate the continuum states of the Morse potential using a dense but discrete set of box states, allowing us to converge the $k_{\text{loss}}^i$ rates numerically.

The BIRD kinetics via radiative bound-continuum transitions is modeled using a Master Equation framework similarly, where the time evolution of vibrational state populations is governed by both radiative transitions between bound states and irreversible loss to the dissociation continuum. In the presence of continuum states, each bound vibrational level can decay not only to other bound levels via stimulated absorption and emission, but also to a continuum of unbound states. This provides an additional decay pathway for each vibrational state, characterized by the continuum loss rate $k_i^{\t{loss}}$ derived from Eq. \ref{kloss}. The transition matrix $\mathbf{J}$ now includes bound-continuum decay rates on the diagonal elements: 

\begin{equation}
\mathbf{J} = 
\begin{pmatrix}
    (k_{12} + \cdots + k_{1n}+ k^{\t{loss}}_{1}) & -k_{21} & \cdots & -k_{n1} \\
    -k_{12} & (k_{21} + \cdots + k_{2n}+k^{\t{loss}}_{2}) & \cdots & -k_{n2} \\
    \vdots & \vdots & \ddots & \vdots \\
    -k_{1n} & -k_{2n} & \cdots &(k_{n1} + \cdots + k_{n.n-1}+k^{\t{loss}}_{n})
\end{pmatrix}.
\end{equation}

The presence of $k_i^{\t{loss}}$ on the diagonal ensures that population is irreversibly removed from each level due to dissociation. The matrix $\mathbf{J}$ remains positive-definite, and its smallest eigenvalue $\lambda_1$ continues to govern the long-time dynamics of the system. Specifically, the solution to Eq.\ref{matrix_form} in the asymptotic limit is:

\begin{equation} [\mathbf{N}(t)] \sim c_1 e^{-\lambda_1 t} [\mathbf{v}_1], 
\end{equation}

indicating that the unimolecular dissociation rate constant is identified with the smallest eigenvalue $\lambda_1$ of the full transition matrix, now including continuum losses.

\begin{figure}
    \centering
    \includegraphics[width=\linewidth]{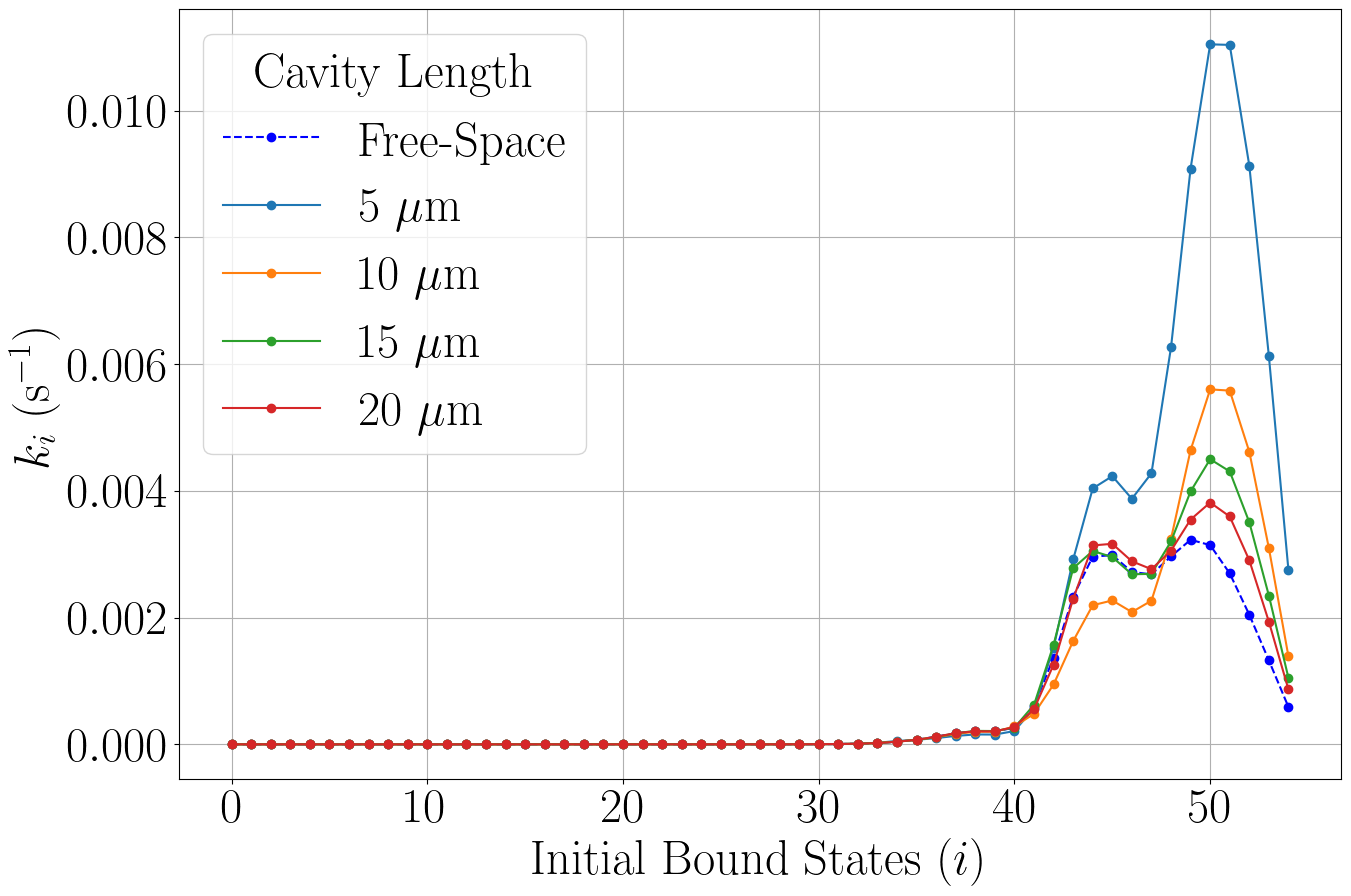}
    \caption{ Decay rates from bound states to the continuum for different cavity lengths. The decay rate $k_i$ increases significantly for higher initial bound states, especially in shorter cavities (5 $\mu$m).}
    \label{fig:kloss}
\end{figure}

% Density of States Derivations
\subsection{Free space thermal radiation}

\par The Hamiltonian for the free photon field, which describes the energy of a system of non-interacting photons, is given by:
\begin{equation}
    H_{0} = \sum_{\boldsymbol{k},\lambda} \hbar\omega_{\boldsymbol{k}}a_{\boldsymbol{k},\lambda}^{\dagger}a_{\boldsymbol{k},\lambda},
\end{equation}
where $\boldsymbol{k}$ is the wave vector, $\lambda = 1,2$ is the polarization, $\omega_{\boldsymbol{k}}$ is the angular frequency of the mode with wave vector $\boldsymbol{k}$, and $a_{\boldsymbol{k},\lambda}^{\dagger}$ and $a_{\boldsymbol{k},\lambda}$ are the creation and annihilation operators for the photon in that mode.

\par The free electromagnetic field partition function $Z(\beta) = \text{tr}(e^{-H/k_\t{B} T})$ allows us to compute all thermodynamic quantities of the bare field. The total radiation thermal energy at inverse temperature $\beta = 1/k_{\t{B}}T$ is given by
\begin{equation}
   E(T) = -\frac{\partial \log Z(\beta)}{\partial \beta}
   = \sum_{\lambda} \dfrac{\hbar \omega_{k}}{e^{\hbar\omega_k/k_{\t{B}}T} - 1},
\end{equation}
and the mean photon number for a mode with frequency $\omega$ is given by
\begin{equation}
    n(\omega,T) = \dfrac{1}{e^{\frac{\hbar\omega_k}{k_B T}} - 1}.
\end{equation}

The density of photon states $D_0(\omega)$ corresponds to the number of available electromagnetic modes per unit frequency per unit volume at a frequency $\omega$. For simplicity, we consider free space to be a three-dimensional periodic box with lengths $L_x = L_y = L_z = L$. The free space DOS is given by\cite{barnes_classical_2020}:
\begin{equation}
    D_0(\omega) = \frac{1}{V} \sum_{\boldsymbol{k}, \lambda} \delta(\omega - c k)
\end{equation}
where $V = L^3$. Using the limit where $L \rightarrow \infty$ and
\begin{equation}
    \frac{1}{V} \sum_{\boldsymbol{k}} \rightarrow \frac{1}{(2\pi)^3} \int d^3k,
\end{equation}
we obtain the free space photon DOS using standard manipulations
\begin{equation}
    D_0(\omega) = 2\frac{4\pi }{(2\pi)^3} \cdot \frac{\omega^2}{c^3} =  \frac{\omega^2}{\pi^2 c^3}.
\end{equation}

The energy density $\rho_0(\omega, T)$ of free-field blackbody radiation can be obtained by multiplying the density of states with the Bose-Einstein distribution function and the energy of each photon:
\begin{equation}
    \rho_0(\omega, T) = D_0(\omega) n(\omega, T) = \dfrac{\omega^2}{\pi^2 c^3} \cdot \dfrac{\hbar \omega}{e^{\frac{\hbar \omega}{k_B T}} - 1}.
\end{equation}

\subsection{Empty microcavity thermal radiation}
Consider an optical Fabry-Perot microcavity composed of two parallel perfect mirrors with area $S$ separated by a distance $L$. The microcavity mode frequencies are given by
\begin{equation}
   \omega_{\text{C}}(m,q) = c \sqrt{q_{x}^2 + q_{y}^2 + \left( \frac{m\pi}{L} \right)^2},
   \label{eq:cavity_modes}
\end{equation}
where $\mathbf{q} = (q_{x}, q_{y})$ is the in-plane wave vector (momentum) corresponding to the in-plane momentum of the cavity modes, and $m$ is zero (transverse magnetic polarization only) or a positive integer (for transverse electric and tranverse magnetic modes) representing the quantization of the wavevector in the $z$-direction (perpendicular to the mirror planes) \cite{zoubi2005microscopic, jackson2021classical}. It follows the empty microcavity DOS is given by
\begin{equation}
    D_{\text{C}}(\omega) = \frac{1}{V} \sum_{m,\mathbf{\textbf{q}},\lambda} \delta[\omega - \omega_{\text{C}}(m,q)],
    \label{eq:dos_initial}
\end{equation}
where $V = LS$ is the resonator volume, $\lambda$ denotes the polarization state, and $m$ and $\mbf{q}$ are as discussed above. 

Using standard manipulations, we can write the total density of microcavity photon states as \cite{barnes_classical_2020}
\begin{equation}
    D_C(\omega) = \frac{\omega}{\pi c^2 L} \left\lfloor \frac{\omega L}{\pi c} \right\rfloor + \frac{\omega}{2 \pi c^2 L},
\end{equation}
where $\lfloor x \rfloor$  equals the greatest integer less than or equal to $x$. It follows the microcavity blackbody radiation energy density $\rho_C(\omega, T)$ is
\begin{equation}
    \rho_C(\omega, T) = D_C(\omega) n(\omega,T) = \frac{\left\{ \left\lfloor \frac{\omega L}{\pi c} \right\rfloor + \frac{1}{2} \right\} \omega}{\pi c^2 L}  \frac{\hbar \omega}{e^{\frac{\hbar \omega}{k_B T}} - 1}.
\end{equation}

% Photon Weighted Density of States Derivations
\subsection{Microcavity thermal radiation and density of states under strong light-matter coupling}
In the strong coupling regime, the Hamiltonian of the system is the sum of the Hamiltonians of the matter, field, and the matter-field interaction:

\begin{equation}
    H_{\t{tot}} = H_{\t{L}} + H_{\t{SM}} + H_{\t{INT}},
\end{equation}
where each term is defined as follows: The free photon field Hamiltonian is
\begin{equation}
    H_{L} = \sum_{\mathbf{q}, m, \lambda} \hbar \omega_C(m,q) \left( a^{\dagger}_{m\mbf{q}\lambda} a_{m\mbf{q}\lambda} + \frac{1}{2} \right)
\end{equation}
with in-plane wavevector $\mbf{q} = (q_x, q_y)$, the mode number $m$, and the polarization index $\lambda$ defined in Eqs. \ref{eq:cavity_modes}, \ref{eq:dos_initial}. The matter Hamiltonian describes a homogeneous dispersionless isotropic material ensemble with significant infrared oscillator strength at a suitable frequency $\omega_M$ with effective Hamiltonian \cite{ashida2021cavity}:
\begin{equation}
H_M = \int_{D_M} \frac{d^3r}{v} \left[ \frac{\bm{\pi}^2(\mathbf{r})}{2m^*} + \frac{m^*\omega_M^2 \bm{\phi}^2(\mathbf{r})}{2} \right],
\end{equation}
where $D_M$ is the region of space occupied by the material, $v$ is the volume occupied by a single local oscillator (``unit-cell volume''), $m^*$ is the effective mass, $ \mbf{r} = (x,y,z) \in D_M$ and \( \bm{\pi}(\mathbf{r}) \) and \( \bm{\phi}(\mathbf{r}) \) represent the conjugate momentum and the matter displacement field respectively. The latter can be rewritten as:
\begin{equation}
\bm{\phi}(\mathbf{r}) = \sqrt{\frac{\hbar}{2m^* N\omega_M }} \sum_{\mathbf{q}, m, \lambda} \left[ b_{m\mbf{q}\lambda}  \mbf{u}_{m\mbf{q}\lambda}(\mathbf{r}) e^{i \mathbf{q} \cdot \mathbf{r_{\parallel}}} + b_{m\mbf{q}\lambda}^\dagger  \mbf{u}^{*}_{m\mbf{q}\lambda}(\mathbf{r}) e^{-i \mathbf{q} \cdot \mathbf{r_{\parallel}}} \right],
\end{equation}
where \( b_{m\mbf{q}\lambda} \) and \( b_{m\mbf{q}\lambda}^\dagger \) are the annihilation and creation operators for the collective matter excitation labeled by ($m\mbf{q}\lambda$), $N = S/v$ denotes the number of molecular modes per unit length, \( \mbf{u}_{m\mbf{q}\lambda}(\mathbf{r}) \) is the mode spatial profile vector and $\mbf{r}_{\parallel} = (x,y)$. Each local oscillator vibration is triply degenerate. The excitations can be treated as bosonic following the standard dilute excitation limit. In this case, the Hamiltonian can be written in a simple way using second-quantization
\begin{equation}
    H_M = \sum_{\mathbf{q}, m, \lambda} \hbar \omega_{M} b_{m\mbf{q}\lambda}^\dagger b_{m\mbf{q}\lambda}.
\end{equation}
We work in the dipole gauge, where the light-matter interaction is given by
\begin{equation}
    H_{\text{int}} = \frac{1}{2\epsilon_0} \int_{D_M} \left[-2\mathbf{D}(\mathbf{r}) \cdot \mathbf{P}(\mathbf{r}) + \mathbf{P}^2(\mathbf{r}) \right] d^3 \mathbf{r},
\end{equation}
where \( \mathbf{P}(\mathbf{r}) \) is the matter polarization density given by:
\begin{equation}
\mathbf{P}(\mathbf{r}) = \frac{eZ^* \bm{\phi}(\mathbf{r})}{v},
\end{equation}
where \( eZ^* \) is the effective charge, and \( \mathbf{D}(\mathbf{r}) \) is the electrical displacement field \cite{craig1998molecular, power1982quantum}
\begin{equation}
\mathbf{D}(\mathbf{r}) = \sum_{\mathbf{q}, m, \lambda} \sqrt{\frac{\epsilon_0 \hbar \omega_C(m,q)}{2V}}  \left( \mbf{u}_{m\mbf{q}\lambda}(\mathbf{r})a_{m\mbf{q}\lambda}e^{i \mathbf{q} \cdot \mathbf{r}_{\parallel}} - \mbf{u}^*_{m\mbf{q}\lambda}(\mathbf{r}) a^{\dagger}_{-m\mbf{q}\lambda}e^{-i \mathbf{q} \cdot \mathbf{r}_{\parallel}} \right),
\end{equation}
and $V=SL_{C}$ is the microcavity volume. Inserting the polarization operator expression in second-quantization representation into the light-matter interaction and separately writing linear and quadratic terms of the interaction Hamiltonian in the matter polarization field\cite{todorov2012intersubband} we obtain 
\begin{align}
& H_{\text{int,1}} = i \sum_{\mathbf{q}, m, \lambda} \frac{\hbar \Omega_R}{2} \sqrt{\frac{\omega_C(m,q)}{\omega_{M}}} \left( a_{m\mbf{q}\lambda}^\dagger - a_{-m\mbf{q}\lambda} \right) \left( b_{m\mbf{q}\lambda} + b_{-m\mbf{q}\lambda}^\dagger \right), \\
& H_{\text{int,2}} = \sum_{\mathbf{q}, m, \lambda} \hbar \omega_{M} b_{m\mbf{q}\lambda}^\dagger b_{m\mbf{q}\lambda} + \frac{\hbar \Omega_R^{2}}{4 \omega_{M}} \left( b_{m\mbf{q}\lambda}^\dagger + b_{m\mathbf{-q}\lambda} \right) \left( b_{m\mathbf{-q}\lambda}^\dagger + b_{m\mbf{q}\lambda} \right)
 %& H_{M} + H_{\text{int,2}} = \sum_{\mathbf{q}, m, \lambda} \hbar \omega_{M} b_{m\mbf{q}\lambda}^\dagger b_{m\mbf{q}\lambda} + \frac{\hbar \Omega_R^{2}}{4 \omega_{M}} \left( b_{m\mbf{q}\lambda}^\dagger + b_{\mathbf{-q}m\lambda} \right) \left( b_{\mathbf{-q}m\lambda}^\dagger + b_{m\mbf{q}\lambda} \right).
\end{align}
where \( \Omega_R = \sqrt{\frac{(eZ^*)^2}{\epsilon_0 m^* v}} \) is the coupling constant, describing the strength of the interaction between the light field and the molecular excitations. 

To find the normal modes of the light-matter system, we first define a new set of operators \( p_{m\mbf{q}\lambda} \) such that:
\begin{equation}
    [p_{m\mbf{q}\lambda}, H_{M} + H_{\text{int,2}}] = \hbar \Omega_{M} p_{m\mbf{q}\lambda}, \quad \Omega_{M}^2 = \omega_{M}^2 + \Omega_R^{2}
\end{equation}
where \( \Omega_{M} \) represents the renormalized material excitation due to light-matter coupling inside the microcavity. The new polarization operators \( p_{m\mbf{q}\lambda} \) are written in terms of the original matter operators \( b_{m\mbf{q}\lambda} \) and their conjugates:
\begin{equation}
    p_{m\mbf{q}\lambda} = \frac{\Omega_{M} + \omega_{M}}{2 \sqrt{\Omega_{M} \omega_{M}}} b_{m\mbf{q}\lambda} + \frac{\Omega_{M} - \omega_{M}}{2 \sqrt{\Omega_{M} \omega_{M}}} b_{m-\mbf{q}\lambda}^\dagger.
\end{equation}
This transformation effectively diagonalizes the quadratic part of the Hamiltonian, allowing us to describe the system in terms of new quasi-particles that reflect the modified nature of the molecular excitations under strong coupling. The total Hamiltonian, now expressed in terms of the new operators, takes the form:
\begin{equation}
\begin{split}
    H_{\text{total}} = \sum_{\mathbf{q}, m, \lambda} \hbar \Omega_{M} p_{m\mbf{q}\lambda}^\dagger p_{m\mbf{q}\lambda} + \sum_{\mathbf{q}, m, \lambda} \hbar \omega_{C}(m,q) \left( a_{m\mbf{q}\lambda}^\dagger a_{m\mbf{q}\lambda} + \frac{1}{2} \right) \\
    + i \sum_{\mathbf{q}, m, \lambda} \hbar\frac{\Omega_R}{2} \sqrt{\frac{\omega_{C}(m,q)}{\Omega_{M}}} \left( a_{m\mbf{q}\lambda}^\dagger - a_{-m\mbf{q}\lambda} \right) \left( p_{m\mbf{q}\lambda} + p_{-m\mbf{q}\lambda}^\dagger \right),
\end{split}
\end{equation}

\par To obtain the polaritonic (hybrid) normal-modes after performing this initial matter Bogoliubov transformations, we note the total light-matter Hamiltonian can be written as: 
\begin{equation}
\begin{split}
    H_{\text{total}} = \sum_{\mathbf{q}, m, \lambda} \hbar \omega(m, q)\Pi_{m\mbf{q}\lambda}^\dagger \Pi_{\mathbf{q} m \lambda} ,
\end{split}
\end{equation}
where $\Pi_{\mathbf{q} m \lambda}^\dagger$ and $\Pi_{m\mbf{q}\lambda}$ are the creation and annihilation operators of polariton modes. In what follows, we simplify the notation by replacing $m\mbf{q}\lambda$ in polariton operators by $k$. Polariton annihilation operators are given by
\begin{equation}
    \Pi_{k} = x_k a_k + y_k a_{-k}^{\dagger} + z_k p_k + t_k p_{-k}^{\dagger}.
\end{equation}
where  \(x_k\) and \(y_k\) are the Hopfield coefficients representing the photonic contributions, and \(z_k\) and \(t_k\) are the coefficients representing the matter contributions. These coefficients are obtained by enforcing the bosonic commutation relations for the polariton operators and the condition $[\Pi_k, H_{\t{tot}}] = \hbar\omega(k) \Pi_{k}$ where $\omega(k)$ is a normal-mode (polariton) frequency. This leads to the set of linear equations.
\begin{align}
 &   |x_k|^2 - |y_k|^2 + |z_k|^2 - |t_k|^2 = 1, \label{eq:norm} \\
& [\omega(k) - \omega_C(m,q)] x_k + G \sqrt{\frac{\omega_C(m,\mbf{q})}{\Omega_M}} (z_k + t_k) = 0,\\
& [\omega(k) + \omega_C(m,q)] y_k + G \sqrt{\frac{\omega_C(m,q)}{\Omega_M}} (z_k + t_k) = 0,\\
&[\omega(k) - \Omega_M] z_k - G \sqrt{\frac{\omega_C(m,q)}{\Omega_M}} (x_k - y_k) = 0,\\
&[\omega(k) + \Omega_M] t_k + G \sqrt{\frac{\omega_C(m, q)}{\Omega_M}} (x_k - y_k) = 0,
\end{align}
where \( \omega_C(m,\mbf{q}) \) is the cavity mode frequency, and \( G = \frac{i \hbar \Omega_R}{2} \) is the light-matter coupling constant. These equations can be used to find the polariton frequencies as the solutions to
\begin{equation}
    [\omega^2(k) - \Omega_M^2] [\omega^2(k) - \omega_C^2(m,q)] - \Omega_R^{2} \omega_C^2(m,q) = 0.
\end{equation}
Specifically, the polariton frequencies $\omega_{\pm}(k)$ are given by
\begin{equation}
    \omega_{\pm}^2(k) = \frac{1}{2} \left[\omega_C^2(m,q) + \omega_M^2 + \Omega_R^{2} \pm \sqrt{[\omega_M^2 + \Omega_R^{2} - \omega_C^2(m,q)]^2 + 4 \omega_C^2(m,q) \Omega_R^{2}} \right].
    \label{eq:polaritonmodes}
\end{equation}
Substituting the expressions for \(x_k\) and \(z_k\) into Eq. \ref{eq:norm}, we get:
\begin{equation}
    y_k = \frac{[\omega(k) - \omega_C(m,q)][\omega^2(k) - \Omega_M^2]}{2 \sqrt{\omega(k) \omega_C(m,q)} \sqrt{[\omega^2(k) - \Omega_M^2]^2 + \Omega_R^{2} \omega_C^2(m,q)}},
\end{equation}
\begin{equation}
    x_k = \frac{[\omega(k) + \omega_C(m,q)][\omega^2(k) - \Omega_M^2]}{2 \sqrt{\omega(k) \omega_C(m,q)} \sqrt{[\omega^2(k) - \Omega_M^2]^2 + \Omega_R^{2} \omega_C^2(m,q)}}.
\end{equation}

We can rewrite the photon and the matter field operators in terms of the polaritonic operators
\begin{align}
   & a_{k} = x^*_{k}\Pi_k - y_{k}\Pi^{\dagger}_{-k}, \\
   & p_{k} = z^*_{k}\Pi_k - t_{k}\Pi^{\dagger}_{-k}.
\end{align}
The photonic part of each polariton mode $k = (m,\mbf{q},\lambda)$ with frequency $\omega$ is given by $P_C(\omega) = |x_k|^2 - |y_k|^2$ and matter part is $P_M(\omega) = |z_k|^2 - |t_k|^2$ \cite{todorov2012intersubband}. In terms of the matter and photon frequencies and the collective light-matter coupling strength, $P_C(\omega)$ is given by
\begin{equation}
        P_C(\omega) %&=|x_k|^2 - |y_k|^2 %\\
        %&= \frac{(\omega^2 - \Omega_M^2)^2}{(\omega^2 - \Omega_M^2)^2 + \omega_C^2 \Omega_R^{2}} \\
        = \frac{(\omega^2 - \omega_M^2 - \Omega_R^{2})^2}{(\omega^2 - \omega_M^2 - \Omega_R^{2})^2 + \omega_C^2(\omega) \Omega_R^{2}},
\end{equation}
where $\omega_C(\omega)$ is the frequency of the photon mode that hybridizes with matter polarization to give a polariton mode with frequency $\omega$ (Eq. \ref{eq:polaritonmodes}). The group velocity $v_g(\omega, q)$ of a polariton with frequency $\omega$ and in-plane wave vector magnitude $q$ can be obtained by taking the derivative of Eq.~\ref{eq:polaritonmodes} with respect to $q$
\begin{equation}
    v_g(\omega, q) = \frac{d\omega_{\pm}}{dq} = \frac{qc^2}{2 \omega} \left( 1 \pm \frac{\omega_C^2 - \omega_M^2 + \Omega_R^{2}}{\sqrt{[\omega_C^2 + \omega_M^2 + \Omega_R^{2}]^2 - 4 \omega_C^2(\omega) \omega_M^2}} \right),
\end{equation}
where we take the positive sign if the polariton mode with frequency $\omega$ is in a UP branch [if $\omega > \omega_C(\omega)]$  and the negative sign otherwise.

The previous results will be employed to obtain the polariton density of states $D(\omega)$ defined as
\begin{equation}
D(\omega) = \frac{1}{V} \sum_{\mathbf{q}, m, \lambda, \alpha = \pm} \delta[\omega - \omega_{\alpha}(m,q)].\label{eq:domegapm_int}
\end{equation}
We obtain a closed-form expression of the polariton density of states by using
\begin{equation}
\delta[\omega - \omega_{\alpha}(m, q)] = \frac{\delta[q - q(m, \omega_\alpha)]}{v_g[\omega,q(m,\omega)]}
\end{equation}
where \( q(m, \omega) \) is the in-plane wave vector magnitude that corresponds to the polariton frequency $\omega = \omega_{\alpha}(m, q)$ for a given mode order \( m \) and polarization \( \lambda \). Substituting this result into Eq. \ref{eq:domegapm_int} and integrating over the space of in-plane wave vectors, we obtain
\begin{equation}
D(\omega) = \frac{1}{L_C} \sum_{m, \lambda} \frac{q(m,\omega)}{2\pi} \frac{1}{v_g(\omega,q)}\theta[\omega_{C}(\omega)-m\pi c/L_C]
\end{equation}
where $\theta[.]$ is the Heaviside step function. 

The rate of spontaneous emission of a material system under weak coupling with the electromagnetic field of a polaritonic device is controlled by the photon-weighted polariton DOS (PDOS) 
\begin{align}
    D_P(\omega) & = P_C(\omega) D(\omega)  \\ 
    & = \frac{1}{L_C} \sum_{m, \lambda} \frac{q(m,\omega) P_C(\omega)}{2\pi v_g(\omega,q)}\theta[\omega_{C}(\omega)-m\pi c/L_C] \label{eq:PDOS}.
\end{align}
This quantity can be split into the sum of a contribution from  \( m = 0 \) and \( m > 0 \). For \( m = 0 \), there is only TM$_{0}$ mode contribution, but for \( m > 0 \), there are both TE and TM modes, resulting in a factor of 2 for the polarization sum. It follows that the PDOS can be written as:
\begin{equation}
D_P(\omega) = \frac{1}{L_C}  \sum_{m>0} \frac{q(m,\omega) P_C(\omega)}{\pi v_g(\omega,q)}\theta[\omega_\t{C} - m\pi c/L] + \frac{q_0(\omega) P_C(\omega)}{2\pi L_C v_g[\omega,q_0(\omega)]},
\end{equation}
where $q_0(\omega) = q(m=0,\omega)$. This expression for $D_P(\omega)$ can be written as a single sum by denoting polariton frequencies by $\omega$ and writing
\begin{equation}
    D_{P}(\omega) = \sum_{m=0}^{\infty}\left(1-\frac{\delta_{0,m}}{2}\right)\frac{q(m,\omega)P_C(\omega)}{\pi  v_g[\omega, q(m,\omega)]L_{C}} \theta[\omega_\t{C}(\omega)-m\pi c/L_C].
    \label{eq:dpomega}
\end{equation}
The thermal radiation energy density \( \rho_P(\omega, T) \) in a polaritonic system can be written as the product of the polariton energy, the Bose-Einstein distribution, and the photon-weighted polariton density of states 
\begin{equation}
    \rho_P(\omega, T) = D_P(\omega) \frac{\hbar \omega}{e^{\frac{\hbar \omega}{k_B T}} - 1}.
\end{equation}
Substituting the expression for \( D_P(\omega) \), we obtain:
\begin{equation}
    \rho_P(\omega, T) = \sum_{m=0}^{\infty}\left(1-\frac{\delta_{0,m}}{2}\right)\frac{q(m,\omega)P_C(\omega)}{\pi  v_g[\omega, q(m,\omega)]L_{C}} \frac{\hbar \omega}{e^{\frac{\hbar \omega}{k_B T}} - 1}\theta[\omega_\t{C}(\omega)-m\pi c/L_C]. \label{eq:rhop}
\end{equation}
This expression can be used to calculate the contribution of polaritonic modes to the overall thermal radiation inside a microcavity, considering both the upper and lower polariton branches. Equation \ref{eq:rhop} is particularly useful for computing the thermal absorption and emission rates of a charged system weakly coupled to a polaritonic material.

% HF and LiH results
\subsection{HF and LiH molecules}

\begin{figure}
    \centering
    \includegraphics[width=\linewidth]{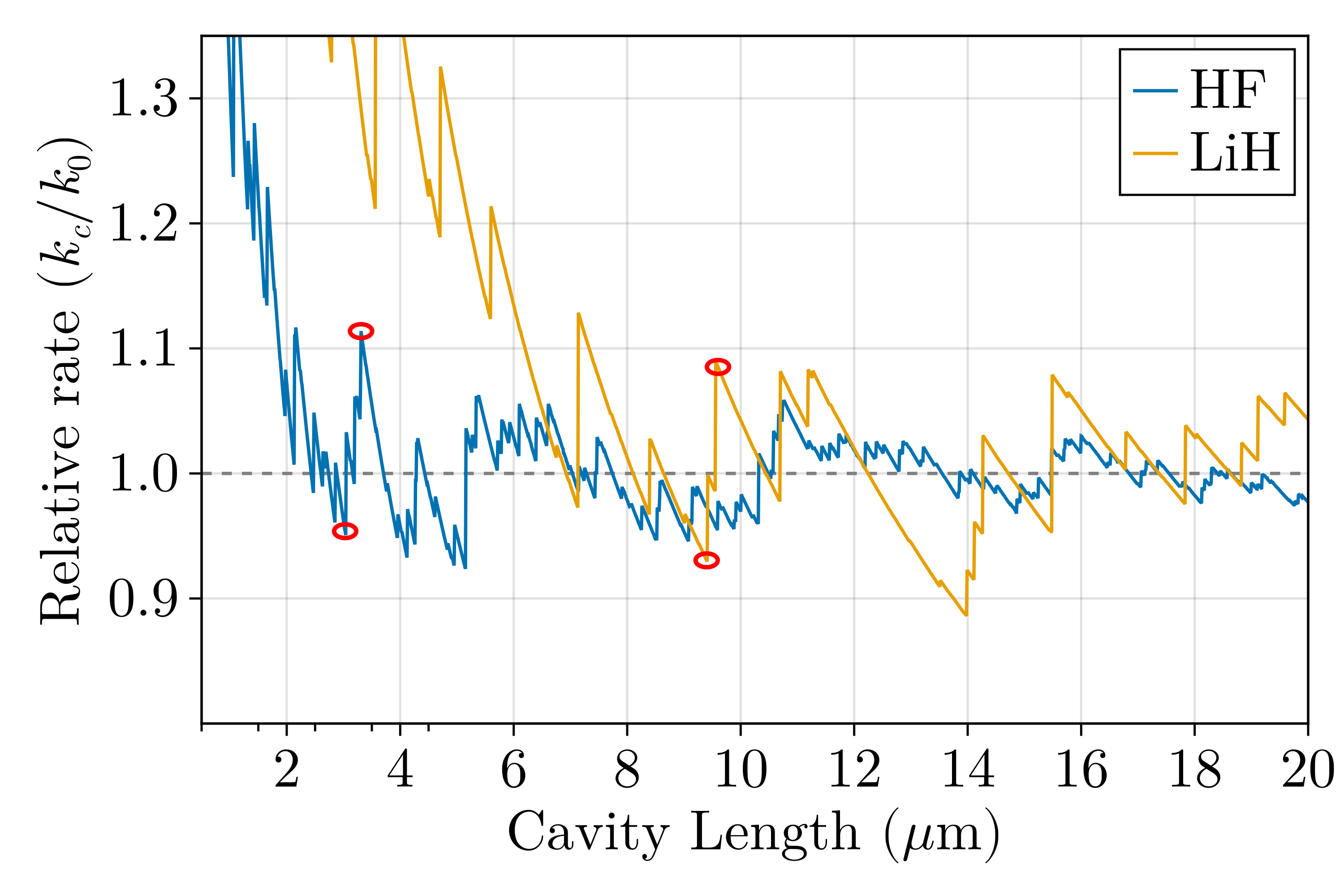}
    \caption{Ratio of BIRD rate inside a microcavity (weak coupling regime) to the free-space rate as a function of microcavity length for HF and LiH molecules. Red circles highlight specific lengths further discussed in Fig. \ref{fig:wc_dos}.}
    \label{fig:wc_rates}
\end{figure}

In Fig. \ref{fig:wc_rates}, BIRD rates in the weak coupling regime are shown relative to the free-space rate. For small microcavity lengths ($L_C < 2$ $\mu$m for HF and $L_C < 5$ $\mu$m for LiH), a mild but consistent enhancement of BIRD rates is observed due to an increased microcavity electromagnetic DOS over a wide frequency range. As $L_C$ increases, the microcavity BIRD rates converge to the free-space rates.

\begin{figure}
    \centering
\includegraphics[width=\linewidth]{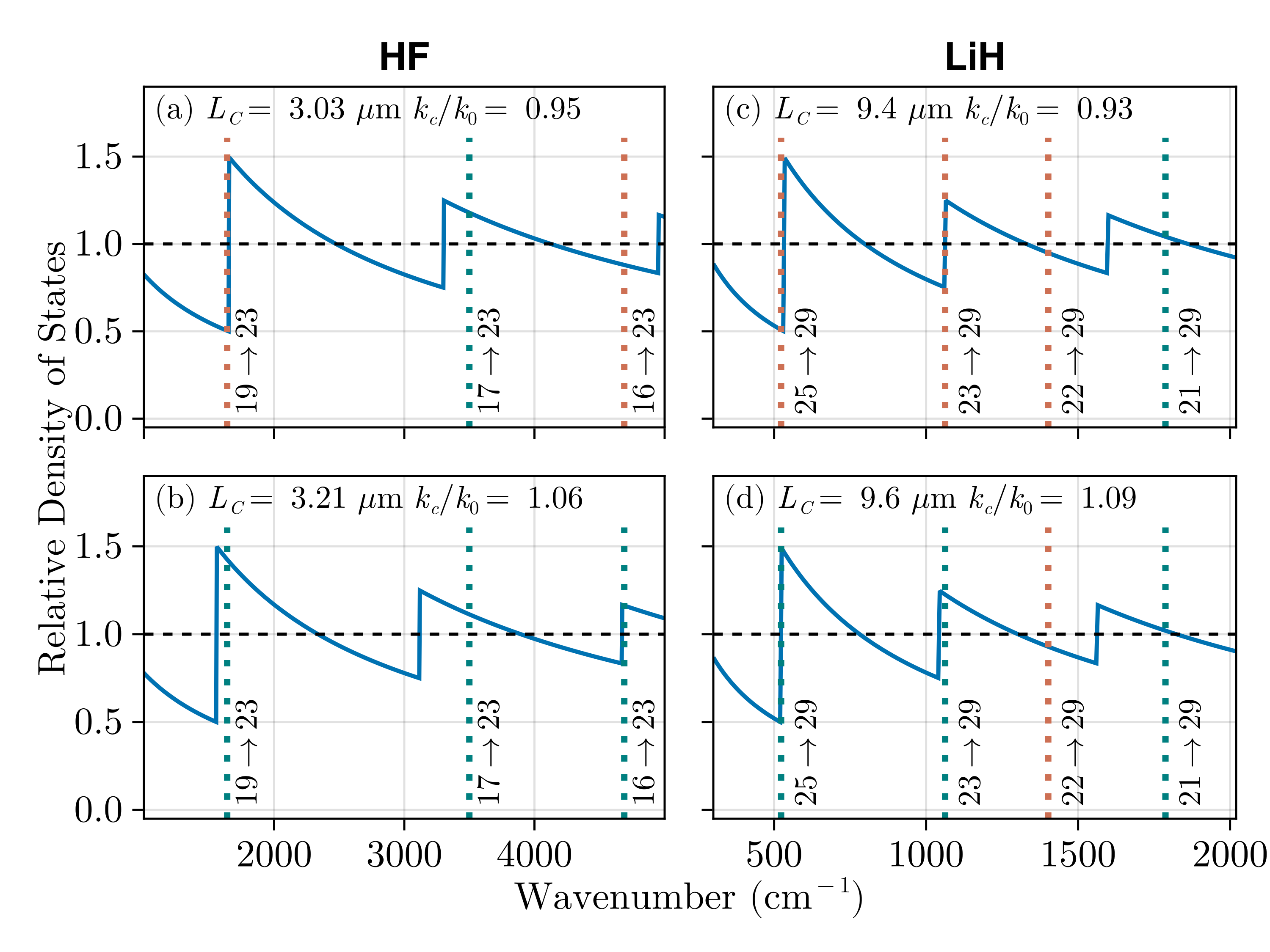}
    \caption{Microcavity photon DOS normalized to the free space DOS for selected lengths ($L_C$) where BIRD suppression and enhancement were observed for HF (left panels) and LiH  (right panels) in Fig. \ref{fig:wc_rates}. The dotted vertical lines indicate the location of the transition frequencies corresponding to the most relevant overtone transitions, with color coding to distinguish between suppression (red) and enhancement (green). Panel (a) shows that the small HF BIRD suppression observed at $L_C = 3.03~\mu\t{m}$ is at least in part caused by the reduced relative DOS at the highlighted overtones.  Panel (b) shows the mild enhancement of HF BIRD in a microcavity with $L_C = 3.21 ~\mu\t{m}$ can be ascribed to the greater microcavity photon DOS at the frequencies of the highlighted overtones. Panels (c) and (d) show the analogous scenario for LiH, where BIRD is observed to be suppressed at $L_C = 9.4~\mu\t{m}$ but enhanced at $L_C = 9.6~\mu\t{m}$, primarily due to the variation of the microcavity photon density of states at the selected $i \rightarrow 29$ overtone frequencies.}
    \label{fig:wc_dos}
\end{figure}

The oscillatory behavior in Fig. \ref{fig:wc_rates} arises from discrete vibrational transitions being enhanced or suppressed depending on $L_C$. Overtones play an essential role, with HF dissociation being most influenced by transitions $i \rightarrow 23$, with $i \in {16, 17, 19}$, and LiH by transitions $i \rightarrow 29$, with $i \in {21, 22, 23, 25}$. These overtones excite the system to the highest energy-bound states, leading to dissociation. Fundamental transitions ($i\rightarrow i+1$) at high excitation levels involve small energies, leading to reduced photon populations and oscillator strengths.

Fig. \ref{fig:wc_dos} presents the microcavity DOS normalized to free space for selected $L_C$ values where BIRD suppression or enhancement was observed. The suppression at $L_C = 3.03$ $\mu$m for HF and $L_C = 9.4$ $\mu$m for LiH correlates with suppressed overtones, while enhancement at $L_C = 3.21$ $\mu$m (HF) and $L_C = 9.6$ $\mu$m (LiH) corresponds to enhanced microcavity photon DOS at relevant overtone frequencies. Finite linewidths in radiative transitions would smooth oscillations and dampen their amplitude, limiting microcavity effects to small $L_C$.

BIRD suppression effects remain minor as multiple dissociative pathways exist. Blocking a single transition does not significantly impact dissociation, reinforcing two key findings: (i) overtone transitions are crucial, and (ii) blocking a single reactive pathway is largely inconsequential.

\begin{figure}[t]
    \centering
    \includegraphics[width=\linewidth]{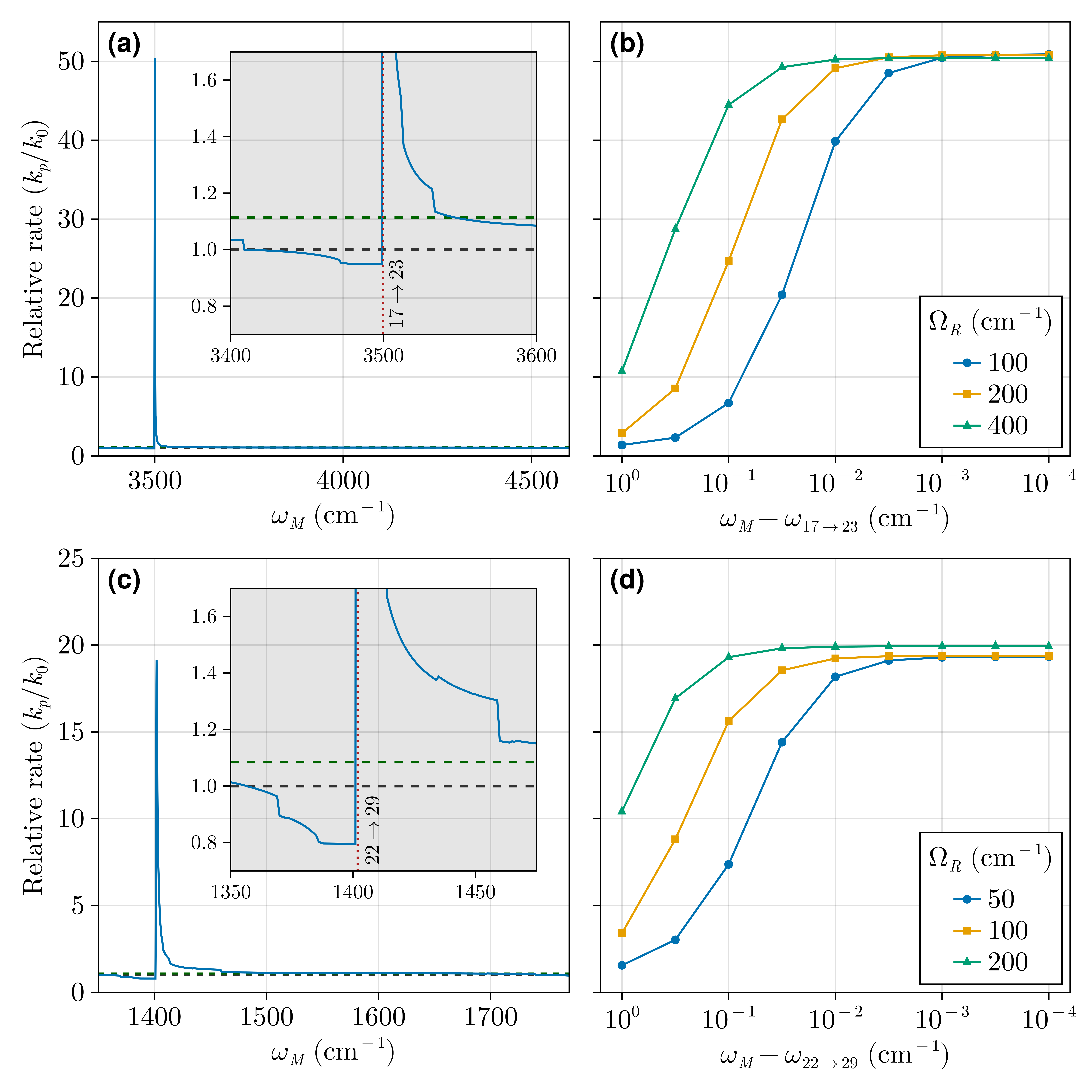}
    \caption{Ratio of polariton-assisted BIRD rates to free space BIRD rates. In the polaritonic case, the diatomic molecule is embedded in a strongly coupled microcavity with variable host material with frequency $\omega_M$. The top panels, (a) and (b), correspond to HF, and the bottom panels, (c) and (d), show results for LiH. In (a) and (c), the collective light-matter interaction strength, $\Omega_R$, is fixed at 400 and 200 cm$^{-1}$, respectively. The horizontal dashed gray and green lines indicate where the dissociation rates equal $k_0$ and $k_c$, respectively. The insets in (a) and (c) show zoomed-in views around the overtones $17 \rightarrow 23$ for HF and $22 \rightarrow 29$ for LiH. In (b) and (d), relative BIRD rates are shown for different Rabi frequencies ($\Omega_R$), depicting the dependency on the detuning between the host molecule and the diatomic transition energies.}
    \label{fig:sc_relative_rates_omegaM}
\end{figure}

Fig. \ref{fig:sc_relative_rates_omegaM} shows polariton-assisted BIRD rates as a function of host matter frequency ($\omega_M$) for HF ($L_C = 3.21$ $\mu$m, $\Omega_R = 400$ cm$^{-1}$) and LiH ($L_C = 9.6$ $\mu$m, $\Omega_R = 200$ cm$^{-1}$). Enhancement peaks occur when $\omega_M$ is slightly above a diatomic overtone frequency ($i = 17 \rightarrow 23$ for HF, $i = 22 \rightarrow 29$ for LiH). Suppression occurs when $\omega_M$ approaches these transitions from the left, placing them in the stopgap region where the electromagnetic DOS is zero.

\begin{figure}[t]
    \centering
    \includegraphics[width=\linewidth]{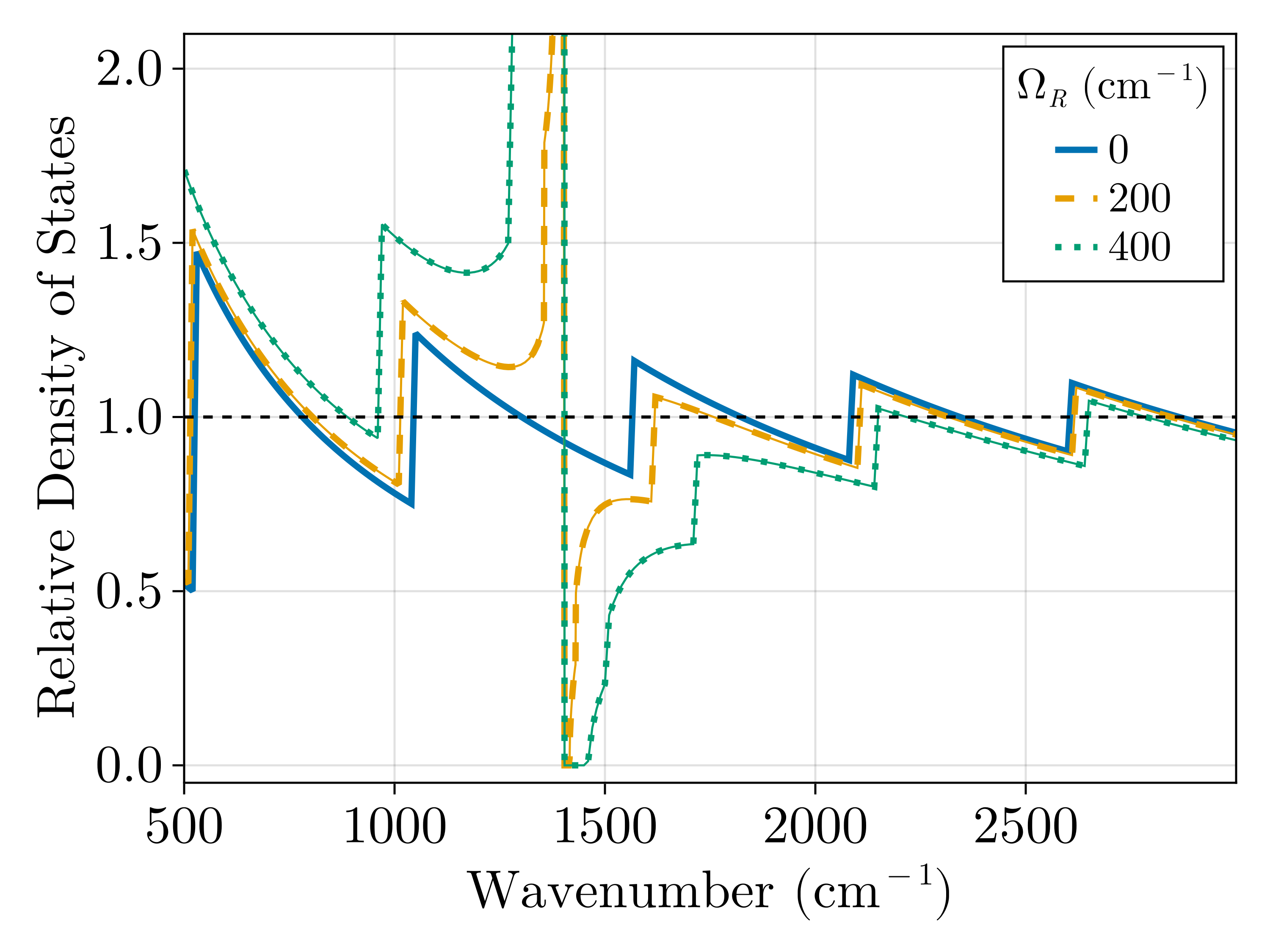}
    \caption{Photon-weighted polariton density of states $D_P(\omega)$, normalized to the free space DOS, for a microcavity with length $L_{C} = 9.6 ~\mu$m strongly coupled to a material with a bright transition at frequency $\omega_M = 1401$ cm$^{-1}$. The curves show the effect of different Rabi splitting values ($\Omega_R$) on the photon-weighted polariton DOS, which is relevant for radiative processes mediated by polaritonic systems. As $\Omega_R$ increases, the frequency range below $\omega_M$, where the photon-weighted polariton DOS significantly exceeds that of free space, widens. This trend explains the impact of the collective light-matter interaction strength on the observed polariton-assisted bond infrared dissociation (BIRD) enhancement shown in Fig. \ref{fig:sc_relative_rates_omegaM}.}
    \label{fig:DOSp}
\end{figure}

Fig. \ref{fig:DOSp} shows the photon-weighted polariton DOS $D_P(\omega)$ for selected $\Omega_R$ at $L_C = 9.6$ $\mu$m and $\omega_M = 1401$ cm$^{-1}$. Strong coupling alters $D_P(\omega)$ near $\omega_M$, influencing vibrational transition rates. The singularity in $D_P(\omega)$ allows a transition to become arbitrarily fast compared to free space, but reaction rates remain finite due to other limiting steps.

Experimentally, imperfections in the electromagnetic device and material disorder smooth out $D_P(\omega)$ near the stopgap, reducing BIRD enhancement compared to theoretical predictions. Including finite linewidths relaxes strict resonance conditions, making enhancement dependent on the spectral overlap between $\omega_{i\rightarrow j}$ and $D_P(\omega)$. Thus, reported results represent upper bounds for polariton-assisted BIRD rates under ideal conditions.

%Lossy Cavities
\subsection{Dissociation Rates in Lossy Cavities}

The preceding analysis assumed ideal optical microcavities with perfectly reflecting mirrors. In practice, the mirrors are either metallic or multilayer dielectric stacks with alternating refractive indices, known as distributed Bragg reflectors (DBRs) \cite{kavokin2017microcavities}. Both devices are imperfect and inevitably introduce losses: finite conductivity limits the reflectivity of metallic mirrors and causes ohmic absorption, whereas DBRs suffer from material absorption and fabrication imperfections. In this section we evaluate how such losses, focusing on weakly coupled microcavities with absorptive metallic mirrors, affect the corresponding BIRD rates obtained assuming perfect mirrors (Sec. III. A). We also provide comments on the polariton-assisted case.

%In lossless systems, electromagnetic modes have real-valued eigenfrequencies, and the density of optical states (DOS), local DOS (LDOS), and partial LDOS (PLDOS) can be expressed using sums over discrete eigenmodes. However, when loss is present, such as with metallic mirrors, these eigenfrequencies become complex, and the delta-function definitions of DOS no longer apply. Instead, a more general and powerful approach involves computing the imaginary part of the electromagnetic Green's function.

\subsubsection{Computational Method.}  In microcavities with lossy mirrors, the Green function formalism described e.g., in Ref.~\cite{barnes_classical_2020} can be adopted to obtain the local electromagnetic density of states in terms of the imaginary part of the electromagnetic Green tensor. %In this formalism, the photon partial LDOS at position $\mathbf{r}$, frequency $\omega$ can be obtained from
%\begin{equation}
%D_p(\hat{\mathbf{e}}_d, \mathbf{r}, \omega) = \frac{2\omega}{\pi c^2} \, \hat{\mathbf{e}}_d \cdot \text{Im}\left[ \mathbf{G}(\mathbf{r}, \mathbf{r}'; \omega) \right] \cdot \hat{\mathbf{e}}_d,
%\end{equation}
%where $\mathbf{G}(\mathbf{r}, \mathbf{r}'; \omega)$ is the dyadic Green function\cite{barnes_classical_2020, kavokin2017microcavities} describing the electric field at $\mathbf{r}$ in the direction $\hat{\mathbf{e}}_d$ generated by a point dipole at $\mathbf{r}'$ directed along $\hat{\mathbf{e}}_d$. Averaging over all dipole orientations yields the isotropically averaged LDOS:
%\begin{equation}
%D_L(\mathbf{r}, \omega) = \frac{2\omega}{\pi c^2} \, \text{Tr} \left[ \text{Im} \, \mathbf{G}(\mathbf{r}, \mathbf{r}; \omega) \right].
%\end{equation}
%These definitions remain valid regardless of whether the system is lossy.
%Finally, the total photon DOS is obtained by integrating the LDOS over all space:
%\begin{equation}
%D(\omega) = \frac{2\omega}{\pi c^2} \int_V %\text{Tr} \left[ \text{Im} \, \mathbf{G}%(\mathbf{r}, \mathbf{r}; \omega) \right] %d^3\mathbf{r}.
%\end{equation}

Following Ref. \cite{barnes_classical_2020}, the photon LDOS relative to free space at position $z_0$ inside a planar microcavity consisting of identical front and back metallic mirrors separated by a distance $L$ is given by
\begin{equation}
\frac{D_C(\omega, z_0)}{D_0(\omega)} = 1 + \frac{1}{2} \text{Re} \int_0^\infty \frac{u}{\sqrt{1 - u^2}} \left[\frac{N_{\text{TE}}(u)}{D_{\text{TE}}(u)} + \frac{N_{\text{TM}}(u)}{D_{\text{TM}}(u)}\right] du,
\label{eq:ldos_integral}
\end{equation}
where $u = \sin \theta$ is the in-plane propagation angle, and $N_{\text{TE}}$, $N_{\text{TM}}$, $D_{\text{TE}}$, and $D_{\text{TM}}$ are related to TE and TM reflection coefficients $r_{\text{TE}}$ and $r_{\text{TM}}$ according to
\begin{align}
N_{\text{TE}} &= r_{\text{TE}} e^{2 i k_z z_0} + r_{\text{TE}} e^{2 i k_z (L - z_0)} + 2 r_{\text{TE}}^2 e^{2 i k_z L}, \\
D_{\text{TE}} &= 1 - r_{\text{TE}}^2 e^{2 i k_z L}, \\
N_{\text{TM}} &= (2 u^2 - 1) r_{\text{TM}} (e^{2 i k_z z_0} + e^{2 i k_z (L - z_0)}) + 2 r_{\text{TM}}^2 e^{2 i k_z L}, \\
D_{\text{TM}} &= 1 - r_{\text{TM}}^2 e^{2 i k_z L},
\end{align}
where  $k_z = \sqrt{\epsilon_1} \omega / c \sqrt{1 - u^2}$ is the transverse component of the wave vector inside the microcavity, and $\epsilon_1$ is the dielectric constant of the background medium (taken as 1 for vacuum). The complex-valued Fresnel reflection amplitudes $r_{\text{TE}}$ and $r_{\text{TM}}$ are evaluated using the dielectric function of the mirrors here approximated by the Drude form
\begin{equation}
\epsilon_2(\omega) = 1 + \frac{\omega_p^2}{\omega^2 + i \gamma \omega},
\end{equation}
where $\epsilon_2$ is the metal complex dielectric function, $\omega_p$ is the plasma frequency and $\gamma$ is the damping rate.

Table~\ref{tab:drude_params} below presents the Drude parameters employed for each metal considered in our study. All parameters were selected based on experimentally reported values corresponding to optimal film thicknesses that maximize plasmonic thermal conductivity \cite{thermal_conductivity_plasmon}.

\begin{table}[h]
\centering
\caption{Drude parameters used for dielectric function calculations.}
\label{tab:drude_params}
\begin{tabular}{lcc}
\hline
Material & \( \omega_p \) (rad/s) & \( \gamma \) (rad/s) \\
\hline
Au & \( 1.22 \times 10^{16} \) & \( 7.11 \times 10^{13} \) \\
Al & \( 1.82 \times 10^{16} \) & \( 1.23 \times 10^{14} \) \\
Pt & \( 7.87 \times 10^{15} \) & \( 1.22 \times 10^{14} \) \\
\hline
\end{tabular}
\end{table}

In our radiative transition rate computations, we employ the isotropically averaged LDOS spatially averaged over all emitter positions inside the microcavity to account for spatial and angular variation of the field distribution inside the cavity and the motion of the diatomic which is appreciable in the macroscopic timescales required for BIRD to happen at moderate temperatures. In this case, the relative LDOS modulates the spontaneous emission rate according to
\begin{equation}
\Gamma_{\text{sp}}^{\text{C}}(\omega) = \Gamma_{\text{sp}}^{0}(\omega) \frac{D_C(\omega)}{D_0(\omega)},
\end{equation}
where $\Gamma_{\text{sp}}^{\text{C}}(\omega)$ 
and $\Gamma_{\text{sp}}^{(0)}$ are dipolar microcavity and free space spontaneous emission rates corresponding to a quantum level transition with frequency $\omega$. Using Einstein coefficient relations described in the main text, the stimulated emission and absorption rates can then be obtained from spontaneous emission rates and the Pauli Master equation employed for BIRD can be implemented as described in the main manuscript. 

\begin{figure}[ht]
    \centering
    \includegraphics[width=0.95\linewidth]{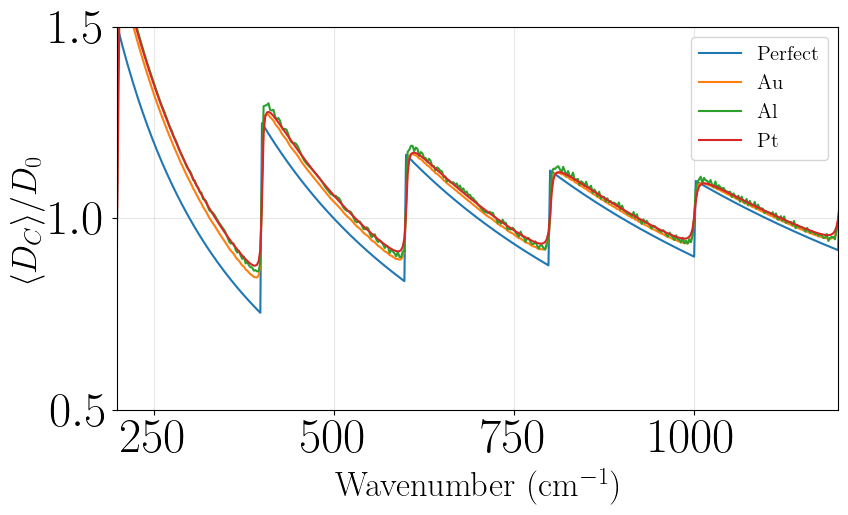}
    \caption{Isotropically and spatially averaged LDOS \( \langle D_C \rangle_{z_0} / D_0 \) at \( L = 25~\mu\text{m} \) for perfect mirrors and three metal mirrors (Au, Al, Pt). Realistic metals suppress and broaden the modal features seen in the ideal case.}
    \label{fig:DOS_metal_compare}
\end{figure}

Prior to examining the effect of microcavity losses on BIRD rates, we present in Fig. \ref{fig:DOS_metal_compare} simulations of the microcavity density of states (normalized by the free-space photon DOS) for a cavity length ( L = 25 $\mu$m). These calculations employ metallic mirrors composed of Au, Al, and Pt using Fresnel reflection coefficients derived from the Drude model as described above.

As shown in Figure~\ref{fig:DOS_metal_compare}, the ideal perfect mirror microcavity exhibits sharp, periodic resonances corresponding to discrete photonic modes. In contrast, realistic metallic mirrors introduce damping and spectral broadening due to absorption losses, resulting in smoother density of states profiles. Notably, both Au and Al demonstrate enhanced local density of optical states at low frequencies despite their lossy characteristics. This enhancement arises from evanescent field contributions near the metal interfaces, where dissipative surface modes can significantly increase the local electromagnetic energy density. While these evanescent components exhibit limited propagation into the resonator interior, their presence near the interfaces enhances the average DOS experienced by molecules in close proximity, particularly in resonators of short length. Consistent with the $1/\sqrt{\omega}$ frequency dependence of the evanescent decay length, a gradual approach to the free-space DOS limit is observed at higher frequencies.

\subsubsection{Results}
\begin{figure}[ht]
    \centering
    \includegraphics[width=0.95\linewidth]{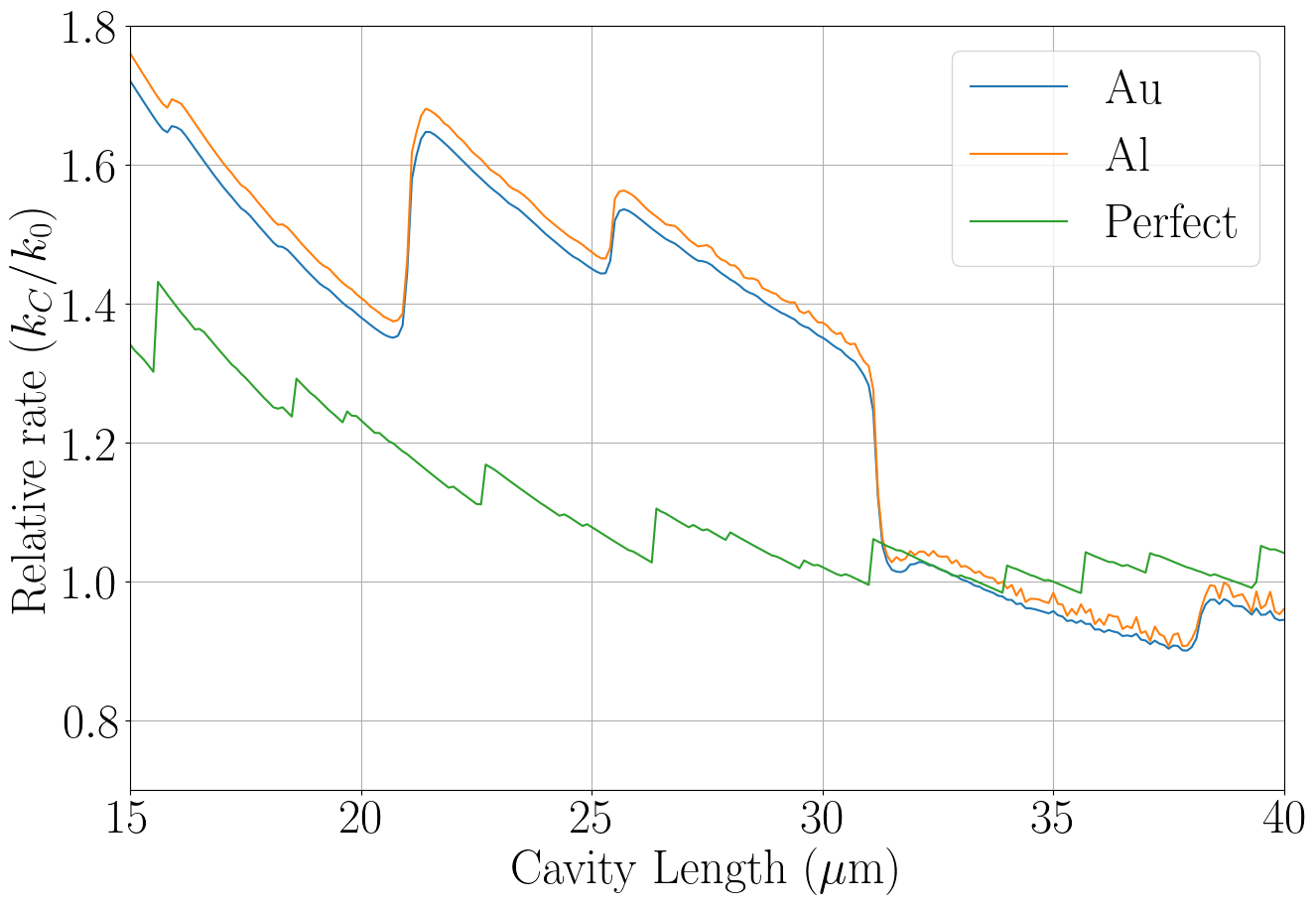}
    \caption{Relative BIRD rate \( k_c/k_0 \) versus cavity length \( L \) at \( T = 400 \)~K. Mirror materials include gold (Au), aluminum (Al), and an idealized perfect reflector. Lossy mirrors broaden and suppress the resonances, but maintain rate enhancements at shorter lengths.}
    \label{fig:lossy_vs_L}
\end{figure}

Figure~\ref{fig:lossy_vs_L} compares the BIRD dissociation rates relative to free space $k_c/k_0$ as a function of microcavity length for perfect mirrors (as in our manuscript) and lossy mirrors made from gold and aluminum. 

Several key inferences can be made from Fig. \ref{fig:lossy_vs_L}. First, the introduction of imperfect absorptive metallic mirrors yields relative BIRD rate enhancements of the same order of magnitude as microcavities with perfect mirrors. Second, simulations with lossy mirrors broaden the sharp resonance features observed in the perfect microcavity length-dependent BIRD rate variation. 

Interestingly, at short microcavity lengths, both metallic mirror types exhibit enhanced BIRD rates relative to perfectly reflecting mirrors. This  enhancement can be attributed to evanescent electromagnetic waves supported at the metal–dielectric interfaces \cite{vinogradov1992, FerroPhase}, as demonstrated in Fig. \ref{fig:DOS_metal_compare}. In lossy resonators, these near-field electromagnetic field components increase the accessible electromagnetic energy density for molecular transitions, leading to enhanced reaction rates. Although these modes are non-propagating and decay exponentially from the interface, their influence remains significant in confined geometries \cite{FerroPhase}. Aluminum mirrors produce slightly higher enhancements due to their marginally greater penetration depth at infrared frequencies \cite{born2013principles}. As the resonator length increases, the contribution of evanescent modes diminishes due to their finite penetration length, and the field intensity depleting absorptive losses imparted by the mirrors become of greater significance. This leads to the weak suppression observed at microcavity lengths greater than 30 $\mu$m. Nevertheless, as $L$ increases, all examined scenarios show slow weak converge toward the free-space dissociation rate limit, as illustrated in Fig.\ref{fig:lossy_vs_L}.

In summary, incorporating realistic mirror losses reveals that microcavity LDOS and BIRD dissociation rate enhancements persist at short microcavity lengths. The enhancement magnitude depends on the mirror material properties. While our model primarily considers radiative contributions, we acknowledge that evanescent near-field modes supported by metallic surfaces may also contribute to the observed LDOS enhancements. Their spatially localized nature and coupling to molecular dipoles can enhance transition rates even in the presence of loss. Nevertheless, the order of magnitude of BIRD rate modifications remains consistent between perfect and imperfect microcavities.

\subsubsection{Comment about polariton-assisted mechanism in lossy resonators}

The treatment of lossy mirror effects introduced above was applied specifically to the empty microcavity case described in Section 3A of the main text. These findings strongly suggest that realistic metallic mirrors would influence polariton-assisted rates in the same qualitative manner: sharp spectral features would become smoother and broader, yet the order-of-magnitude BIRD enhancements predicted for microcavities with perfect mirrors are very likely to persist as they do in the weak coupling analysis. This is expected, especially as, our polariton-assisted simulations already incorporate an effective light-matter coupling cutoff that eliminates the  highly off-resonant couplings responsible for the stop gap singularity observed in the polariton-weighted photon density of states. 

Our introduction of a high energy cutoff effectively removes the polariton stop gap singularity from consideration. Therefore, incorporating microcavity losses is not expected to alter our conclusions regarding the order of magnitude of the observed polariton-assisted rate enhancement. Nevertheless, we recognize that a comprehensive treatment of microcavity losses in the strong coupling regime represents a relevant and important consideration for future investigations.

% Sensitivity analysis
\subsection{Sensitivity Analysis}

Due to the modified DOS, microcavities may enhance or suppress multiple vibrational transitions (relative to free space) in both weak and strong coupling regimes. We performed a sensitivity analysis to find which transitions are the most impactful to the BIRD rate. This consists of the following procedure for each fundamental or overtone transition frequency $\omega_{ij}$.

We perturb the free space photon DOS at $\omega_{ij}$ by  scaling this quantity by a factor of $\delta$
\begin{align}
    D_0(\omega_{ij}) \rightarrow \delta\cdot D_0(\omega_{ij}).
\end{align}
Since transition rates are proportional to the DOS, the transitions associated with the absorption or emission of photons with frequency $\omega_{ij}$ are effectively scaled by the same factor, while all other transitions remain with free space rates. The corresponding perturbed transition matrix, $\mathbf{J}$ is then constructed, and from its lowest eigenvalue, a perturbed BIRD rate is obtained and assigned a sensitivity score $S_{ij}$ based on the definition
\begin{align}
    S_{ij} = \frac{\text{free space BIRD rate after perturbation}}{\text{free space BIRD rate with no perturbation}} - 1  \;.\label{eq:Sij}
\end{align}

\begin{figure}
    \centering
    \includegraphics[width=\linewidth]{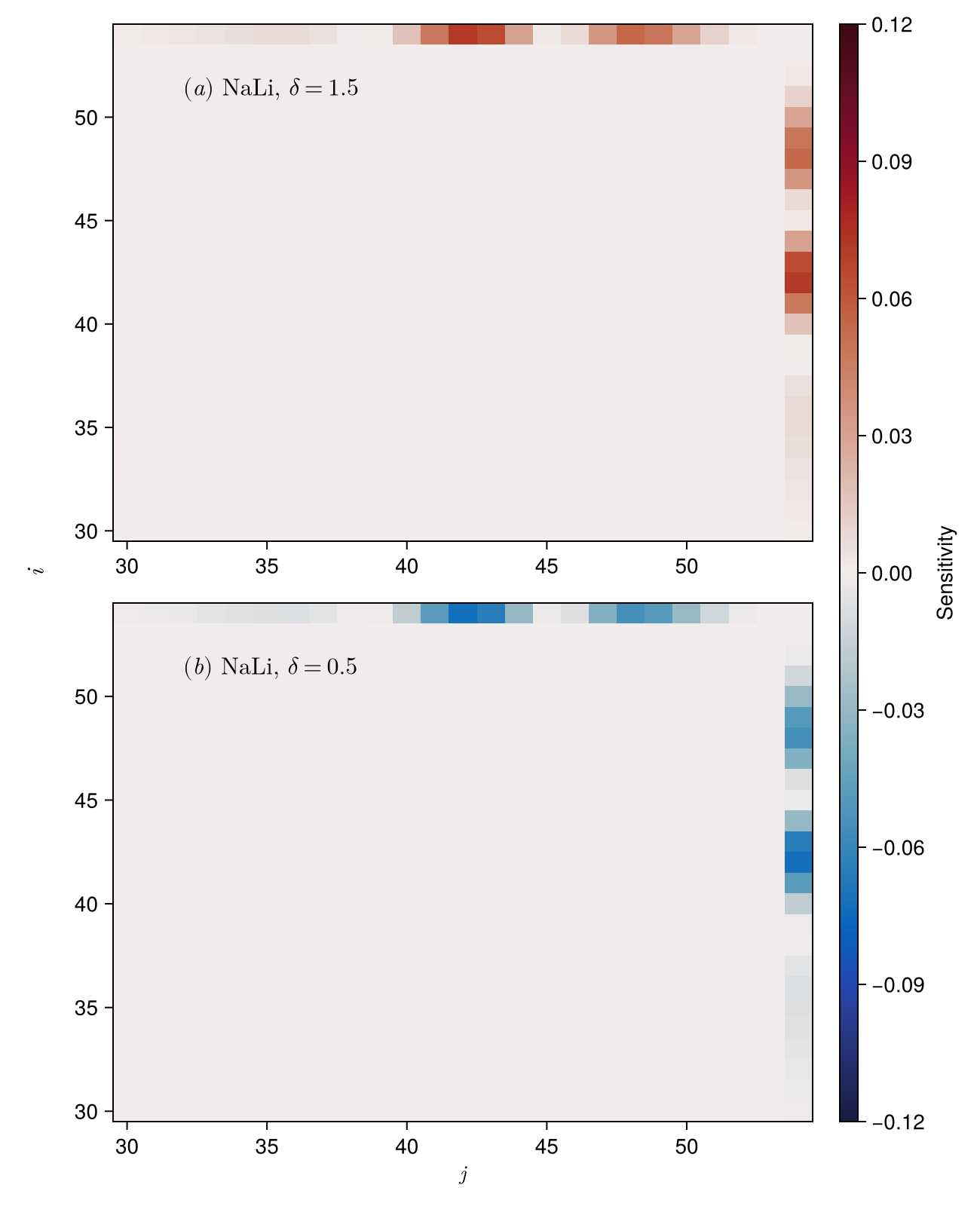}
    \caption{Sensitivity analysis for NaLi under enhancement ($\delta=1.5$) and suppression ($\delta=0.5$). The sensitivity was calculated using Eq. \ref{eq:Sij} following the procedure described in this section. Transitions not shown in the plot have negligible sensitivity.}
    \label{fig:sensitivitynali}
\end{figure}

Repeating this procedure for all $i \neq j$ combinations yields a sensitivity matrix $\mathbf{S}$ shown as a heat map in Fig. \ref{fig:sensitivitynali}. The $S$ matrix is symmetric given that emission and absorption depend on the photon DOS at the same frequency, i.e., it is impossible to enhance/suppress one process without also affecting the reverse process.

From Fig. \ref{fig:sensitivitynali}, we see that the most important transitions are overtones that reach the final bound state from relatively high energy vibrational levels ($i \rightarrow 54$). We list these transitions, along with their energies and sensitivity scores in Table
\ref{tab:NaLi_sensitivity}. The fact that these overtones dominate the dissociation dynamics may be understood from the observation that, in anharmonic systems, fundamental transitions $i \rightarrow i\pm 1$ are increasingly slowed as we climb the energy level ladder. For example, fundamental transitions starting at $n = 40$ involve energies under 50 cm$^{-1}$. Such transitions have, in general, weak oscillator strength and correspond to small DOS and thermal photon populations. Hence, overtones play a dominant role at these higher energy levels and act as bottlenecks for dissociation from the highest energy-bound state.

\begin{table}[h]
\centering
\renewcommand{\arraystretch}{1.5} 
\caption{Most important transitions according to their sensitivity analysis score for NaLi.}
\label{tab:NaLi_sensitivity}
\begin{tabular}{|c|c|c|c|}
\hline
\textbf{Transition} & \textbf{Energy (cm$^{-1}$)} & \textbf{Sensitivity} ($\delta = 1.5$) & \textbf{Sensitivity} ($\delta = 0.5$) \\
\hline
\hline
$42 \rightarrow 54$ & 379.8 & 0.072 & -0.073 \\
\hline
$43 \rightarrow 54$ & 322.6 & 0.065 & -0.066 \\
\hline
$48 \rightarrow 54$ & 106.1 & 0.054 & -0.055 \\
\hline
$49 \rightarrow 54$ & 76.8 & 0.049 & -0.05 \\
\hline
$41 \rightarrow 54$ & 441.7 & 0.048 & -0.048 \\
\hline
$47 \rightarrow 54$ & 140.1 & 0.036 & -0.036 \\
\hline
$44 \rightarrow 54$ & 270.0 & 0.031 & -0.031 \\
\hline
$50 \rightarrow 54$ & 52.1 & 0.029 & -0.029 \\
\hline
$40 \rightarrow 54$ & 508.3 & 0.017 & -0.017 \\
\hline
$51 \rightarrow 54$ & 32.1 & 0.011 & -0.012 \\
\hline
\end{tabular}
\end{table}

%High temperature limit
\subsection{Temperature Dependence}
Diatomic molecules such as HF and LiH are bound by a strong covalent bond that is stable even at temperatures much higher than room temperature. This means their dissociation from the most energetic bound state by thermal radiation can only happen at very high temperatures. In this section, we discuss the temperature dependence of the reported relative rates and demonstrate that the qualitative analysis presented in the main manuscript holds for all examined temperatures.

\begin{figure}
    \centering
    \includegraphics[width=0.7\linewidth]{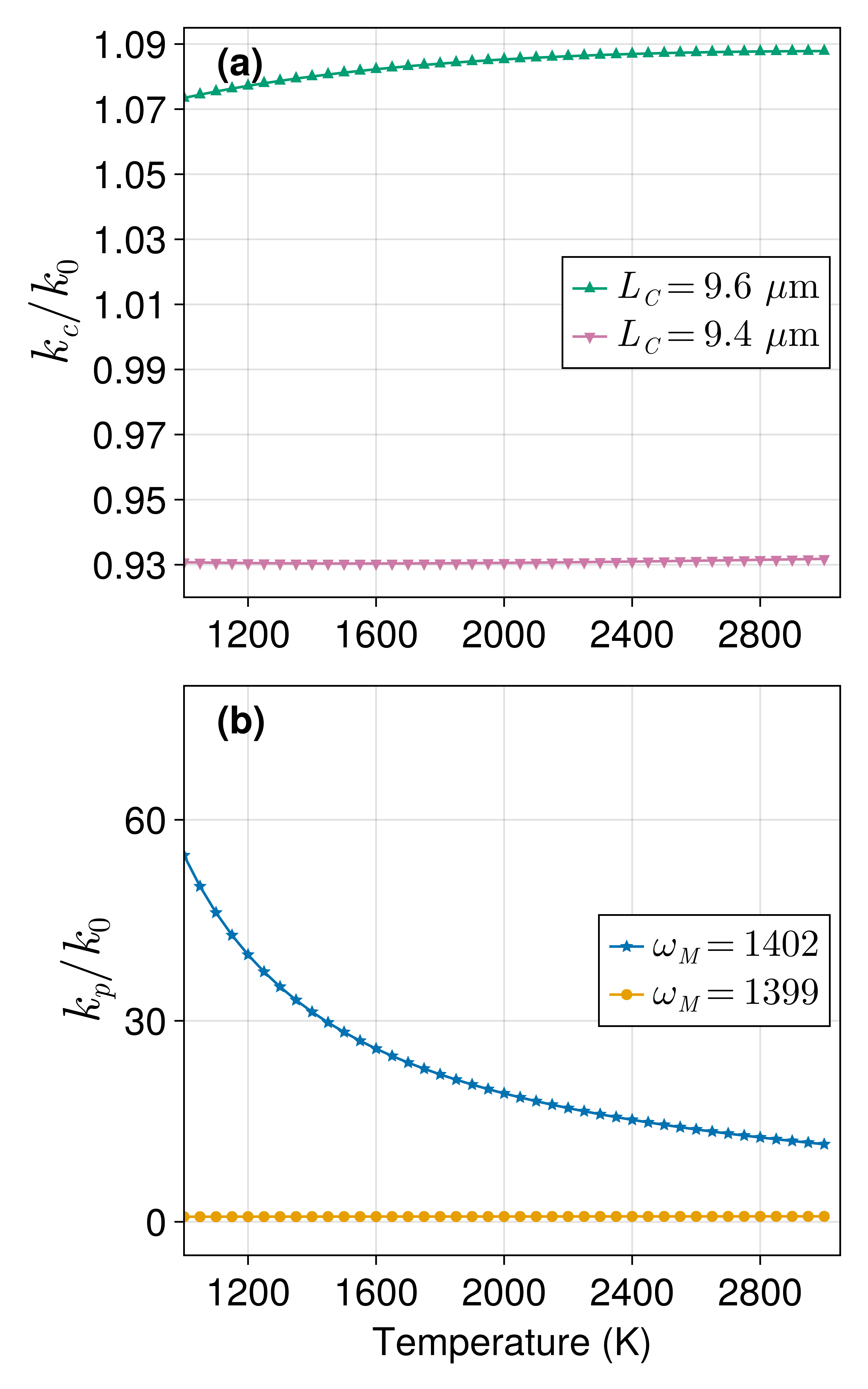}
    \caption{Temperature dependence of the relative BIRD rates for LiH in the \textbf{(a)} weak and \textbf{(b)} strong coupling regimes. In \textbf{(b)},  $\Omega_R = 200$ cm$^{-1}$ and $L_C = 9.6\;\mu$m}
    \label{fig:temperature}
\end{figure}

Achieving numerically stable BIRD rate constants (lowest eigenvalue of transition matrix $\mathbf{J}$) is straightforward at sufficiently high temperatures (e.g., comparable to or larger than the vibrational temperatures of the modeled diatomics). At such temperatures (e.g., 2000 K for LiH and 4000 K HF), the rate constants for absorption and emission are large enough that floating point arithmetic errors are negligible. In Fig. \ref{fig:temperature}, we present relative BIRD rates for temperatures as low as 1000 K in weak and strong coupling regimes. To mitigate floating point errors, these computations were performed using quadruple precision (\textsc{Float128}).

In previous sections of this SI, we explain that a small change in the cavity length (9.4 to 9.6 $\mu$m) leads to a measurable change in the LiH dissociation rate ($k_c/k_0$ shifts from 0.93 to 1.09) - the same qualitative result is also present for NaLi in the main text. In Fig. \ref{fig:temperature}(a), this observation remains valid for a large range of temperatures. Although at lower temperatures the enhancement (at $L_C = 9.6\;\mu$m) is smaller,  the suppression effect ($L_C = 9.4\;\mu$m) is virtually temperature-independent. Moreover, the sudden change in $k_c/k_0$ between the two examined cavity lengths is still present for all temperatures. 

Relative BIRD rates for LiH in the strong coupling regime are shown in Fig. \ref{fig:temperature}(b) for a range of temperatures.  Here we present results for two specific values of matter host frequency:  $\omega_M$ = 1399 cm$^{-1}$, which causes the $22 \rightarrow 29$ overtone to be completely suppressed and $\omega_M = 1402$ cm$^{-1}$ where the $22 \rightarrow 29$ overtone transition rate is greatly enhanced leading to a increased BIRD rate. From Fig. \ref{fig:temperature}
(b), we observe a moderate temperature dependence when $\omega_M = 1402$ cm$^{-1}$. However, the qualitative picture remains the same: when $\omega_M = 1399$ cm$^{-1}$ , the suppression of the $22 \rightarrow 29$ overtone causes a mild reduction of the BIRD rate, whereas at $\omega_M = 1402$ cm$^{-1}$, this overtone transition rate is enhanced and polariton-assisted BIRD via the highest energy bound state is faster by one order of magnitude relative to free space.

Selected numerical values of $k_0$, $k_c$, and $k_p$ are shown in Table \ref{tab:k0vals}.  Although polariton-assisted enhancements ($k_p/k_0$) become more significant at lower temperatures, Table \ref{tab:k0vals} shows BIRD rates are extremely low at the smallest examined temperatures. For example, at T = 1000 K $k_0$ is on the order of $O(10^{-12}/s)$. Hence, we focus our analysis on temperatures comparable to the LiH and HF vibrational temperatures. While a more comprehensive examination of thermal effects is left for future work, in the next section we explore a simple toy model to gain insight into the origin of this temperature dependence on polariton enhancements.

\begingroup
\renewcommand{\arraystretch}{1.2}
\begin{table}
    \centering
    \setlength\extrarowheight{2mm}
    \begin{tabular}{|c|c|c|c|} \hline 
 \textbf{Temperature (K)}& $k_0$ (s$^{-1}$)& $k_c$ (s$^{-1}$)&$k_p$ (s$^{-1}$)\\ \hline 
         1000&  5.96$\times 10^{-12}$&  6.39$\times 10^{-12}$& 3.26$\times 10^{-10}$\\ \hline 
         1500&  1.18$\times 10^{-7}$&  1.27$\times 10^{-7}$& 3.33$\times 10^{-6}$\\ \hline 
         2000&   1.64$\times 10^{-5}$&  1.78$\times 10^{-5}$&  3.14$\times 10^{-4}$\\ \hline 
         2500&   3.15$\times 10^{-4}$&  3.42$\times 10^{-4}$&  4.56$\times 10^{-3}$\\ \hline 
         3000&   2.23$\times 10^{-3}$&  2.43$\times 10^{-3}$&  2.58$\times 10^{-2}$\\ \hline
    \end{tabular}
    \caption{Absolute BIRD rates for LiH in free space ($k_0$), weak coupling regime ($k_c$), and strong coupling regime ($k_p$). In all cases, $L_C = 9.6\;\mu$m, $\Omega_R = 200$ cm$^{-1}$, and $\omega_M = 1402$ cm$^{-1}$.}
    \label{tab:k0vals}
\end{table}
\endgroup

\subsubsection{Temperature Dependence of a Model System}

\begin{figure}
    \centering
    \includegraphics[width=0.8\linewidth]{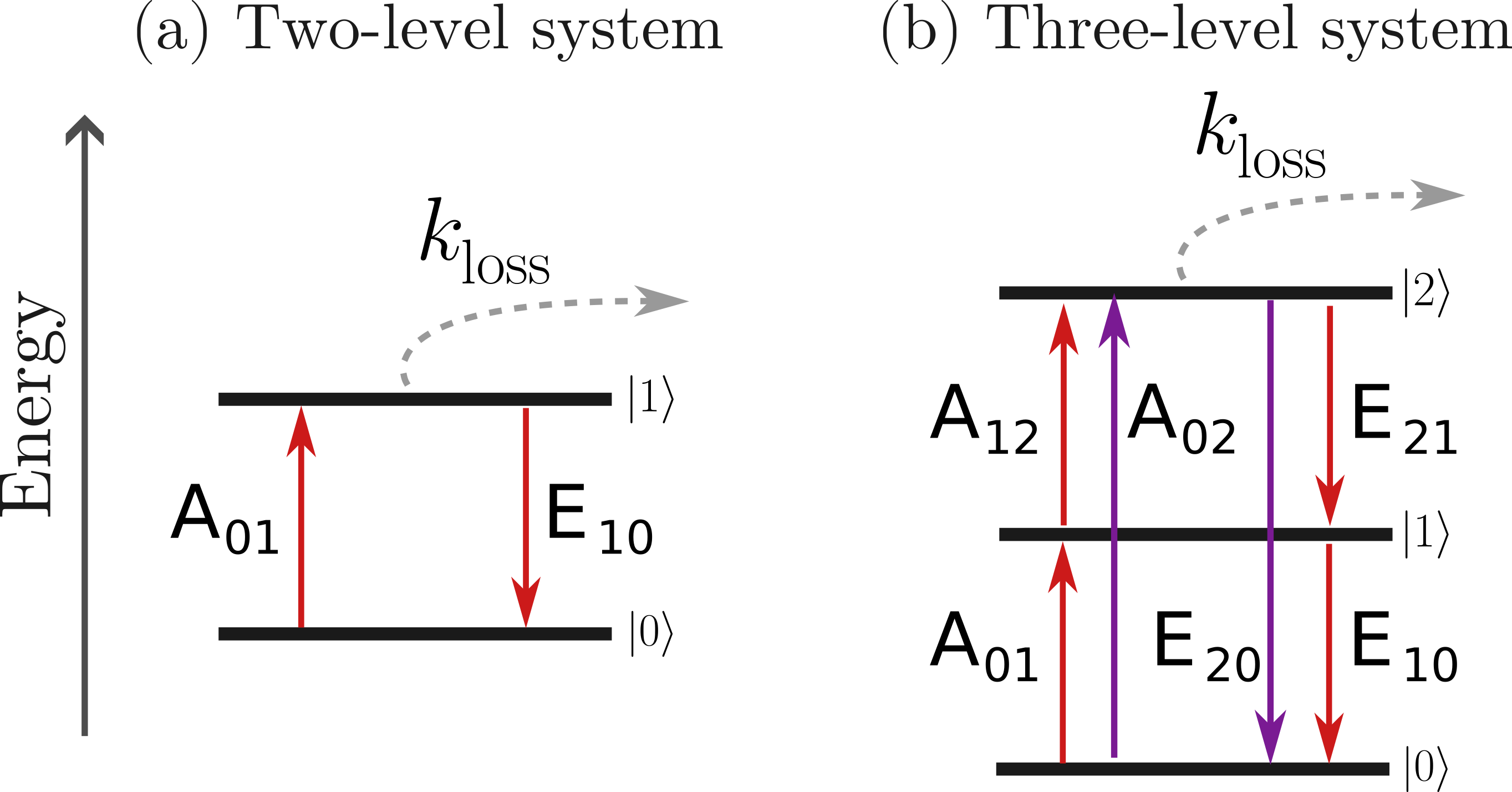}
    \caption{Minimal systems with two and three levels for BIRD rates are analytically examined.}
    \label{fig:modelsys}
\end{figure}

In this section, we examine a minimal BIRD system with very few degrees of freedom that affords an analytical examination of its temperature dependence. The smallest system that can be studied is a Morse oscillator with two bound states [Fig. \ref{fig:modelsys} (a)] with the transport matrix given by
\begin{align}
    \mathbf{J} = \left[\begin{array}{c c}
         A_{01} & -E_{10}  \\
         -A_{01} & k_\text{loss} 
    \end{array}\right]\;,
\end{align}
where $A_{ij}$ and $E_{ij}$ are the absorption of emission rates from $i$ to $j$, respectively. Since $k_\text{loss}$ is much greater than any other rates, the lowest eigenvalue, and therefore BIRD rate is $A_{01}$. Hence, the temperature dependence of the BIRD rate for this trivial case is given by the Bose-Einstein factor. A more interesting picture emerges when three energy levels are considered, represented in Fig. \ref{fig:modelsys} (b). The transport matrix in this case is written as
\begin{align}
    \mathbf{J} = \left[\begin{array}{c c c}
         A_{01}+A_{02} & -E_{10} & -E_{20}  \\
         -A_{01} & E_{10} + \alpha A_{12} & -\alpha E_{21} \\
         -A_{02} & -\alpha A_{12} & k_\text{loss} 
    \end{array}\right]\;, \label{transport3x3}
\end{align}
where the parameters $\alpha$ is introduced here to model the change due to polaritonic enhancement. In general, the polariton-modified density of states will affect multiple transitions if they are close enough to the polariton resonance $\omega_M$. However, for simplicity we will consider here that only one transition is significantly enhanced and its enhancement is quantified by $\alpha$. The characteristic polynomial of this matrix is
\begin{align}
    0 &= (A_{01}+A_{02}-\lambda)\cdot(E_{10}+\alpha A_{12}-\lambda)\cdot k_\text{loss} \\ 
    &-\alpha E_{10}E_{21}A_{01} - \alpha E_{20}A_{01}A_{12} \\
    &-\alpha^2(A_{01}+A_{02}-\lambda)E_{21}A_{12} - E_{10}A_{01}k_\text{loss} \\
    &-(E_{10}+\alpha A_{12}-\lambda)E_{20}A_{02} \;.
\end{align}
Dividing this equation by $k_\text{loss}$ and recognizing that $k_\text{loss} \gg A_{ij},\; E_{ij}\; \forall i,j$, we can discard any terms that do not contain $k_\text{loss}$. Thus, we obtain the quadratic equation
\begin{align}
    \lambda^2 - (A_{01}+A_{02}+E_{10}+\alpha A_{12})\lambda +(A_{01}+A_{02})(E_{10}+\alpha A_{12}) - E_{10}A_{01} = 0 \;.
\end{align}
Since all rates are positive and the lowest eigenvalue of the transport matrix must be positive, $\lambda$ can be computed as
\begin{align}
    \lambda = \frac{1}{2}\left( -b - |b|\sqrt{1 - \frac{4ac}{b^2}}\right) \;, \label{lambdafromquadeq}
\end{align}
where $a = 1.0$, $b = - (A_{01}+A_{02}+E_{10}+\alpha A_{12})$, and $c = (A_{01}+A_{02})(E_{10}+\alpha A_{12}) - E_{10}A_{01}$. We are going to mostly be concerned with the behavior of this equation at low temperatures where spontaneous emission is much faster than absorption of stimulated emissions. Hence, $\frac{4ac}{b^2} \ll 1$ and we can use $\sqrt{1-\frac{4ac}{b^2}} \approx 1 - \frac{2ac}{b^2}$. Eq. \ref{lambdafromquadeq} becomes
\begin{align}
    \lambda \approx \frac{ac}{b} = \frac{(A_{01}+A_{02})(E_{10}+\alpha A_{12}) - E_{10}A_{01}}{(A_{01}+A_{02}+E_{10}+\alpha A_{12})} \label{lambdabeforeT}
\end{align}

\subsubsubsection{Low Temperature Free-space limit}
In free-space there is no polaritonic enhancement, thus $\alpha = 1.0$. At low temperatures, spontaneous emission is the dominant process. This allow us to simplify the denominator of Eq. \ref{lambdabeforeT} as $A_{01}+A_{02}+E_{10}+\alpha A_{12} \approx E_{10}$. A similar simplification is applied to the numerator and we are left with
\begin{align}
    \lambda_{1} \approx A_{02} \;. \label{lambdalowTlowA}
\end{align}
This result can be rationalized as follows: any population in state $|1\rangle$ will be quickly depleted by spontaneous emission before it can move to the doorway state ($|2\rangle$), hence the overtone $0\rightarrow2$ is the only viable way to reach the final state that reacts rapidly before it can emit back to state $|1\rangle$, because $k_\text{loss} \gg E_{21}, \; E_{20}$. The validity of this approximation is shown numerically in Fig. X, where we see that the qualitative temperature dependence is well captured and at low enough temperatures there is also a good quantitative agreement. 

\begin{figure}
    \centering
    \includegraphics[width=0.5\linewidth]{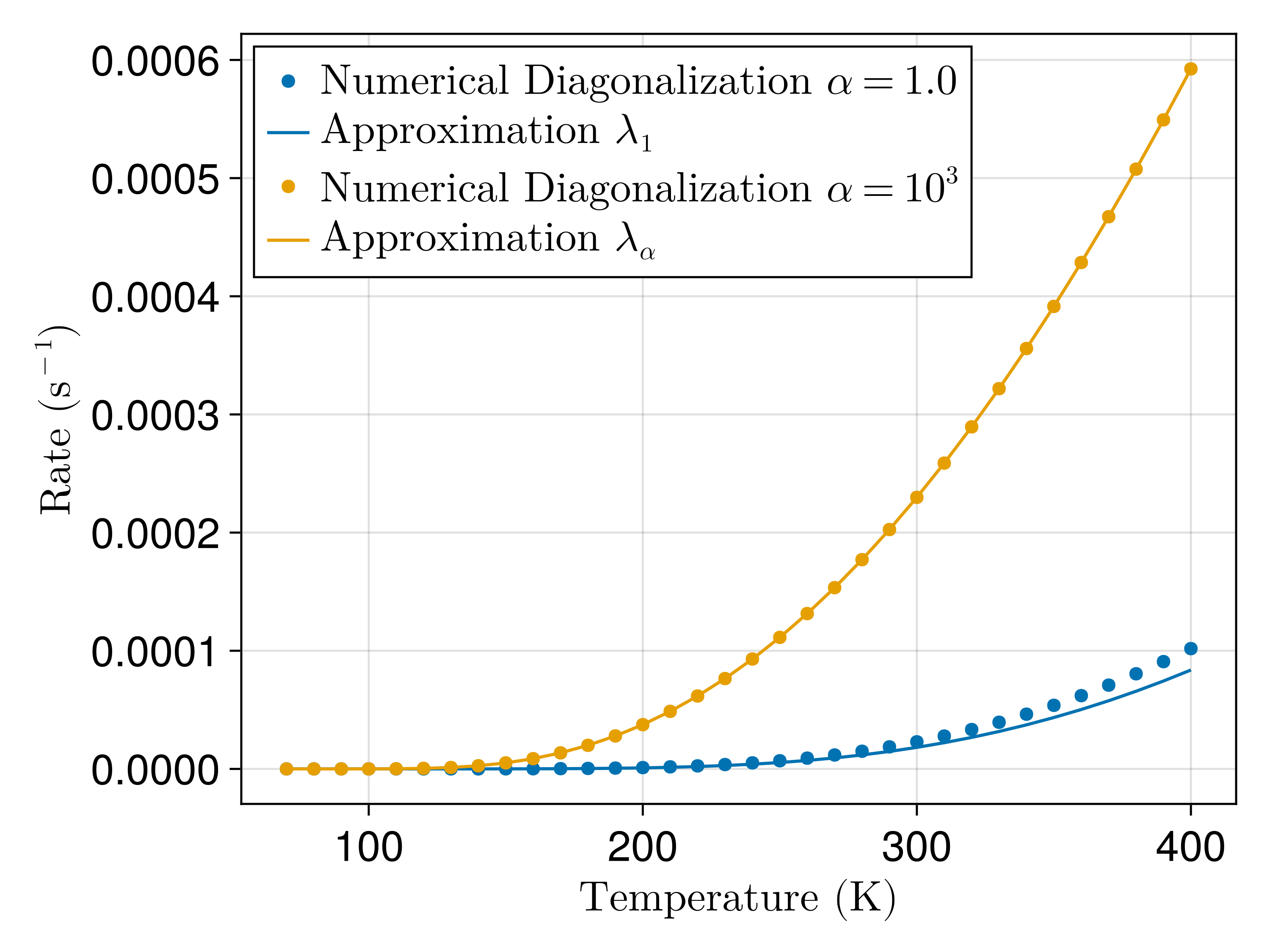}
    \caption{Comparison between BIRD rates obtained via numerical diagonalization and approximations derived in Eqs. \ref{lambdalowTlowA} and \ref{lambdalowThighA}.}
    \label{fig:3lsA}
\end{figure}

\subsubsubsection{Low Temperature Polaritonic limit}
In this scenario, we will examine the polaritonic case where $\alpha$ is a large number, e.g. $\alpha \approx 10^3$. In this case, we can apply the same simplifications used above, except that we also present any rates that multiply $\alpha$. For example, $A_{01}+A_{02}+E_{10}+\alpha A_{12} \approx E_{10}+\alpha A_{12}$. Using this strategy we obtain
\begin{align}
    \lambda_{\alpha} \approx \frac{\alpha(A_{01}A_{12}+A_{02}A_{12}) + E_{10}A_{02}}{E_{10}+\alpha A_{12}} \;. \label{lambdalowThighA}
\end{align}
This is an excellent approximation, as can be seen in Fig. \ref{fig:3lsA}. 

\begin{figure}
    \centering
    \includegraphics[width=0.8\linewidth]{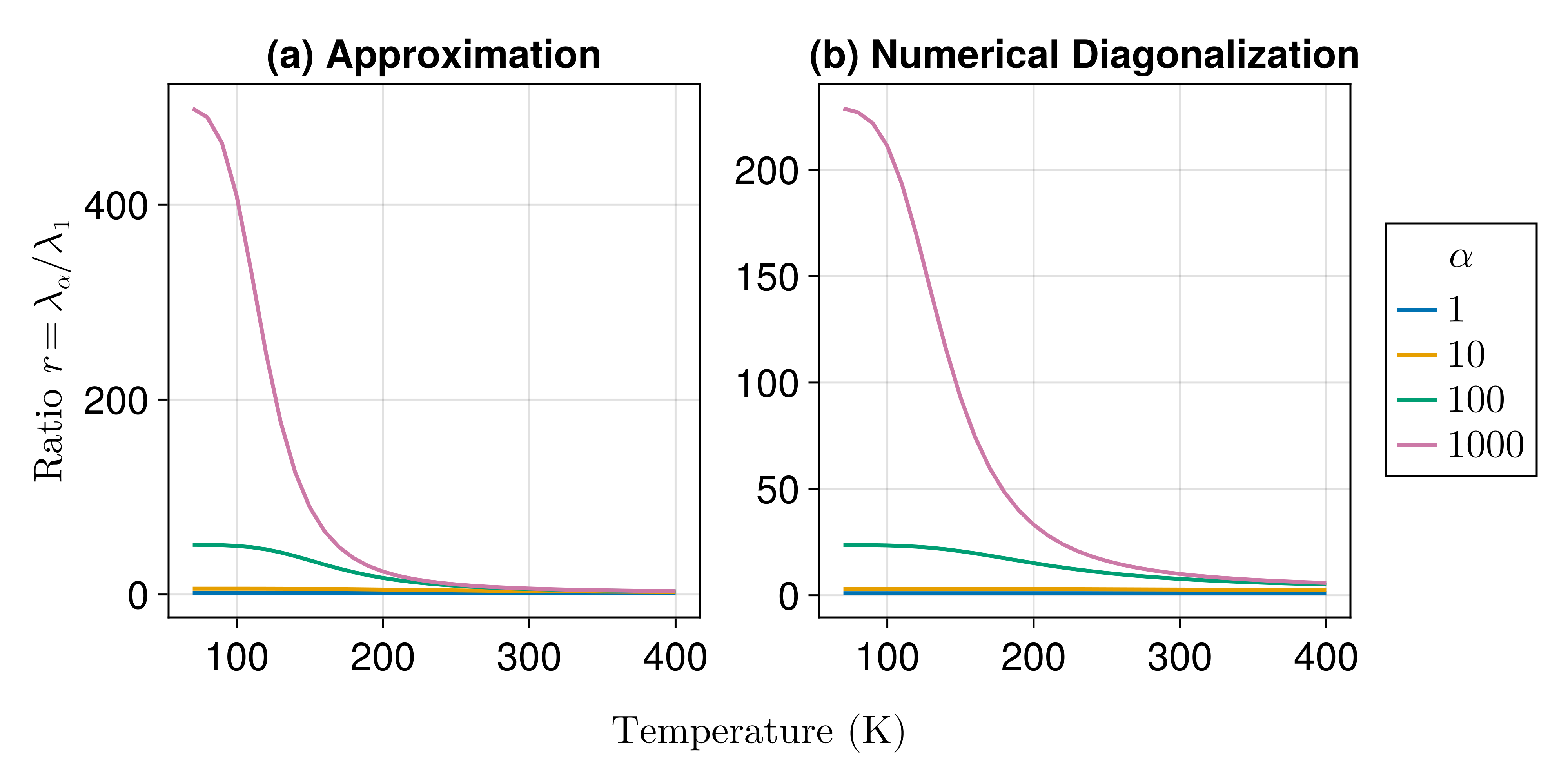}
    \caption{Polariton enhacement factors measured by the ration between rates with $\alpha \gg 1$ and $\alpha =1$ computing using (a) approximate expression given in Eq. \ref{finalr} and (b) numerical diagonalization of the transport matrix described in Eq. \ref{transport3x3} }
    \label{fig:e3lsB}
\end{figure}

\subsubsubsection{Temperature dependence of rate ratios}
Equipped with Eqs. \ref{lambdalowTlowA} and \ref{lambdalowThighA}, we can write an expression for the rates ratio
\begin{align}
    r = \frac{\lambda_\alpha}{\lambda_1} \approx \frac{\alpha(A_{01}A_{12}+A_{02}A_{12}) + E_{10}A_{02}}{A_{02}(E_{10}+\alpha A_{12})} \;.
\end{align}
Each rate, except for spontaneous emissions, depends on the temperature through the Bose-Einstein factor. That is,
\begin{align}
    A_{ij}(T) &=  \frac{W_{ij}}{e^{\beta\epsilon_{ij}}-1} \;,\\
    E_{ij}(T) &= W_{ij}\left(1+ \frac{1}{e^{\beta\epsilon_{ij}}-1}\right) \;,
\end{align}
where $\beta = \frac{1}{kT}$ and $W_{ij} = \frac{\omega_{ij} |\mu_{ij}|^2 \pi}{3\epsilon_0 \hbar} D(\omega_{ij})$. For our qualitative analysis, we will employ the low-temperature approximation $(e^{\beta\epsilon_{ij}}-1)^{-1} \approx e^{-\beta\epsilon_{ij}}$. Hence,
\begin{align}
    A_{ij}(T) &=  W_{ij}e^{-\beta\epsilon_{ij}} \;,\\
    E_{ij}(T) &= W_{ij}\left(1+ e^{-\beta\epsilon_{ij}}\right) \;, \label{emission}
\end{align}
where the $1$ in the parenthesis of Eq. \ref{emission} comes from the spontaneous emission part, which is temperature-independent. The polaritonic enhancement (i.e., the ratio of reaction rates) can be modeled as
\begin{align}
    r = \frac{\alpha W_{12}(W_{01}e^{-\beta\epsilon_{01}}e^{-\beta\epsilon_{12}}+W_{02}e^{-\beta\epsilon_{02}}e^{-\beta\epsilon_{12}}) + W_{02}W_{10}e^{-\beta\epsilon_{02}}(1+e^{-\beta\epsilon_{10}})}{W_{02}e^{-\beta\epsilon_{02}}(W_{10}+W_{10}e^{-\beta\epsilon_{10}}+W_{12}\alpha e^{-\beta\epsilon_{12}})} \;.
\end{align}
Dividing every terms by $e^{-\beta\epsilon_{02}}$ and recognizing that $e^{-\beta\epsilon_{01}}e^{-\beta\epsilon_{12}}e^{\beta\epsilon_{02}} = 1$ we get
\begin{align}
    r &= \frac{\alpha W_{01}W_{12} + W_{02}W_{10}(1 + e^{-\beta\epsilon_{10}}) + W_{12}W_{02}\alpha e^{-\beta\epsilon_{12}}}{W_{02}W_{10}(1 + e^{-\beta\epsilon_{10}}) + W_{12}W_{02}\alpha e^{-\beta\epsilon_{12}}} \\[2mm]
    &= 1 + \frac{\alpha W_{01} W_{12}}{W_{02}(W_{10} + W_{10}e^{-\beta\epsilon_{10}} + \alpha W_{02}e^{-\beta\epsilon_{12})}}
\end{align}
For simplicity, we make the qualitative approximation $W_{01}\approx W_{12} \approx W_{02}/2$ to get the final expression
\begin{align}
        r \approx 1 + \frac{\alpha}{2[1 + e^{-\beta\epsilon}(1 + 2\alpha)]} \;, \label{finalr}
\end{align}
with $\epsilon = \frac{1}{2}(\epsilon_{01}+\epsilon_{12})$.
As $T \rightarrow 0$, $r$ approaches $1+\alpha/2$. For large $a$ and large $T$, the expression approaches a constant value of $r = 1+\frac{1}{4}$. Note, however, that this is not a good approximation for the ratio at high-temperatures, since Eq. \ref{finalr} was derived for low temperatures where spontaneous emission is the dominant process. Nevertheless, this Eq. \ref{finalr} gives us a qualitative picture of how polaritonic enhancement changes when temperatures are decreased. This behavior is illustrated in Fig. \ref{fig:e3lsB}  where we verify that this expression sucessfully captures the qualititive temperature dependence of the reaction ratios. 

The model derived here predicts that polaritonic enhancement is maximized at low temperatures. At high enough temperatures or small enough values of $\alpha$ (such as in the weak coupling regime), Eq. \ref{finalr} will be fairly constant and rate ratios will be perceived as temperature independent.  While the proper bounds of polariton enhancement will depend on the exact energetics and transition dipole moments of the system, we expect that the temperature dependence to hold. In fact, a similar trend where $r$ increases rapidly at low enough temperatures is also observed for the LiH, HF, and NaF molecules. 

% Overtones discussion
\subsection{Rates without Overtones}
\begin{figure}
    \centering
    \includegraphics[width=0.8\linewidth]{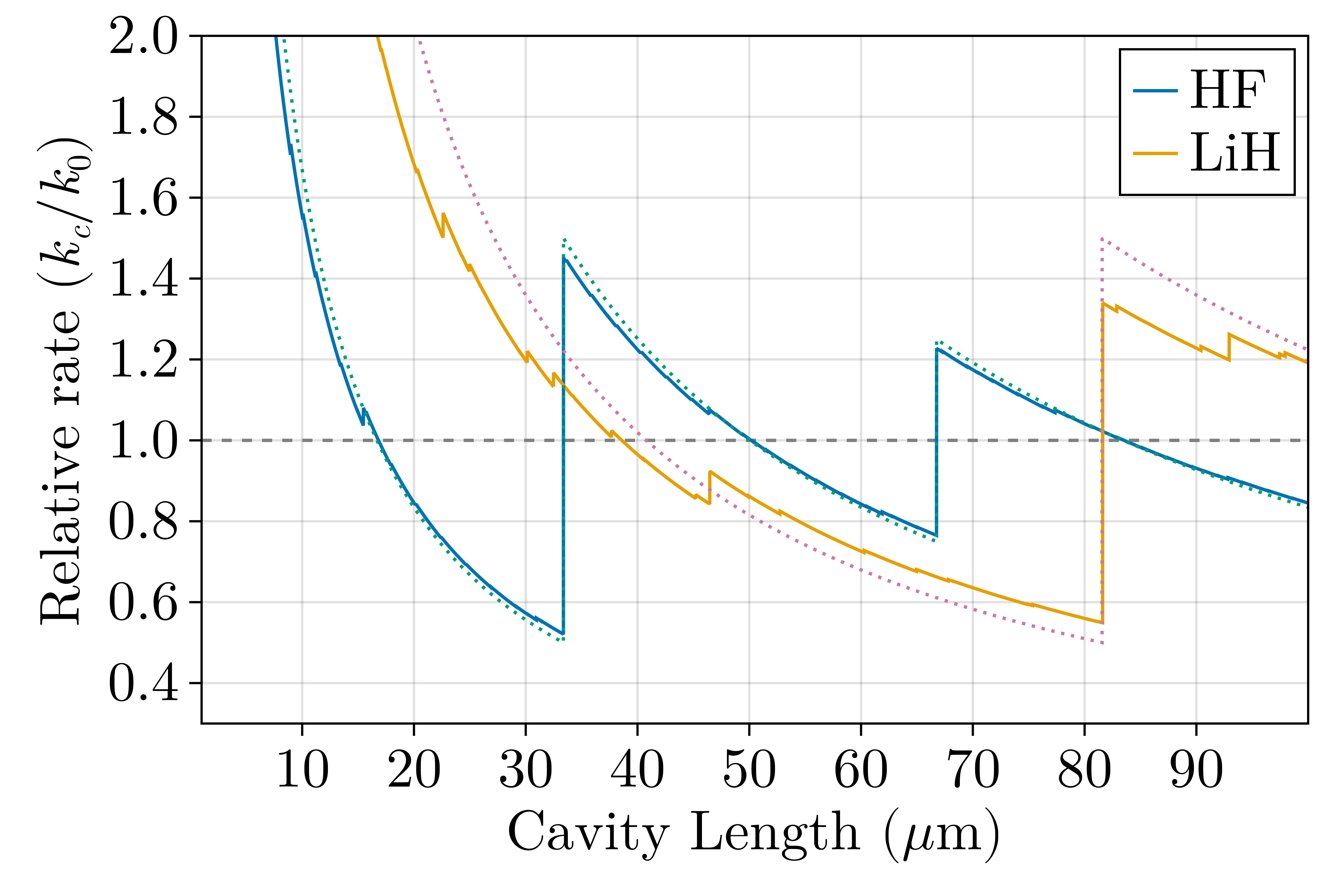}
    \caption{Ratio of BIRD rate, computed without overtones, inside a microcavity (weak coupling regime) to the free-space rate as a function of microcavity length for HF and LiH molecules. Dotted green and pink lines represent the relative density of states, $D_C(\omega)/D_0(\omega)$, at $\omega = 149.8$ cm$^{-1}$ and $\omega = 61.3$ cm$^{-1}$, respectively.}
    \label{fig:noovertones}
\end{figure}

The sensitivity analysis in Sec. 4 revealed that specific overtone transitions act as bottlenecks for the investigated diatomic BIRD from the most energetic bound state. In this section, we provide computations in the weak coupling regime that include only transitions between nearest energy levels $i \rightarrow i \pm 1$ and ignore the effects of overtones. The results are shown in Fig. \ref{fig:noovertones} and Table \ref{tab:noovert}, which show that overtones are essential for accurately capturing the BIRD kinetics inside and outside a microcavity.

Fig. \ref{fig:noovertones} shows relative rates for a range of microcavity lengths in the weak coupling regime. Note that Fig. \ref{fig:noovertones}
is similar to Fig. 2 of the main text, but only transitions between neighboring levels are allowed here. The results are qualitatively different. The complicated pattern seen in Fig. 2 is substituted by a much simpler function that can be fully understood by analyzing a single transition. A sensitivity analysis reveals that the most important transitions for HF and LiH are $22 \rightarrow 23$ and $28 \rightarrow 29$, respectively. These transitions correspond to $i_{\t{max}}-1 \rightarrow i_{\t{max}}$ and have very small frequencies, 149.8 for HF and 61.3 cm$^{-1}$ for LiH. 
Thus, oscillations in the relative rates ($k_c/k_0$) with changes in microcavity length occur over much larger periods (compare the axes of Fig. 2 and Fig. \ref{fig:noovertones}). Dotted green and pink lines in Fig. \ref{fig:noovertones} represent the relative microcavity DOS at 149.8 and 61.3 cm$^{-1}$, demonstrating the effect on the BIRD rate depends only on the photon DOS at the $i_{\t{max}}-1 \rightarrow i_{\t{max}}$ transition frequency. Finally, we also emphasize the quantitative effect of neglecting overtones. The free space rates ($k_0$) are reduced by three orders of magnitude when the model includes only fundamental transitions (Table \ref{tab:noovert}). 

\begin{table}
    \centering
    \setlength\extrarowheight{2mm}
    \begin{tabular}{|c|c|c|} \hline 
         & $k_0$ without overtones (s$^{-1}$)& $k_0$ with overtones (s$^{-1}$)\\ \hline 
         HF & $5.68 \times 10^{-9}$& $2.70 \times 10^{-6}$ \\ \hline 
         LiH&  $7.73 \times 10^{-8}$ & $1.64 \times 10^{-5}$\\ \hline
    \end{tabular}
    \caption{Computed free space BIRD rates ($k_0$) with and without overtone transitions at $T = 2000$ and $4000$ K for LiH and HF, respectively.}
    \label{tab:noovert}
\end{table}

\end{spacing}
\bibliography{lib}